\documentclass[12pt,a4paper]{article}   	

\usepackage[utf8]{inputenc}
\usepackage[UKenglish]{isodate}

\usepackage[a4paper,margin=2cm, marginparwidth=2cm]{geometry}
\usepackage{multirow}
\usepackage{amsmath,amssymb,amscd}

\usepackage{graphicx}
\graphicspath{{./figures/}{figures/}{./figures/SPT/}{figures/SPT/}}

\usepackage{caption}
\usepackage{rotating}

\newcommand{\tsedef}[1]{\textsc{#1}} 
\newcommand{\tseemph}[1]{\emph{#1}} 

\usepackage[colorlinks,allcolors=blue]{hyperref}

\providecommand{\href}[2]{{#2}}
\providecommand{\url}[1]{\texttt{#1}}

\newcommand{\bea}{\begin{eqnarray}}
\newcommand{\eea}{\end{eqnarray}}
\newcommand{\beq}{\begin{equation}}
\newcommand{\eeq}{\end{equation}}

\newcommand{\tref}[1]{(\ref{#1})}
\providecommand{\eqref}[1]{(\ref{#1})}
\renewcommand{\eqref}[1]{Eq.~(\ref{#1})}

\newcommand{\figref}[1]{Figure~\ref{#1}}

\newcommand{\tabref}[1]{Table~\ref{#1}}

\newcommand{\secref}[1]{Section~\ref{#1}}

\newcommand{\appref}[1]{Appendix~\ref{#1}}

\newcommand{\tsevec}[1]{\mathbf{#1}}
\newcommand{\tsemat}[1]{{\mathbf{\textsf{#1}}}}

\newcommand{\Bmat}{\tsemat{B}}

\newcommand{\uvec}{\tsevec{u}}
\newcommand{\vvec}{\tsevec{v}}
\newcommand{\wvec}{\tsevec{w}}

\newcommand{\Ecal}{\mathcal{E}}
\newcommand{\Gcal}{\mathcal{G}}

\newcommand{\Tcal}{\mathcal{T}}
\newcommand{\Vcal}{\mathcal{V}}

\usepackage[dvipsnames]{xcolor} 

\newcommand{\tnote}[1]{} 
\newcommand{\tcomment}[1]{} 

\newcommand{\zbar}{\bar{z}}
\newcommand{\zbarfit}{\bar{z}^{\mathrm{(fit)}}}
\newcommand{\zbarnum}{\bar{z}^{\mathrm{(num)}}}
\newcommand{\zbarrnd}{\bar{z}^{\mathrm{(rnd)}}}
\newcommand{\betafit}{\beta^{\mathrm{(fit)}}}
\newcommand{\betanum}{\beta^{\mathrm{(num)}}}
\newcommand{\betarnd}{\beta^{\mathrm{(rnd)}}}

\newcommand{\ellmax}{L}
\newcommand{\ellav}{\texpect{\ell}}
\newcommand{\texpect}[1]{\langle #1 \rangle}
\newcommand{\kav}{\texpect{k}}

\newcommand{\chisqred}{\chi^2_\mathrm{r}}

\begin{document}


\renewcommand{\thefootnote}{\fnsymbol{footnote}}
\begin{center}
{\Large\textbf{Linking the Network Centrality Measures}}\\
{\Large\textbf{Closeness and Degree}} \\[\baselineskip]
{\large \href{https://orcid.org/0000-0003-3501-6486}{Tim S.\ Evans\footnote{Corresponding author. \texttt{T.Evans@imperial.ac.uk}}}}, 
{\large \href{https://orcid.org/0000-0003-3595-3285}{Bingsheng Chen}}
\\[0.5\baselineskip]
\href{http://complexity.org.uk/}{Centre for Complexity Science}, and \href{http://www3.imperial.ac.uk/theoreticalphysics}{Theoretical Physics Group},
\\
Imperial College London, London, SW7 2AZ, U.K.
\\
13th June 2022
\\
Published as \emph{Communications Physics} \textbf{5} 172 (2022), 
DOI: \href{http://doi.org/10.1038/s42005-022-00949-5}{\texttt{10.1038/s42005-022-00949-5}}
\end{center}
\renewcommand{\thefootnote}{\alph{footnote}}

\bigskip

\renewcommand{\abstractname}{Abstract}
\begin{abstract}
Measuring the importance of nodes in a network with a centrality measure is a core task in any network application. 
There are many measures available and it is speculated that many encode similar information.
We give an explicit non-linear relationship between two of the most popular measures of node centrality: degree and closeness. 
Based on a shortest-path tree approximation, we give an analytic derivation that shows the inverse of closeness is linearly dependent on the logarithm of degree.
We show that our hypothesis works well for a range of networks produced from stochastic network models and for networks derived from 130 real-world data sets.
We connect our results with previous results for other network distance scales such as average distance.
Our results imply that measuring closeness is broadly redundant unless our relationship is used to remove the dependence on degree from closeness. The success of our relationship suggests that most networks can be approximated by shortest-path spanning trees which are all statistically similar two or more steps away from their root nodes.
\end{abstract}



\setcounter{footnote}{0}
\renewcommand{\thefootnote}{\arabic{footnote}}

\section*{Introduction}

Network science has proved to be an exceptionally useful tool, especially with the large scale data sets now available in every discipline.
Many systems are driven by pairwise interactions at the microscopic level and the network formalism is ideal as
it represents these bilateral relationships as edges between nodes.
The real power of network analysis is that it gives us tools to look at the emergent behaviour such system even when driven by interaction's on meso- and  macro-scopic scales \cite{BRMW13}.

A good example is the key task of finding the most important nodes in the system.
In network science this is done by using a centrality measure which assigns
nodes a centrality value with larger values indicating greater importance. Centrality has been developed over the last seventy years \cite{B50,S66,F78,HH95,WS03,HK04,KS08,KB08,YD09,LFH10,NSJ11,WMWJ11,BH14,DSP18} and is a core part of many introductory texts on Network Science \cite{WF94,N10,LNR17,C21}
The simplest centrality measure is degree, the number of neighbours each node has, probably the first node property examined in any study.
However, the degree is insensitive to the wider network structure which is the primary goal of network analysis.

In order to probe the broader network structure, many centrality measures are based on the distance between nodes defined as the length of the shortest path between nodes. For instance, our focus is on one of the oldest and most widely used centrality measures, closeness \cite{B50}, the inverse of the average distance from a node to all others. Closeness is an important indicator in many different contexts.
It has been used to investigate the role of different academic disciplines when academics choose their journals and bibliographies \cite{NSJ11}.
Closeness centrality measures the impact of an author on a field and their social capital \cite{YD09}. When used to select potential leads in customer data, closeness led to a significant gain in the success rate \cite{KB08}.
In an air transport network, it has been shown that the closeness of a city is highly correlated with socio-economic indicators such as gross regional domestic product \cite{WMWJ11}. Closeness has been applied to biological networks \cite{WS03} and closeness measures were able to identify more than 50\% of the global regulators within the top 2\% of the ranked genes \cite{KS08}. Essential genes have been found to have higher closeness than nonessential genes in protein-interaction networks \cite{HK04}. In a metabolic network the closeness of nodes can identify the most important metabolites \cite{MZ03a}. Centrality measures such as closeness are an important tool to analyse many types of data.

However, there are a vast number of centrality indices available, as visualised nicely by Schoch \cite{S15a,S16}, suggesting that there is a lot of redundancy.
Many different centrality measures encode similar information as seen in the strong correlations between centrality indices \cite{B88,RPWDMK95,F97,L06b,VCLC08,BN14,LLBM15,SVB17,LNR17,OFPASF19,BEEKSWWS19,APB20}. In particular Pearson correlation coefficients are invariably used which are most sensitive to linear correlations between centrality measures. That is we measure $\rho_{cd} = \texpect{(c_u-\bar{c})(d_u-\bar{d})}/(\sigma_c\sigma_d)$ where $c_u$ and $d_u$ are two centrality measures for node $u$, the average is over all nodes with the means ($\bar{c}$, $\bar{d}$) and standard deviations ($\sigma_c$, $\sigma_d$) found using the same ensemble average.
There seems to be no clear consensus from these studies other than there are often strong relationships between centrality measures but these vary from network to network. For instance, in the introduction of Schoch et al.\ \cite{SVB17} it is suggested that ``Reported results, however, are often inconsistent with regard to the similarity of centrality indices''. Of particular interest here is the conjecture made in the Introduction section of Valente et al. \cite{VCLC08} where it is stated that ``We expect that measures of degree and closeness centrality will be more highly correlated with each other than with other measures, because they are both based on direct ties.'' Later, in the discussion of results by Valente et al. \cite{VCLC08}, the authors conclude that ``The amount of correlation between degree, betweenness, closeness, and eigenvector indicates that these measures are distinct, yet conceptually related'' and the closeness-degree pair is only the third most correlated pair of centrality indices in their study.

In this paper we will focus on the relationship between closeness centrality and degree. Our result is that closeness centrality and degree have a non-linear relationship, namely that the inverse of closeness  (`farness' \cite{OAS10}, a normalised version of `status' \cite{H59,HH95,WS03}) is linearly dependent on the logarithm of the degree.
This explains why linear correlation measures often link degree and closeness centrality but at the same time no general pattern has been seen before. Equally our results suggest that studies based on linear correlation measures may well miss important other features in the landscape of centrality measures.

This non-linear relationship means that measuring closeness gives the same broad information as closeness making the computationally expensive calculation of the latter redundant to a first approximation. To extract useful information from closeness the dependence on degree must first be removed and only then will closeness reveal important and distinct information on individual nodes.

Our starting point is that the shortest paths from any one node to all other nodes can be arranged as a spanning tree. We then conjecture that the branches of this tree are statistically similar, implying that the closeness of a node can only depend on the number of such branches, i.e.\ the degree of the node. If we also assume that the number of nodes in each branch grows exponentially, we find that the inverse of closeness is linearly dependent on the logarithm of degree centrality. The success of this non-linear relationship in a wide range of real-world networks suggests that both these conjectures are broadly true, an insight that can be exploited in a much wide range of network studies.

\section*{Results and Discussion}
\subsection*{Theory}

\subsubsection*{General Definitions}

For simplicity, we will assume throughout this paper that we are analysing a simple graph $\Gcal$ with just one component. We will denote the degree of each node $v$ as $k_v$.
A path in a network of length $\ell$ is a sequence of $(\ell+1)$ distinct nodes
such that each consecutive pair of nodes in the path is connected by an edge.
We will define the distance between two nodes $u$ and $v$ in a network to be the length of a shortest path between two nodes, denoted here as $d_{uv}$.

The \tsedef{closeness} $c_v$ \cite{B50,WF94,N10,LNR17,C21} of a vertex $v$ is then defined to be the inverse of the average distance from $v$ to every other vertex in the graph, so
\beq
 \frac{1}{c_v} =    \frac{1}{(N-1)}\sum_{u \in \Vcal \setminus v} d_{uv}
 \label{e:closenessdef}
\eeq
where $\Vcal$ is the set of nodes and $N=|\Vcal|$ is the number of nodes.
Clearly the closer a vertex $v$ is to other vertices in the network, the larger the closeness. Thus closeness mimics the properties of points in a geometric shape where those points closest to the geometric centre will have highest closeness.

Trees \cite{WF94,N10,LNR17,C21}  are connected networks with no loops so the number of edges is always one less than the number of nodes.
Here we use a \tsedef{spanning tree} \cite{S21b} which is a connected subgraph of the original graph $\Gcal$  containing all the original vertices $\Vcal$ but a subset of $(N-1)$ edges that are just sufficient to keep every node connected to all others. We are also going to work with \tsedef{rooted trees} $\Tcal(r)$ in which we have singled out one special node, the \tsedef{root} $r$ of the tree.

\subsubsection*{Estimate of Closeness}\label{s:closenessrand}

We start from the idea that some of the statistical properties of real-world networks may be captured by spanning trees \cite{S21b}.
Here we are interested in closeness which uses the lengths of shortest paths between nodes so the most useful trees for this work are the \href{https://en.wikipedia.org/wiki/Shortest-path_tree}{\tsedef{shortest-path trees}}, $\Tcal(r)$, that contain one shortest path from a root node $r$ to each remaining node in the network.
As our networks are unweighted, the shortest-path trees always exist and are easily defined as part of a \href{https://en.wikipedia.org/wiki/Breadth-first_search}{breadth-first search} algorithm,
see the Supplementary Note 2 on the Shortest-Path Tree Algorithm for more details.
Every node can act as a root node so there is at least one shortest-path tree, $\Tcal(r)$, for every node $r$.
These trees are not unique as there can be many shortest paths between a pair of nodes.

\begin{figure}[htb]
	\begin{center}
		\includegraphics[width=0.65\linewidth]{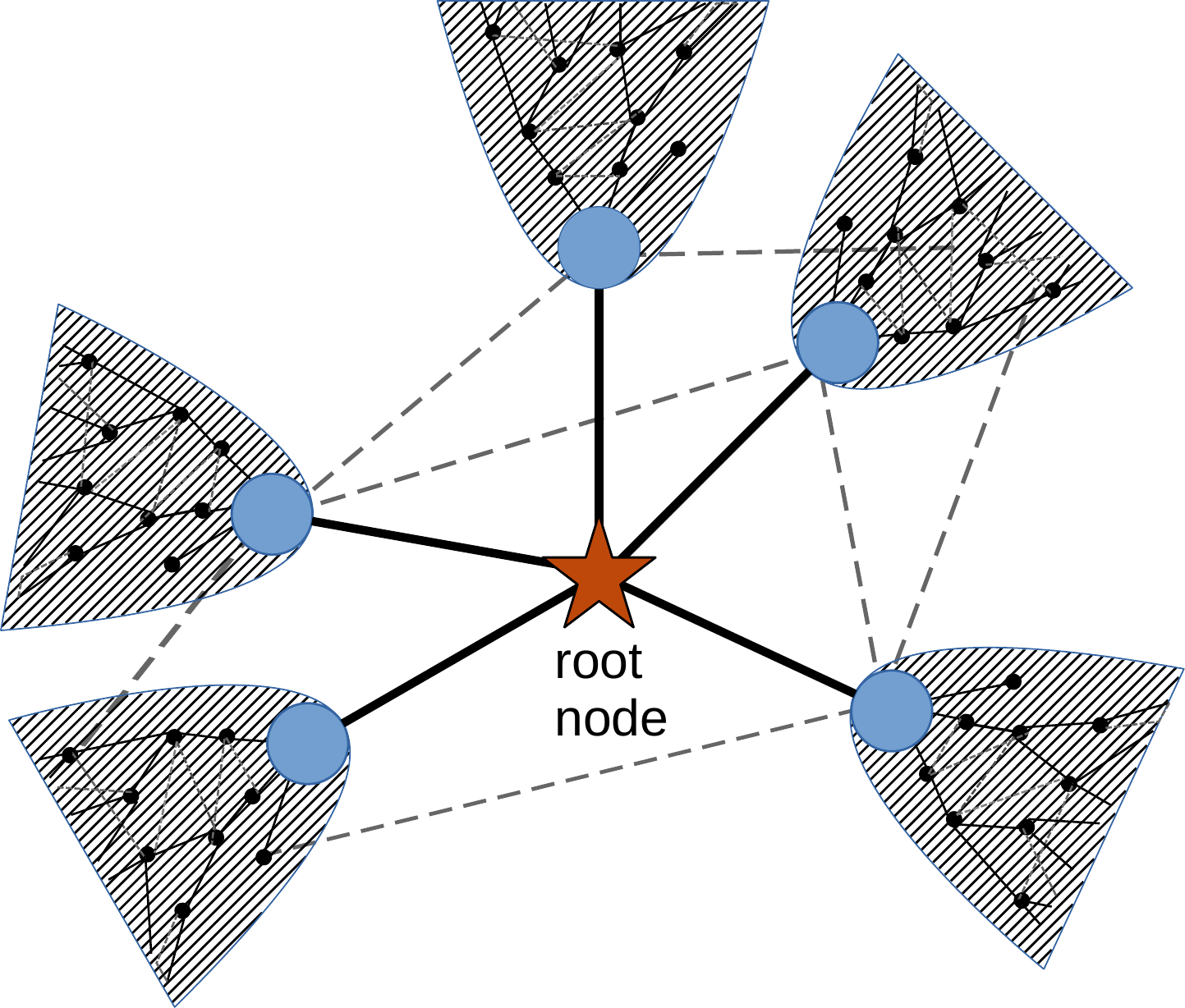}
	\end{center}
	\caption{\textbf{An illustration of the shortest-path tree approximation.}
Every node $r$, here the red star, is considered to be the root node of a shortest-path tree $\Tcal(r)$. This has $k_r$ nearest neighbours, as indicated by the five solid black lines. Each of the neighbouring nodes, here the blue circles, is treated as the root of a branch of the shortest-path tree. These branches are treated as being statistically identical with a branching number $(1+\zbar)$ as indicated here through the use of the same shaded shape rooted on each neighbouring node. The grey dashed lines represent some of the many edges in the graph $\Gcal$ that are not included in the shortest-path tree.}
	\label{f:sptapprox}
\end{figure}

Our picture for these shortest-path trees is shown in \figref{f:sptapprox}. We start with the observation that close to the root node the structure of these shortest-path trees will vary and in particular, the number of nearest neighbours $k_r$ of the root vertex $r$ will vary.
However, as we move further away from the root node, the number of nodes $n_r(\ell)$ at some distance $\ell$ from root node $r$ grows exponentially with each step in most networks.
This is the origin of the small-world effect seen in many networks, where the distance between nodes is typically much smaller than is found in similar size networks that are constrained by Euclidean geometry, such as a regular square grid of streets or a random geometric graph. Regardless of the local context of a root node, the shortest-path trees quickly access a similar set of nodes in the main bulk of the network provided there is no large scale inhomogeneity in the network. Thus we conjecture that the structure and statistical properties of these trees away from the root node are similar for all possible root nodes. The contribution to closeness of each node in the bulk is bigger as they are further from the root and more numerous. So we expect that the largest contributions to closeness always come from the same bulk regions where we can expect statistical similarity.

The most important difference when comparing different root nodes therefore comes from the local structure of each root node.  In particular, the initial value for the exponential growth in the number of nodes at distance $\ell$ from the root will have a large effect. This will depend on the local structure and the simplest effect comes from the number of immediate neighbours the root node has, i.e.\ the degree of the root node $k_r$. The simplest approximation for the growth of these shortest-path trees is therefore $n_r(\ell)=k_r \zbar^\ell$, where $\zbar$ is some measure of the rate of exponential growth of the shortest-path tree. Note that our assumption of statistical similarity suggests that the branching factor of these trees is, on average, the same so we use a single parameter $\zbar$ to represent the exponential growth from any root node $r$.

Our crude approximation is clearly going wrong when we look at large distances from the root as eventually any real network will run out of vertices so $n(\ell)=0$ for large $\ell$. One can model the end of the exponential growth in different ways but we will use the simplest.  Namely, we will define a sharp upper cutoff ${\ellmax}_r$ and assume that $n_r(\ell)=0$ for $\ell > {\ellmax}_r$ which gives
\beq
 N =  1 + \sum_{\ell=1}^{\ellmax_r}  k_r \,\zbar^{\ell-1}  = 1 + k_r \frac{(\zbar^{\ellmax_r}-1)}{(\zbar-1)}\, .
\label{e:Nfromn2}
\eeq
We can invert this expression to express this cutoff ${\ellmax}_r$ in terms of parameters $N$ and $k_r$, giving us ${\ellmax}_r={\ellmax}(N,k_r)$ where for large $N$
\beq
 {\ellmax}(N,k)\approx \frac{\ln\left(N(\zbar-1)/k \right) }{\ln(\zbar)} \, .
 \label{e:lmaxexp}
\eeq
Individual distances are integers but it is clear from the form in \eqref{e:lmaxexp} that we need $\ellmax$ to be a real number; in some sense $\ellmax$ is an average over the actual distances from the root to the leaves (nodes with degree one) of the tree. This also tells us that the distance scale associated with a node depends on the logarithm of that node's degree. Since the inverse closeness is a sum of distances, this distance scale $\ellmax$ controls the result and that explains why $\ln(k)$ appears in our expression for inverse closeness. As an aside, the $\ln(N)$ dependence of this distance scale $\ellmax$ in \eqref{e:lmaxexp} reflects the small-world effect seen in most networks.

We now have that $n_r(\ell) = k_r \zbar^{\ell-1}$ for $\ell< \ellmax(N,k_r)$ and zero for larger $\ell$ which depends on local parameters $k_r$, the degree of each root node, and two global parameters, the total number of nodes $N$ and some measure of the growth rate of the shortest-path trees $\zbar$.
We can now rewrite the closeness $c_r$ \eqref{e:closenessdef} of a vertex $r$ in terms of $n_r(\ell)$ as $1/{c_r}= {(N-1)}^{-1} \sum_{\ell=1}^{L_r} \ell n_\ell $ to find that
\bea
\frac{1}{c_r}
&=&
\frac{1}{(N-1)} \sum_{\ell=1}^{\ellmax_r}  \ell k_r \zbar^{\ell-1}
\label{e:fcalc1}
=
\frac{k}{(N-1)}
\left(
\frac{(\ellmax_r +1)\zbar^{\ellmax_r}}{\zbar-1}
-
\frac{(\zbar^{\ellmax_r + 1}-1)}{(\zbar-1)^2}
\right)
\eea
By using \eqref{e:Nfromn2} and \eqref{e:lmaxexp} we can eliminate $\ellmax_r$ to find that
\beq
\frac{1}{c_r}  = -\frac{1}{\ln(\zbar)}\ln(k_r) + \beta \, .
\label{e:farnesskform}
\eeq
Our calculation shows the parameter $\beta$ is also independent of the root vertex $r$ chosen but it is a function of the global network parameters $N$ and $\zbar$ so that $\beta=\beta(\zbar,N)$ where
\beq
\beta(\zbar,N)
=
\left( \frac{1}{(\zbar-1)}
+ \frac{\ln(\zbar-1)}{\ln(\zbar)} \right)
+ \frac{1}{\ln(\zbar)} \ln(N ) \, .
\label{e:betadef}
\eeq
Our prediction is that the inverse of closeness $c_r$ of any node $r$ should show a linear dependence on the logarithm of the degree $k_r$ of that node with a slope that is the inverse of the log of the growth parameter $\zbar$.

In our analysis we will not assume the parameter $\beta$ is given by \eqref{e:betadef}. By adding one additional fitted parameter we lose a little predictive power since many parameters are needed to characterise a network. This leaves us with a conjecture based on the number of nodes $N$ and degree of each nodes $k_r$ which are usually known. Then in principle we have two unknown global parameter values which we find from a linear fit to our data for $c_r$ and $k_r$ giving $\zbarfit$ and $\betafit$.

\subsection*{Numerical Results for Theoretical Models}\label{s:models}

We looked at the relationship between closeness and degree using simple networks produced from three different theoretical models \cite{N10,C21,LNR17}:
the Erd\H{o}s-R\'{e}yni (ER) model \cite{ER59}
the Barab\'{a}si-Albert model with pure preferential attachment \cite{BA99}
and the configuration model \cite{MR95}
network starting from a network generated with the same Barab\'{a}si-Albert model.
In the first and third model, the edges are completely randomised so there are no vertex-vertex correlations.  The last two models both have fat-tailed degree distributions.
Our networks built from artificial models were created using standard methods in the \texttt{networkx} package \cite{HSS08}.

\begin{figure}[htb!]
	\begin{center}
		\includegraphics[width=1\linewidth]{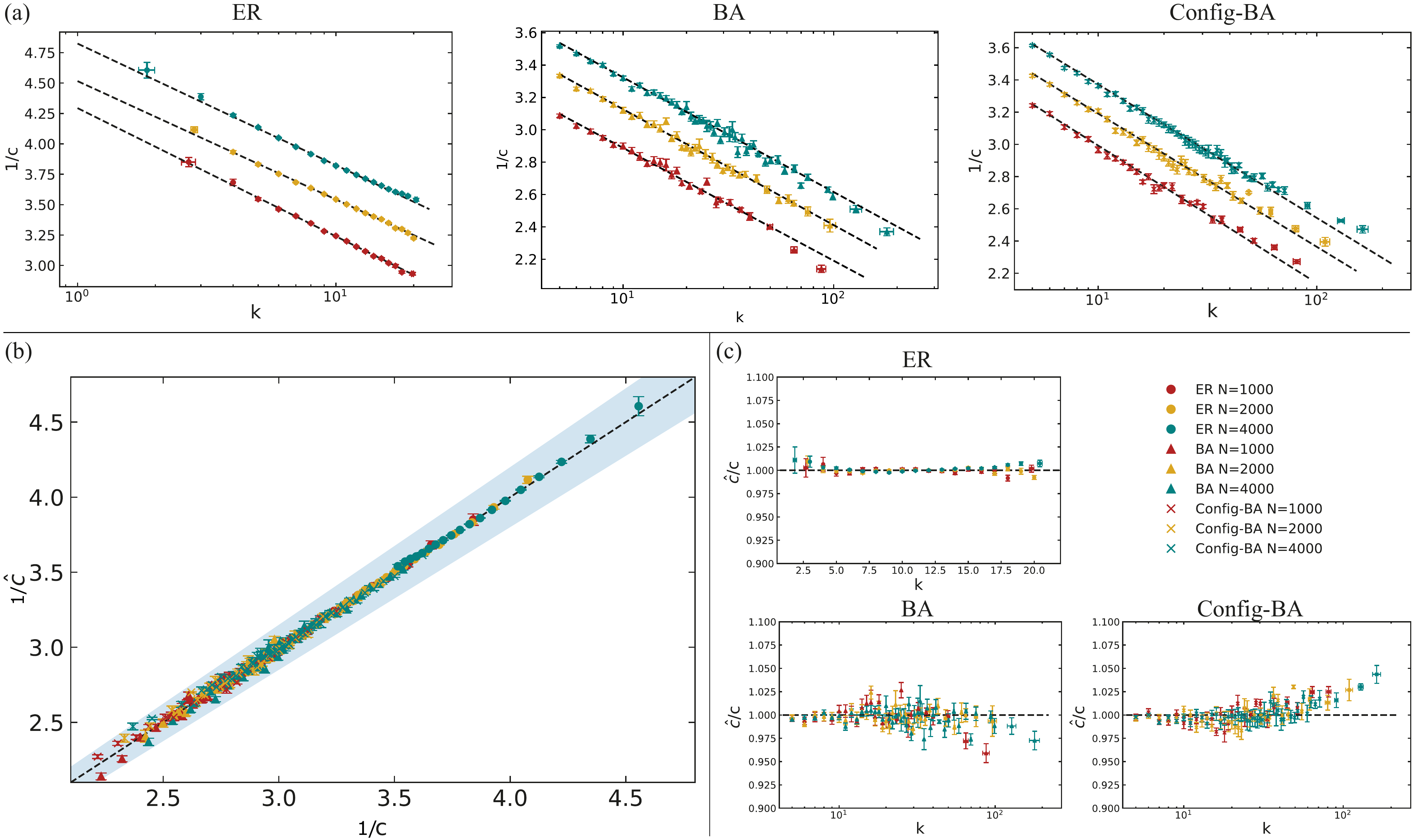}
	\end{center}
	\caption{\textbf{Closeness and degree for artificial networks.}
In panel (a), each plot shows results for networks formed from one artificial model: the Erd\H{o}s-R\'{e}yni (ER) model, the Barab\'{a}si-Albert (BA) model, and the configuration model network starting from a Barab\'{a}si-Albert model (Config-BA). The dashed lines shows the best linear fit of $1/c$ to $\ln(k)$ using \eqref{e:farnesskform}.
The same data from all nine artificial networks is shown in the scatter plot in panel (b) with data $1/\hat{c}$ against predicted value $c$ obtained from the best fit \eqref{e:farnesskform} and the shaded region corresponds to a 5\% deviation from the theoretical prediction.
Panel (c) shows the fractional error, the fitted value of closeness divided by data value. The results are for three different sized networks: $N=1000$ (red points) $N=2000$ (blue points) and $N=4000$ (yellow points) where $N$ is the number of nodes. All networks have average degree $10.0$ and 100 realisations were taken for each case. The values of closeness for each value of degree are binned, the mean is shown as the data point with error bars the standard error of the mean. The results show that the non-linear correlation of closeness and degree predicted in \eqref{e:farnesskform} works most of the time within a 2\% variation. There are some hints of small but systematic at higher degree value  but the data is sparse and less reliable here.
    }
	\label{f:fkERmodel}
\end{figure}

For a single network, we get several nodes with the same degree and we use this variation to find a mean and standard error in the mean shown. The fit is done using  \eqref{e:farnesskform} with two free parameters $\zbarfit$ and $\betafit$ and the goodness of fit measures in \tabref{t:modelresults} show this is a good fit, confirmed visually by the plots in \figref{f:fkERmodel}a.
Note that a good linear fit of status ($(N-1)/c$) to the logarithm of degree for one example of each of the  Erd\H{o}s-R\'{e}yni and Barab\'{a}si-Albert models was found in Fig.~1 of Wuchty \& Stadler \cite{WS03}.

Roughly speaking we find that mean inverse closeness values for any one degree are typically within 2\% of the prediction made from the best fit as we can see in Results are shown in \figref{f:fkERmodel}b. The small deviations seen in \figref{f:fkERmodel}c, especially for the Config-BA model, are for higher degree values where the data is sparse and uncertainties are large. This means no firm conclusions can be drawn from \figref{f:fkERmodel}c about the presence of higher order corrections to our form \eqref{e:farnesskform}.

\begin{table}[htb!]
	\begin{tabular}{cc|cc|ccccc}
		\hline
		Network type               & N    & $1/\ln(\zbarfit)$ & $\betafit$      & $\zbarfit$      & $\beta(\zbarfit,N)$ & $\rho(c,k)$ & $\chisqred$ \\ \hline
		\multirow{3}{*}{ER}        & 1000 & $0.46\pm 0.01$    & $4.29\pm 0.01$  &  $8.87\pm 0.20$ & $4.23\pm 0.03$                & 0.94  & 1.02       \\
		& 2000 & $0.42\pm 0.01$    & $4.52\pm 0.01$  &  $10.64\pm 0.18$ & $4.28\pm 0.03$                & 0.93  & 1.02       \\
		& 4000 & $0.43\pm 0.01$    & $4.82\pm 0.01$  &  $9.99\pm 0.12$ & $4.67\pm 0.02$                & 0.93  & 1.03       \\ \hline
		\multirow{3}{*}{BA}        & 1000 & $0.30\pm 0.01$    & $3.59\pm 0.02$  & $28.03\pm 3.11$ & $3.09\pm 0.13 $               & 0.75  & 1.16       \\
		& 2000 & $0.32\pm 0.01$    & $3.86\pm 0.01$  & $22.76\pm 2.22$ & $3.46\pm 0.07$                & 0.70  & 1.29       \\
		& 4000 & $0.31\pm 0.01$    & $4.03\pm 0.01$  & $25.17\pm 2.61$ & $3.60\pm 0.08$                & 0.65  & 1.16        \\ \hline
		\multirow{3}{*}{Config-BA} & 1000 & $0.35\pm 0.01$    & $3.76\pm 0.02$  & $17.41\pm 1.42$ & $3.46\pm 0.14$                & 0.75  & 1.19       \\
		& 2000 & $0.36\pm 0.01$    & $4.01\pm 0.02$  & $16.08\pm 1.24$ & $3.78\pm 0.15 $               & 0.70  & 1.28       \\
		& 4000 & $0.35\pm 0.01$    & $4.19\pm 0.01$  & $17.41\pm 1.41$ & $3.94\pm 0.08$                & 0.66  & 1.19       \\ \hline
	\end{tabular}
	\caption{\textbf{Results for artificial networks.}
Table of results for one example of a simple graph with average degree $10.0$ produced using one of three artificial models with the same average degree $\texpect{k}=10.0$ but with a different number of nodes, $N$. Each `ER' network is a standard Erd\H{o}s-R\'{e}yni network, a `BA' network is produced using pure preferential attachment in the Barab\'{a}si-Albert model, and the `Config-BA' network is a configuration model version of a Barab\'{a}si-Albert model network. The results for $1/\ln(\zbarfit)$ and $\betafit$ come from linear fits of inverse closeness, $1/c_v$ to the logarithm of degree, $\ln(k_v)$ for each vertex $v$,
		i.e.\ $1/c_v = (\ln(\zbar))^{-1}\ln(k_v) + \beta$ \eqref{e:farnesskform}. The value of $\beta$ derived from $\zbarfit$ and $N$ using \eqref{e:betadef} is also shown for comparison. The fits are very good as indicated by the column for the reduced chi-square $\chisqred$.
The Pearson correlation measure between closeness and degree $\rho(c,k)$ is given for comparison with earlier work.
	}
	\label{t:modelresults}
\end{table}

We now turn to look at the actual values obtained from these fits of data on closeness and degree from the artificial networks to \eqref{e:farnesskform}. As \tabref{t:modelresults} shows there is a small amount of variation in value of $\zbarfit$, the fit for the shortest-path tree growth factor, with the size of the network. What is of more interest are the differences in values between these three types of artificial networks.
All these networks had an average degree of about $10.0$ and an infinite tree with constant degree $10$ (a Bethe lattice) would have a growth factor $\zbar=9$, one less than the average degree. So the best fit values for the growth factor $\zbar$ in the Erd\H{o}s-R\'{e}yni networks are a little higher than this while the Barab\'{a}si-Albert network and its randomised version are a lot bigger.

Another possible reference value for the shortest-path tree growth factor $\zbar$ is the average degree of a neighbour in a random graph with the same degree distribution which is $\kav_\mathrm{nn} = {\texpect{k^2}}/{\texpect{k}}$. This is the relevant value for diffusive processes on a random graph.
For our finite Erd\H{o}s-R\'{e}yni networks we have that $\kav_\mathrm{nn} \approx \kav$ so again the growth factor found to give the best fit, $\zbarfit$, in actual  Erd\H{o}s-R\'{e}yni networks is still a bit higher than this estimate. For the Barab\'{a}si-Albert networks and their randomised versions, the $\kav_\mathrm{nn}$ is around twenty-two to twenty-five for the networks in \tabref{t:modelresults}.  This value is much closer but still not in complete agreement. This suggests our shortest-path trees are sampling nodes in a different ways from diffusion but still with a bias to higher degree nodes.

Since spanning trees have many fewer edges than the original graphs, it is perhaps somewhat surprising that we find that the growth factors are comparable with any measures of the average degree in the original network. So the high values of $\zbar$ are telling us that the shortest-path trees are sampling the nodes of their networks with a large bias towards high degree nodes in the parts of the tree close to the root node and that is why we need such a high growth rate $\zbarfit$ when we fit our data for closeness. That way when we prune the edges to produce a tree we will still have high degrees in the tree close to the root node. The corollary is that the outer parts of shortest-path trees are dominated by leaves (degree one nodes) and other low degree nodes, and these also correspond to low degree nodes in the original network.

It is also clear that node correlations play an important role as these are present in the Barab\'{a}si-Albert model but absent in the randomised version. The large difference in $\zbar$ values for these two cases show such node correlations are important and yet, the non-linear relationship \eqref{e:betadef} still holds well in these artificial networks, with or without these correlations.

The $\beta$ parameter in \eqref{e:farnesskform} is harder to interpret but \tabref{t:modelresults} shows a comparison between the two values of $\beta$. The first is $\betafit$ derived from a two-parameter fit of the data to \eqref{e:farnesskform}. The second value is $\beta(\zbarfit,N)$ the value predicted using \eqref{e:betadef} where we use the $\zbar$ value obtained from the same two parameter fit and the number of nodes $N$. What we can see is that the values derived using \eqref{e:betadef}, $\beta(\zbarfit,N)$, are consistently poorer than the values $\betafit$ derived from a two-parameter fit. It highlights that the details of our theoretical form, such as the precise formula for $\beta$, here \eqref{e:betadef}, can be improved.  However, our simple calculation has captured the important features of the problem so that the form \eqref{e:farnesskform} does work in these theoretical models provided we treat both $\zbar$ and $\beta$ in \eqref{e:farnesskform} as free parameters to be determined.

\subsection*{Numerical Results for Real Data}

For networks representing real-world data, we used data which is open access and easily obtained \cite{pajek,KONECT,SNAP,P20}.
We aimed for a wide range of networks both in terms of size and in terms of the type of interaction encoded in these real-world networks.

The first set of eighteen data sets we refer to as the Konect-SNAP networks. These were derived from real-world data and were chosen to reflect five broad categories of network:
social networks (\texttt{social-\ldots}),
communication networks (\texttt{commun-\ldots}), citation networks (\texttt{citation-\ldots}), co-author networks (\texttt{coauth-\ldots}), and hyperlink networks (\texttt{hyperlink-\ldots}).
A more detailed description of these Konect-SNAP networks is given in Supplementary Note 3 on data sets.

Summary statistics are given in \tabref{t:friendshipresults}.
The reduced chi-square $\chisqred$  measure is between $1.05$ and $1.61$ for ten, more than half, of our examples and another four networks have values between $2.09$ and $2.86$. Given the wide range of both size and nature of these networks and the simplicity of our theoretical derivation, this level of agreement may not have been expected. We also give the Pearson correlation measure between closeness and degree, $\rho(c,k)$, and this is generally high as has been noted before \cite{B88,RPWDMK95,F97,L06b,VCLC08,BN14,LLBM15,SVB17,OFPASF19,BEEKSWWS19}. The success of our non-linear relationship between closeness and degree is not incompatible with high $\rho(c,k)$ values.


\begin{table}[htb!]
	\centering\resizebox{\linewidth}{!}{%
		\begin{tabular}{cc|cc|cccc}
			\hline
			Network                      & N     & $1/\ln(\zbarfit)$  & $\betafit$             & $\zbarfit$       & $\beta(\zbarfit,N)$ & $\rho(c,k)$ & $\chisqred$ \\ \hline
			\texttt{social-karate-club}    & 34    & $0.460\pm 0.066$   & $2.997\pm 0.095$       & $8.81\pm 2.76$   & $2.69\pm 0.25$      & $0.77$      & $1.23$     \\
			\texttt{social-jazz}  & 198   & $0.367\pm 0.015$   & $3.349\pm 0.048$       & $15.28\pm 1.72$  & $2.98\pm 0.08$      & $0.86$      & 13.15      \\
			\texttt{social-hamster}        & 1788  & $0.353\pm 0.009$   & $4.129\pm 0.020$       & $17.05\pm 1.22$  & $3.68\pm 0.07$      & $0.68$      & $1.20$     \\
			\texttt{social-oz}             & 217   & $0.403\pm 0.010$    & $3.492\pm 0.038$       & $11.96\pm 1.02$  & $3.22\pm 0.08$      & $0.89$      & $2.99$     \\
			\texttt{social-highschool}     & 70    & $0.561\pm 0.039$   & $3.734\pm 0.079$       & $5.95\pm 0.74$   & $3.48\pm 0.17$      & $0.87$      & $1.25$     \\
			\texttt{social-health}         & 2539  & $0.537\pm 0.008$   & $5.605\pm 0.016$       & $6.43\pm 0.17$   & $5.31\pm 0.06$      & $0.75$      & $1.07$     \\ \hline
			\texttt{commun-email}          & 1133  & $0.394\pm 0.007$   & $4.309\pm 0.014$       & $12.64\pm 0.54$  & $3.83\pm 0.05$      & $0.84$      & 1.06       \\
			\texttt{commun-UC-message}     & 1893  & $0.264\pm 0.003$   & $3.526\pm 0.008$       & $43.92\pm 2.16$  & $3.01 \pm 0.03$      & $0.72$      & 2.28      \\
			\texttt{commun-EU(core)-email} & 986   & $0.259\pm 0.004$   & $3.324\pm 0.012$       & $47.63\pm 2.67$  & $2.80\pm 0.03$      & $0.84$      & 2.48      \\
			\texttt{commun-DNC-email}      & 1833  & $0.222\pm 0.010$   & $3.499\pm 0.012$       & $91.16\pm 18.84$ & $2.67\pm 0.08$      & $0.41$      & 1.38       \\
			\texttt{commun-DIGG-reply}     & 29652 & $0.388\pm 0.002$   & $5.078\pm 0.003$       & $13.12\pm 0.18$  & $5.05\pm 0.02$      & $0.61$      & 1.60       \\ \hline
			\texttt{citation-DBLP}         & 12494 & $0.361\pm 0.003$   & $4.856\pm 0.004$       & $15.98\pm0.35 $  & $4.45\pm 0.03 $     & $0.54$      & 1.17       \\
			\texttt{citation-Cora}         & 23166 & $0.503 \pm 0.004$  & $6.639 \pm 0.008$      & $7.31 \pm 0.13$  & $6.14\pm 0.04 $     & 0.48        & 1.10       \\ \hline
			\texttt{coauthor-astro-ph}     & 14845 & $0.441 \pm 0.004$  & $5.735 \pm 0.010$      & $9.67 \pm 0.21$  & $5.30 \pm 0.04$     & 0.61        & 1.07      \\
			\texttt{coauthor-netsci}       & 379   & $0.382 \pm 0.080$  & $ 6.553    \pm 0.119$  & $13.74 \pm 7.51$ & $3.31 \pm 0.48$     & 0.35        & 1.27       \\
			\texttt{coauthor-pajek}        & 6927  & $0.259 \pm  0.002$ & $ 3.894     \pm 0.002$ & $47.45 \pm 1.33$ & $3.30 \pm 0.02$     & 0.64        & 12.56      \\ \hline
			\texttt{hyperlink-polblog}     & 1222  & $0.240 \pm 0.004$  & $3.316 \pm 0.011$      & $64.84 \pm 4.44$ & $2.72\pm 0.03$      & 0.72        & 2.15       \\
			\texttt{hyperlink-blogs}       & 1222  & $0.239 \pm 0.004$  & $3.316 \pm 0.011$      & $65.07 \pm 4.48$ & $2.71\pm 0.03$      & 0.72        & 2.18       \\ \hline
	\end{tabular}}
	\caption{\textbf{Results for the largest components the Konect-SNAP networks.}
These networks are derived from a variety of derived from real-world data,
see the description of Konect-SNAP networks in Supplementary Note 3 on data sets for more details. 
The results for $1/\ln(\zbarfit)$ and $\betafit$ come from linear fits of inverse closeness, $1/c_v$ to the logarithm of degree, $\ln(k_v)$ for each vertex $v$,
		i.e.\ $1/c_v = (\ln(\zbar))^{-1}\ln(k_v) + \beta$ \eqref{e:farnesskform}. The value of $\beta$ derived from $\zbarfit$ and $N$ using \eqref{e:betadef} is also shown  as $\beta(\zbarfit,N)$ for comparison. The fits are very good as reduced chi-square $\chisqred$ values show.
The Pearson correlation measure between closeness and degree $\rho(c,k)$ is given for comparison with earlier work.
}
	\label{t:friendshipresults}
\end{table}

\begin{figure}[htb!]
	\centering
		\includegraphics[width=0.85\linewidth]{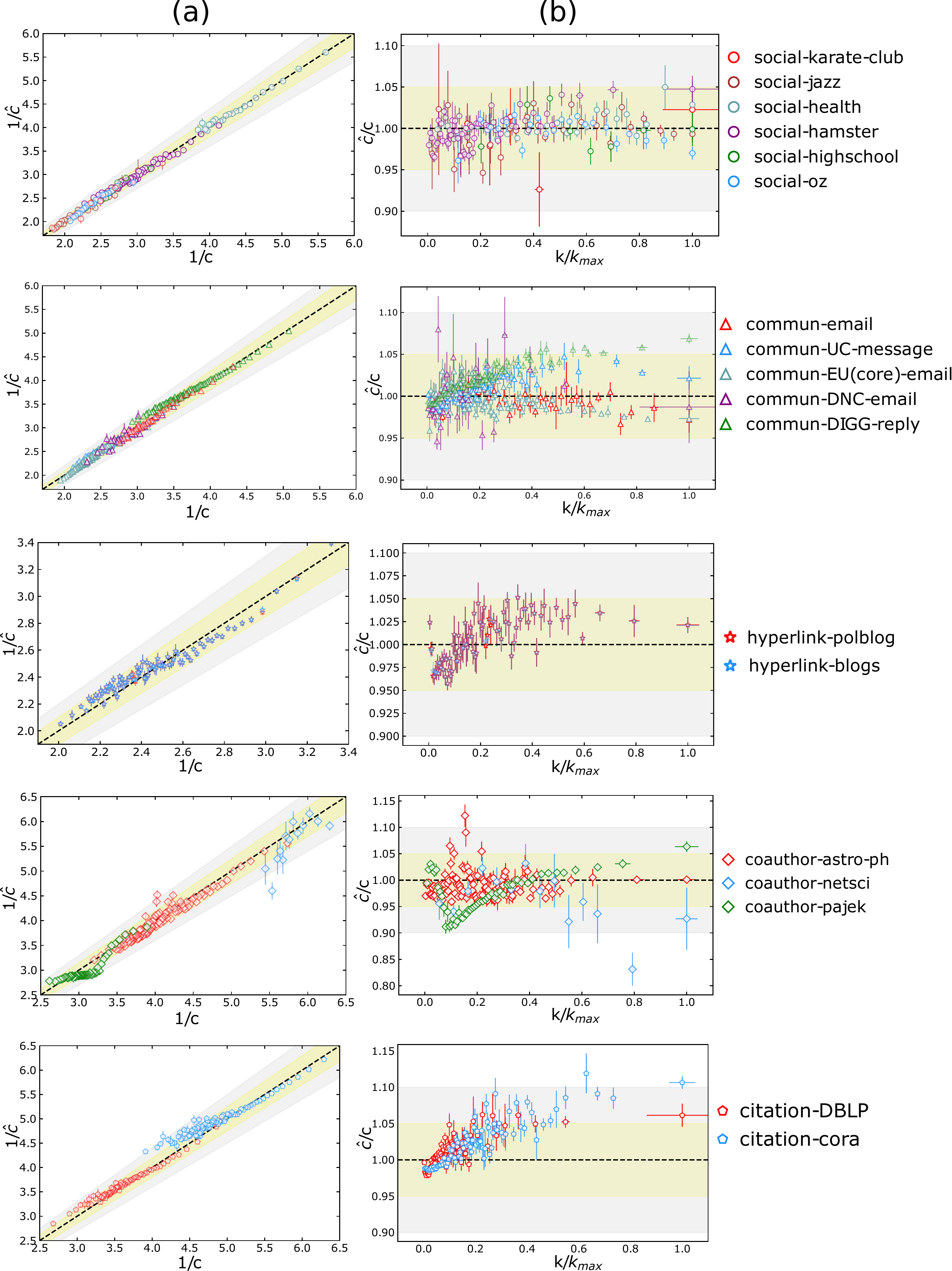}
	\caption{\textbf{Closeness and degree for Konect-SNAP networks.}
    Results for the largest components of eighteen Konect-SNAP networks derived from real-world data, see \tabref{t:friendshipresults} for the statistics of each dataset. The yellow shaded region corresponds to $5\%$ deviation and grey region corresponds to a $10\%$ deviation. Plots in column (a) show the inverse of the predicted result $1/\hat{c}$ from the best fit against the inverse of the mean measured value $1/c=1/\texpect{c}_k$ averaged over nodes with the same degree $k$. Both axes are essentially $\ln(k)$. Column (b) shows the ratio of best fit value $\hat{c}$ over measured value $c$  as a function of degree $k$ rescaled by the largest degree in each network $k_\mathrm{max}$. If the prediction matched data perfectly, points will lie on the dashed lines. The error bars represent from standard error of mean of the inverse closeness. For majority of points, we can see our prediction \eqref{e:farnesskform} captures the relation between closeness and degree, usually with within a $5\%$ margin.}
	\label{f:realnetwork}
\end{figure}

The data for each network is shown in more detail in \figref{f:realnetwork}. Again, we can see that within the error bars the average closeness at each degree generally follows the form we predict within 5\% when the best fit parameters are used. Further, the uncertainties estimated for these data points suggest that the vast majority of average closeness values are statistically consistent with the predicted value for that degree, something already captured by the reduced chi-square values in \tabref{t:friendshipresults}.

Finally we looked at our relationship between degree and closeness in 112 networks taken from the \href{https://networks.skewed.de/}{Netzschleuder} archive of 276 network data sets \cite{P20}. Our only selection criterion was that we could automatically download a network and that it could be analysed successfully by our standard code without further work. This excluded several examples in this archive such as  those with multiple networks (e.g.\ \href{https://networks.skewed.de/net/amazon_copurchases}{\texttt{amazon\_copurchases}}), or some which were too large for our code (e.g.\ \href{https://networks.skewed.de/net/academia_edu}{\texttt{academia\_edu}}).
Supplementary Note 3 contains further information on these \href{https://networks.skewed.de/}{Netzschleuder} networks.
As a result, our sample contains many for which we would not expect much success: some are very small, some are very dense, some have add additional known structure e.g.\ a bipartite network. At the same time there are many for which we would expect to be successful, typically anything sparse and large. We represented each as a simple graph and analysed the largest connected component which had up to 40,000 nodes and an average degree of less than 300.

\begin{figure}
    \centering
    \includegraphics[width=0.7 \textwidth]{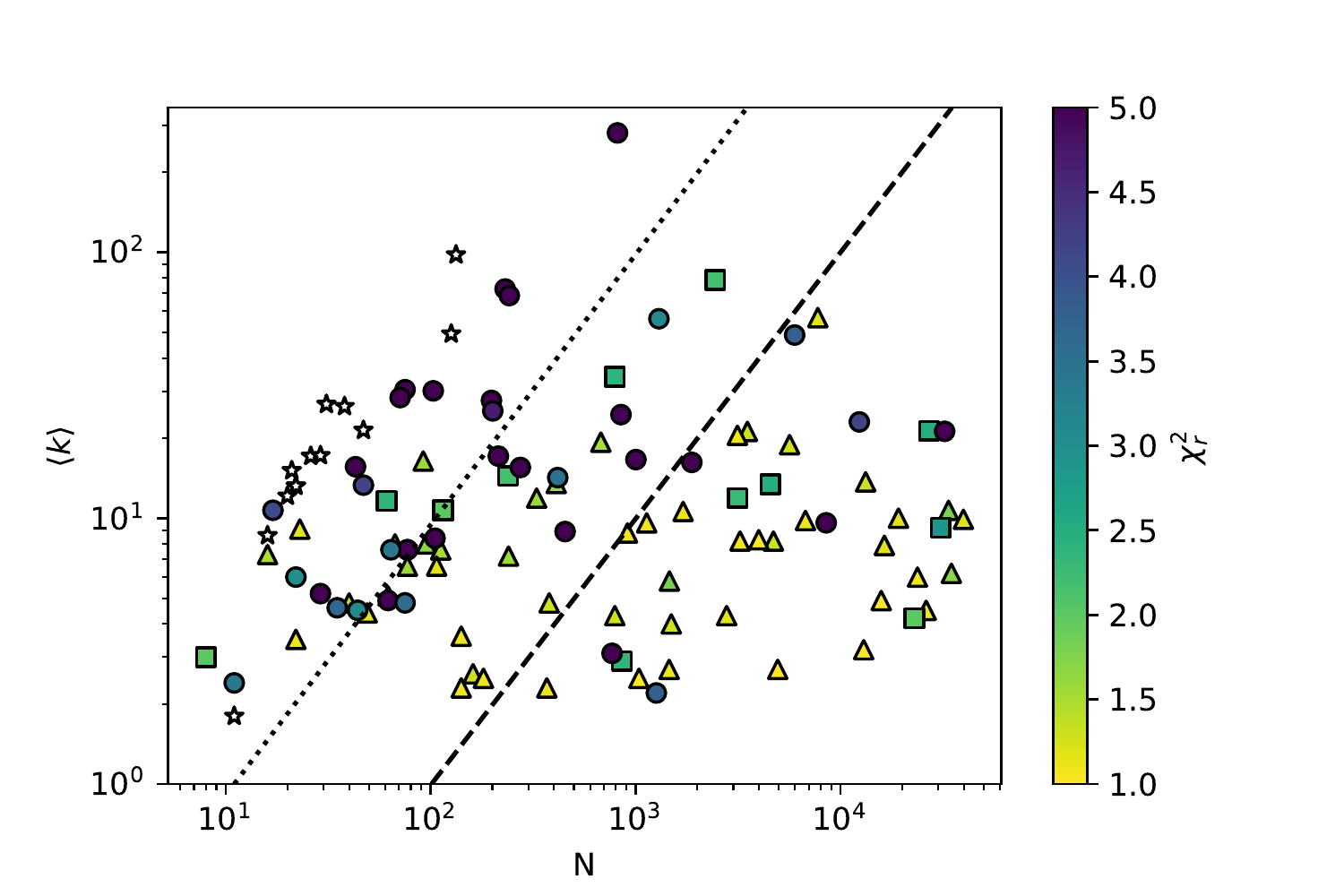}
    \caption{\textbf{Closeness and degree for Netzschleuder real-world networks.}
    Results for the largest components of an additional 112 networks taken from the \href{https://networks.skewed.de/}{Netzschleuder} archive \cite{P20}. Each point represents one network, plotted using the number of nodes $N$  and the average degree $\kav$ as coordinates.  The symbol colour indicates reduced chi-square ($\chi_r^2$) value with yellow being the best results (closest to 1.0). The symbol shape also indicates $\chi_r^2$ values with triangles representing best results where $\chi_r^2<2.0$, squares are for $2.0\leq\chi_r^2<3.0$ and circles represent poor performance with $3.0\leq\chi_r^2$. The stars (white fill) are networks where our $\chi_r^2$ calculation failed due to a lack of an error estimate on each point which occurs when we have one node per $k$ value. To the right (left) of the dashed (dotted) line, networks have a density $\kav/(N-1)$ of less than 0.01 (more than 0.1).}
    \label{f:Netzschleuder}
\end{figure}

The results are shown in \figref{f:Netzschleuder} with additional information and a table of results on the \href{https://networks.skewed.de/}{Netzschleuder} networks provided in Supplementary Note 4.
We found 50 networks had an excellent fit with a reduced chi-square $\chi_r^2$ close to 1.0 and always less than 2.0. Another 13 networks gave a reasonable fit $2.0\leq \chi_r^2 < 3.0$ and 9 networks had $3.0\leq \chi_r^2 < 4.0$. For the remaining networks, our code did not find a $\chi_r^2$ for 13 of these as there was at most one node for each degree value and our code could not estimate the uncertainty in the measurement of closeness. Overall, of the 99 networks where we had a $\chi_r^2$ result, our degree-closeness relationship was very successful ($\chi_r^2<2.0$) in these arbitrary networks 50\% of the time.

However, from \figref{f:Netzschleuder} we can see that the density of a network, the number of edges divided by the number of node pairs, has an impact on the success rate. Our sample has included several very dense networks where there was a lot of additional information on the edges so the existence of an edge was much less important.  Indeed from \tabref{t:Netzschleuder} we see that for \href{https://networks.skewed.de/}{Netzschleuder} networks with densities below $0.04$ we have around $70\%$ fitting with a reduced chi-square of less than $2.0$, rising to roughly $80\%$ for $\chi_r^2<3.0$.
\begin{table}[!htb]
\begin{center}
\begin{tabular}{cccc}
Upper   & Number	& \% Networks    & \% Networks \\
Density & Networks  & $\chi_r^2<2.0$ &$\chi_r^2<3.0$ \\ \hline
0.01	& 32	    & 66\%	         & 84\% \\
0.02	& 44	    & 70\%	         & 84\% \\
0.03	& 48	    & 69\%	         & 81\% \\
0.04	& 55	    & 67\%	         & 80\% \\
\end{tabular}
\end{center}
\caption{\textbf{The success rate for sparse Netzschleuder networks.}
These results are for sparse networks taken from the 112 \href{https://networks.skewed.de/}{Netzschleuder} networks analysed in this paper. Each row gives the statistics for all networks whose largest connected component has a density less than or equal to the density given. The percentage of these networks whose reduced chi-square is less than two or less than three is given in the last two columns. }
    \label{t:Netzschleuder}
\end{table}

\subsection*{Using Closeness}

Our numerical results confirm our analytical work that the inverse of closeness depends linearly on the logarithm of degree \eqref{e:farnesskform} for most networks.  This is a correlation, true on average but not an exact result for every node. It has long been known that nodes with larger degrees tend to have smaller closeness which leads to significant correlation measures \cite{B88,RPWDMK95,F97,L06b,VCLC08,BN14,LLBM15,SVB17,LNR17,OFPASF19,BEEKSWWS19,APB20} but this is often discussed in terms of the Pearson correlation coefficient which is most sensitive to linear correlations. A non-linear relationship between degree and closeness is not discussed in previous studies and our relationship puts this intuitive understanding of a degree-closeness relationship on a firm footing.


Our work suggests that in the majority of networks, to a first approximation, closeness of individual nodes captures little more information on average than is contained in the degree. There is no point spending time calculating closeness if you only want a rough measure of centrality, you may as well just use degree.
However, closeness measurements can give useful information on a network if used correctly.

First, closeness and degree measurements yield our fitted parameters, $\betafit$ and $\zbarfit$. These characterise each network and can be used to compare different types of networks and even networks of different sizes. In particular, $\zbarfit$ characterises the exponential growth that is a feature of most complex networks and which is behind many well-known phenomena as we shall note later in this discussion.

There are a number of ways to determine $\zbar$ and $\beta$ from our relation \eqref{e:farnesskform}. We have found the most effective approach is the simpler method where we determined  $\zbar$ and $\beta$ from a linear fit of our data to \eqref{e:farnesskform}.
There are alternatives but these throw light on various possible approximations rather than being of practical use, see the sections on determining $\zbar$ and $\beta$ in Supplementary Note 1 and results for higher-order polynomial fits in Supplementary Note 4.

For networks where degree and closeness are linked by our relationship \eqref{e:farnesskform}, our work shows that useful information from closeness centrality for individual nodes can \tseemph{only} come by comparing values for individual nodes against the expected value derived using our relationship \eqref{e:farnesskform} using the logarithm of degree.
The only useful information in closeness values is the deviation from their expected value.
In these situations, we could start by examining the degree centrality of every node. This would be the primary measure of centrality. We then fit our closeness values using \eqref{e:farnesskform} to produce an expected value of closeness $c^{(\mathrm{fit})}_v$ for each node. Finally, we use this fit to find nodes which are noticeably more (or less) central than expected. One way to do this is to look at the normalised closeness
\beq
 c^{(\mathrm{norm})}_v
 =    \frac{c_v}{c^{(\mathrm{fit})}_v} \, .
\eeq
Our normalised closeness measure $c^{(\mathrm{norm})}_v$ will highlight the outliers which would then be of most interest.
See Supplementary Note 4 to see some examples of the fluctuations in closeness values around our predicted value.

One could also compare closeness to degree by running the configuration model, which keeps node degree constant, and measuring the closeness of each node in such a null-model network. However our method, requiring a simple linear fit to data already acquired, will be much faster than running the configuration model.


\subsection*{General Network Insights}\label{s:insights}

The success of our conjecture also suggests that most networks satisfy two key assumptions built into our derivation.

First we assumed that the number of nodes a distance $\ell$ from any node grows exponentially. Such exponential growth is common in networks as it is the mechanism behind the concepts of the ``six degrees of separation'' and the ``small world'' effect \cite{WS98} often reported in networks. More formally, this is linked to length scales in networks models with $N$ vertices which grow as $O(\ln(N))$. This is to be contrasted with a network controlled by the geometry of a $d$-dimensional Euclidean space (such as Random Geometric Graphs) where the number of nodes at distance $\ell$ from a node grows as a power law $\ell^{d-1}$ and length scales in such networks grow as $O(N^{1/d})$.
We will consider the length scales in more detail in the Network Length Scales subsection below.

This non-Euclidean behaviour of most networks highlights one situation where closeness is a useful centrality measure independent of and uncorrelated with degree, so a situation where our relationship \eqref{e:farnesskform} will fail. That is for graphs embedded in Euclidean space.  This is somewhat ironic as this type of planar network, such as a network representing the connections between intersections of streets in a city, is the prototypical example used to motivate the idea that the network measure closeness is related to our intuition about the concept of centrality in a network. Indeed, Bavelas \cite{B50} only used planar graph examples to develop closeness so a link between closeness and degree in most networks was never an issue in the original motivation for the closeness measure.

The second assumption that our work supports is that the branches of the shortest-path trees are statistically similar as illustrated in \figref{f:sptapprox}. The success of our analysis suggests this assumption works well whenever we are looking at measurements that depend on the bulk of the network.
This simple approximation may therefore help analyse other network measurements, and we consider some of these in the Network Length Scales subsection below.

\subsection*{Closeness and Real-World Networks}\label{s:realworldnet}

Given the simple assumptions and approximations used in deriving our relationship \eqref{e:farnesskform}, it is perhaps surprising to find that this closeness-degree relation works well for so many networks based on real world data. So how often is our relationship a success?  Also, can we understand when our closeness-degree relationship may succeed and when it is likely to fail?

For the first set of eighteen Konect-SNAP networks shown in \figref{f:realnetwork} and \tabref{t:friendshipresults}, we had eleven (61\%) with a reduced chi-square of less than $2.0$ and all but two (89\%) have a reduced chi-square with less than $3.0$.
The success rate for the 112 \href{https://networks.skewed.de/}{Netzschleuder} networks, as measured by reduced chi-square, is lower. Of the ninety-nine \href{https://networks.skewed.de/}{Netzschleuder} networks where we had a reduced chi-square measurement, fifty (51\%) had a reduced chi-square of less than $2.0$ while sixty-three (64\%) had a a reduced chi-square of below $3.0$. The lower success rate for the \href{https://networks.skewed.de/}{Netzschleuder} networks can be understood as we are sampling without bias a varied collection of network data sets and, in many cases, we have information on the networks which suggests we might never have expected success.

The simplest issue is network density, the fraction of possible edges that are actually present in a network.  If the density is very high, the distances between nodes will be low, perhaps just one or two. This leads to very small variations in the distances encoded in the closeness of each node making it harder to distinguish meaningful patterns in closeness measurements so our relationship is more likely to struggle with high density networks. This trend is clear in \figref{f:Netzschleuder}. Of the two Konect-SNAP networks with poor chi-square, one, the \texttt{social-jazz} network, has a high density of $0.13$. Conversely, we find that the success rate is much higher in our \href{https://networks.skewed.de/}{Netzschleuder} networks for networks with density below 0.04: roughly two-thirds of these low density networks
fit our degree-closeness relationship well with $\chi_r^2<2.0$ and this rises to 80\% if we accept anything with $\chi_r^2<3.0$, see \tabref{t:Netzschleuder}.

More generally, a failure of our relationship is a powerful tool to highlight where there is additional structure in the network. Our theoretical analysis assumes a generic network where every node sees the same exponential growth in the number of nodes at distance $\ell$ away
as discussed in the General Network Insights subsection.
If this generic structure is not present, we have no reason to expect our relationship to work well.

To see this structural issue, consider some of the notable exceptions to our low density criterion. Looking at \figref{f:Netzschleuder}, the low density networks with $N$ between $700$ and $2000$ and $\kav$ between $2.0$ and  $3.5$, you can see two good networks with good fits (triangles) and three with poor fits (two circles and a square). One circle comes from the \href{https://networks.skewed.de/net/crime}{\texttt{crime}} network with $N=1263$, $\kav=2.2$ for which we find a reduced chi-square of $\chi_r^2 = 3.8$. The description of the \href{https://networks.skewed.de/net/crime}{\texttt{crime}} network mentions that the data was obtained using ``snowball sampling from five initial homicides'' so we are looking at a biassed sample of a much larger network. We suspect this sampling produces a different structure from a typical complex network leading to a failure of our relationship in this case. The other circle is the
\href{https://networks.skewed.de/net/plant_pol_kato}{\texttt{plant\_pol\_kato}} network with $N=768$, $\kav=3.1$ and  $\chi_r^2=34.1$ which is a ``bipartite network of plants and pollinators''  from a forest. Our analysis does not allow for a bipartite structure and in this case the difference between the two types of nodes is large with 91 plant species nodes but 715 nodes representing species of insects. The square represents the
\href{https://networks.skewed.de/net/unicodelang}{\texttt{unicodelang}} network with $N=858$, $\kav=2.9$ and $\chi_r^2=2.4$. This is still a good fit but it is also a ``bipartite network of languages and the countries'' so again, a better result might be achieved if we adapted our approach for two-mode networks.

In some cases, the meta-data we have for a network may already tell us about large-scale structure, such as a bipartite network, that will invalidate our relationship \eqref{e:farnesskform}. In simple cases, it may be possible to adjust our derivation to find a more appropriate relation that matches the known structure. For instance, for bipartite networks we could represent our shortest-path trees using two growth rate parameters, $\zbar_a$ and $\zbar_b$, for odd and even distances from the root node.

In other cases failure may indicate the presence of structure that is not already known.  For example, the network may have strong inhomogeneities such as high degree nodes clustering together in a dense core. That would invalidate our assumption that all the branches of the shortest-path trees look statistically similar.  We can see this type of problem in discussions of the average path length in random graph models \cite{CL02,DMS03a}. There it was noted that the typical $\ln(N)$ dependence of length scales in complex networks is not seen when these models have a degree distribution of the form $p(k) \sim k^{-\gamma}$ with $2 < \gamma < 3$. In these cases this failure is linked to the network taking on a rather different structure, perhaps an  ``octopus'' \cite{CL02}, a star-like structure with one dense core connected to many legs. This type of network invalidates our assumptions and we would not expect to see our closeness-degree relationship.

\subsection*{Network Length Scales}\label{s:lengthscales}


We can compare our analytical approach to that of other theoretical work on length scales in networks. The vast majority of the analytical results on distance in networks is for well defined simple theoretical models, typically the Erd\H{o}s-R\'{e}yni graph \cite{CL02,BL06a,BGHJ07,N10}, the Barab\'{a}si-Albert model \cite{BR04,ECV20}, or scale-free random graphs \cite{CH03} (our BA-Configuration model). The focus in the literature tends to be on characteristic length scales for a graph, such as the average path length $\ellav$ or the diameter, and not on length scales associated with each vertex, such as closeness that we consider.

What is distinctive about our approach is that we focus on generic network properties that appear to hold in many networks. So our approach can be used on a much wider range of networks and in particular work for real-world networks not just for one simple model.  The downside is that we do not have the mathematical rigour of those working with simple models. So, let us consider how our approach can be related to other length scale calculations in the literature.

We have already noted that the concepts of the ``six degrees of separation'' and the ``small world'' effect \cite{WS98} are linked to a network models where the length scale grows as $O(\ln(N))$, much slower that for $O(N^{1/d})$ behaviour expected for networks embedded in $d$-dimensional Euclidean space. One common way to study this more precisely is to consider the average distance between nodes $\ellav$, where each node pair contributes equally to the average. This is what we obtain if we take half the average of the inverse closeness \eqref{e:closenessdef} over all vertices, $\ellav = (2N)^{-1} \sum_r (c_r)^{-1}$.
From our results for closeness, we can see that the $N$ dependence in $\ellav$ comes from the $\ln(N)$ term in the expression for $\beta$ of \eqref{e:betadef}, see the section on average shortest path length in Supplementary Note 1 for more details.
So our result is consistent with the behaviour found more rigorously for the average path length in random graph models \cite{CL02,DMS03a}. The success of our method in terms of this average path length result shows that small world behaviour can be linked to the simple network features built into our method, those discussed in the General Network Insights subsection.

Our theoretical cutoff $L(N,k)$ of \eqref{e:lmaxexp} gives another length scale for each node and again this has the same small world $\ln(N)$ behaviour. One might conjecture that this cutoff length scale $L(N,k)$ could be linked to the eccentricity length scale. The eccentricity \cite{WF94,WS03} $e_r$ of a node $r$ is the largest distance from $r$ to any other node, $e_r = \max\{ d_{rv} | v \in \Vcal\}$. In that case we are making a new conjecture that eccentricity should also be linearly dependent of $\ln(k)$ with a gradient of $-1/\ln(\zbarfit)$ to match our theoretical cutoff $L(N,k)$.   As seen in Fig.~1 of Wuchty \& Stadler \cite{WS03} and confirmed in our own analysis, eccentricity does depend linearly on $\ln(k)$ but the gradient does not seem to match the prediction from our theoretical cutoff $L(N,k)$. Roughly speaking our cutoff $L(N,k)$ represents a typical average large scale, not an extremal value of a distribution, so we should never expect a close link with eccentricity.
For more details see the section on eccentricity and $L(N,k)$ in Supplementary Note 1.

As the largest eccentricity is the diameter of a graph, we can see the $O(\ln(N))$ behaviour of our $L(N,k)$ expression as matching this behaviour seen analytically in the diameter in simple models. However, just as for eccentricity and $L(N,k)$, we don't expect to be able to get a precise handle on diameter in our approach.


We also see strong similarities between our approach and that used in several papers \cite{CH03,BL06a} where the network is reduced to a set of rings of nodes, each ring containing all the nodes at the same distance $\ell$ from a root node.
For instance, the mean first-passage time of random walkers on a network to a given vertex $v$ is just the inverse of closeness where now closeness is defined in terms of a new distance function $d_{uv}$ in \eqref{e:closenessdef}, where $d_{uv}$ is the average first-passage time for a random walker to move from vertex $u$ to vertex $v$ (this is also known as Markov centrality \cite{WS03a} or \href{https://en.wikipedia.org/wiki/Random_walk_closeness_centrality}{Random walk closeness centrality}).
Mean first-passage time has been observed to be proportional to degree \cite{BL06a} and applying our simple approximations to the ``ring'' method of Baronchelli \& Loreto \cite{BL06a} quickly reveals this feature, see the section on the ring calculations of first passage times in Supplementary Note 1 for more details.

Overall, our approach can give insights in the the behaviour of many network lengths scales, sometimes only very roughly, sometimes with more precision. It can never match the precision of the analytic calculations done for the simplest models but our approach can be used in a much broader range of networks.

\subsection*{Improvements}\label{s:improvements}

We have already noted one simple improvement when working with bipartite networks. That is to use two growth rate parameters, $\zbar_a$ and $\zbar_b$ for odd and even distances from the root node in the shortest-path tree of bipartite networks.

Other extensions are suggested by probing our numerical results in more detail. For our smaller set of eighteen Konect-SNAP networks, beyond the two poorly fitting cases, our relationship \eqref{e:farnesskform} is very successful for most individual nodes within a 5\% margin, a success which may not be expected given the simple analytical derivation. However we can see some clear if small trends in the deviations in \figref{f:realnetwork}.
We suggest these trends highlight the limitations of our analytical approach but it is possible to improve our theoretical methods.

At the simplest level, we could replace the sharp cutoff used for $n_\ell(r)$ where $n_\ell(r)=0$ for $\ell>L$. There are examples of these distributions for some simple models in Baronchelli \& Loreto \cite{BL06a}. A better cutoff may well lead to better predictions for $\beta$ allowing one to fit a function with one independent parameter rather than two that we used by keeping $\beta$ as a fitted parameter. However, while fitting one rather than two parameters may be theoretically satisfying, it does not seem much of a gain for analysis of real-world networks.

Another option might be to calculate a different network parameter, namely the second degree $k^{(2)}_r= n_{\ell=2}(r)$ \cite{FLAOYMEC20} for each node $r$. By finding the number of nodes two steps away from every node, we can make a better approximation for  $n_\ell(r)$, that is $n_0(r)=1$, $n_1(r)=k_r$, and $n_\ell(r)=k^{(2)}_r \zbar^{\ell-2}$ for $2 \leq \ell \leq \ellmax_r$ and $n_\ell(r)=0$ for $\ell>\ellmax_r$.  This approach cannot be worse than the method used here as the latter is included as a special case where the second degree $k^{(2)}_r=\zbar k_r$ for all nodes $r$.
To leading order we get the same type of result, namely that $1/c_r = (\zbar)^{-1} \ln(k^{(2)}_r) + \beta$ since the degree $k_r$ now only  contributes a small number of terms to closeness. So in this approach using second degree we need to measure a different set of $N$ parameters, the second degree of each node. Finding second degree is slower numerically than degree but both scale in the same way with increasing network size. The success of our simpler method here points to the idea that second degree and degree may often be correlated so it is likely that using second degree may only enhance results in a few cases.

More serious changes will be needed to the calculation if other effects neglected here, such as community structure or degree assortativity, are to be included.

\subsection*{Distance and Logarithm of Degree}\label{s:distlnk}

The logarithm of degree $\ln(k)$ has been found to play an important role in network analysis before. A large fraction of papers on networks will show degree distributions where the horizontal axis is the dependent variable $\ln(k)$ and not simply the degree $k$.  A more specific example comes from Zhou et al.\ \cite{ZMS20} where the ratio of the degrees of nodes at the two ends of each edge (largest value in the numerator) is used to assign a `distance' $\eta(u,v)$ to each edge $(u,v)$. This is equivalent to defining $\lambda(u,v) = \ln(\eta(u,v)) = | \ln(k_u)-\ln(k_v) |$. In fact one can quickly see that while both $\eta$ and $\lambda$ are semi-distances on the set of edges in the formal mathematical sense, only $\lambda$ is also a semi-metric and so $\lambda$ is in some sense the more natural `distance' measure in a qualitative sense. Our work suggests that an alternative view is to replace the logarithm of degree by the inverse of closeness. Since $(c_u)^{-1}$ is the actual average of the shortest-path distances from $u$ to all other nodes, we can immediately see it is natural to work with inverse closeness when considering distances. For instance Zhou et al.\ \cite{ZMS20} we could look at a different edge measure $\tilde{\lambda}(u,v) = | (c_u)^{-1}-(c_v)^{-1} |$. While the inverse closeness is a more natural distance, the degree is much easier to calculate in practice.  Our work allows researchers to move between these two pictures.

\subsection*{Closeness and Gromov Centrality}\label{s:geom}

It has been noted \cite{BDL22} that the inverse of closeness is related to another centrality measure based on the average of the Gromov product. This centrality measure captures the extent to which the triangle inequality is not saturated between three nodes in a network, and so this is deeply connected to the geometry of a network. Our result leads to a natural prediction for this Gromov Centrality $G_r^D$ of node $r$ defined on the scale of the network's diameter $D$ \cite{BDL22}. This Gromov Centrality is defined on other network length scales, $G_v^\ell$, and Babul et al.\ \cite{BDL22} suggest there are useful generalisation of closeness.
Our approximations will prove just as effective for such generalisations of closeness.


\section*{Conclusions}\label{s:concl}

We have derived a non-linear relationship between degree and closeness \eqref{e:farnesskform}, two of the most important centrality measures in Network Science. This was achieved by assuming that every node can be seen as the root of a shortest-path spanning tree where for all nodes these trees are statistically similar two or more steps away from the root.
We have shown that this degree-closeness relationship works on a large number of artificial and real-world networks, particularly low density networks where there is no special macroscopic structure.

Our results mean that measuring closeness is of little use unless our relationship is used to remove the dependence on degree from closeness. Further, most networks can be seen as shortest-path spanning trees which are statistically similar two or more steps away from their root nodes.
Equally, if our relationship fails, it is an indication that our basic assumptions about network structure are wrong and so there are strong inhomogeneities and macroscopic structure in the data.

\section*{Acknowledgements}

TSE would like to thank Max Hart, Oskar Hogburg and Luke Melville for initial investigations on this topic.

\section*{Data Availability}

All the data used in this paper is publicly available from several repositories. Our copies were downloaded from one of the following: {KONECT}
 \cite{KONECT}, SNAP \cite{SNAP}, and the {Netzschleuder} \cite{P20} repositories.
More detailed information on some of the networks used along with additional references are also given in Supplementary Note 3 on data sets.
The networks used in this study along with tables of many of the results are available in \cite{EC22}.

\section*{Author Contributions}

TSE performed the analytical research and BC performed the numerical research. TSE and BC designed the research, analysed data and wrote the paper.

\section*{Competing Interests}

The authors declare no competing interests.

\appendix
\begin{center}
\LARGE\textbf{Appendices}
\end{center}
\renewcommand{\thesection}{\Alph{section}}
\setcounter{equation}{0}
\renewcommand{\theequation}{\thesection.\arabic{equation}}
\renewcommand{\thefigure}{\thesection.\arabic{figure}}
\renewcommand{\thetable}{\thesection.\arabic{table}}
\numberwithin{equation}{section}
\numberwithin{figure}{section}
\numberwithin{table}{section}
\setcounter{section}{0}
\renewcommand{\theHequation}{\thesection.\arabic{equation}}
\renewcommand{\theHfigure}{\thesection.\arabic{figure}}
\renewcommand{\theHtable}{\thesection.\arabic{table}}



\section{Calculations}\label{a:closenessrand}

\begin{figure}[htb]
	\begin{minipage}[c]{0.5\textwidth}
		\includegraphics[width=\textwidth]{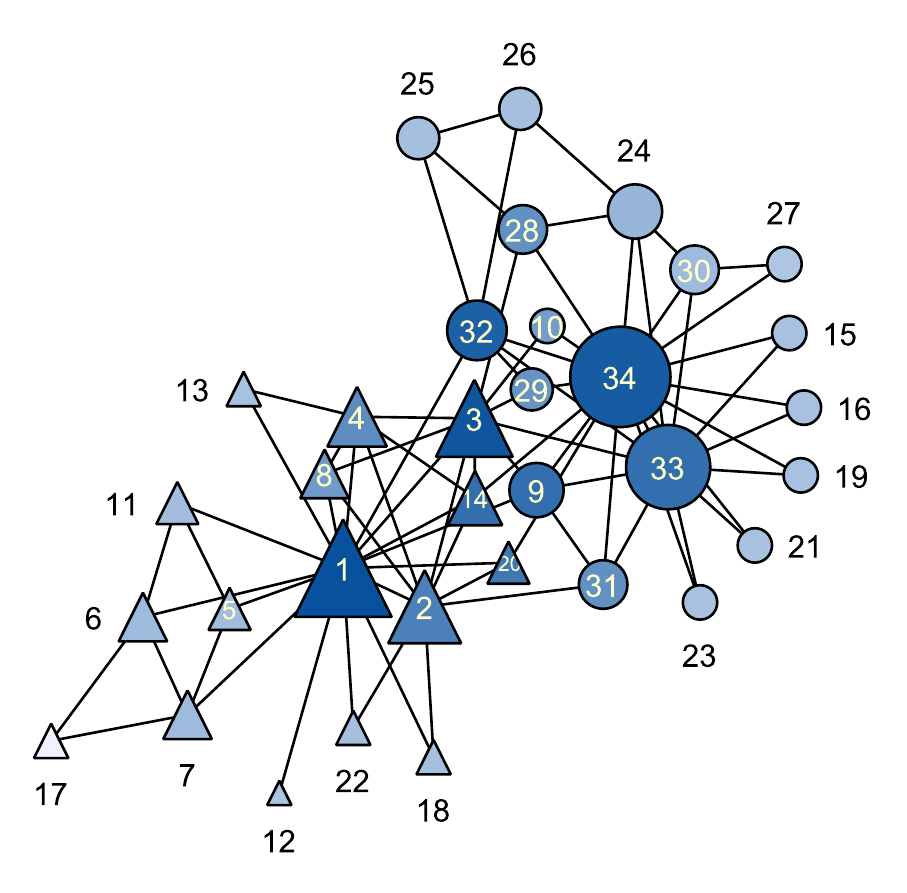}
	\end{minipage}
	\hfill
	\begin{minipage}[c]{0.5\textwidth}
		\caption{The Zachary karate club network \cite{Z77}.  The nodes are coloured so that the smaller the closeness, the darker the colour while the size of the node is proportional to the degree.
			Node labels correspond to those used in \cite{Z77}. The largest nodes tend to be darkest, the smallest nodes lightest indicating that there is significant correlation between degree and closeness. The node shape indicates the two communities found by Zachary \cite{Z77} using the Ford-Fulkerson binary community algorithm.}
		\label{f:Zacharyck}
	\end{minipage}
\end{figure}

In this Supplementary Note, we will give greater detail on the calculations referred to in the main text. As we have used the Zachary karate club network \cite{Z77} as an example to illustrate some key principles, e.g.\ in \figref{f:Zachary}, we start by illustrating the correlation between degree and closeness in this network in \figref{f:Zacharyck}.

\subsection{Estimate of Closeness}\label{a:closenessest}

In this section we will look at our calculation of closeness in more detail.  This enables us to highlight in more detail the approximations and assumptions made, and to show the type of terms we are neglecting in our expansions.

\begin{figure}[htb!]
  \begin{minipage}[c]{0.5\textwidth}
		\includegraphics[width=\textwidth]{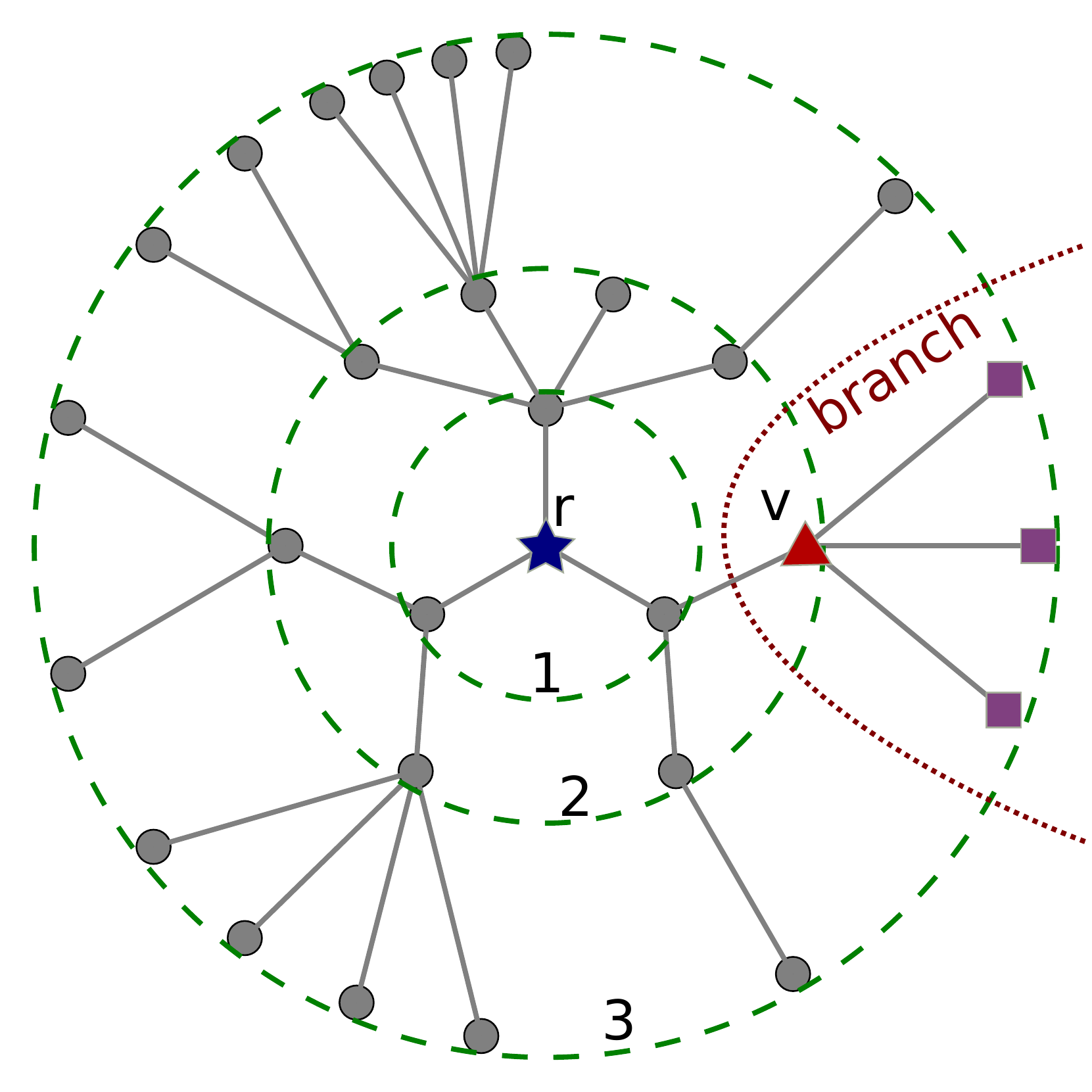}
  \end{minipage}
  \hfill
  \begin{minipage}[c]{0.5\textwidth}
	\caption{An example of a rooted tree $\Tcal(r)$ defined in terms of a root node $r$, the blue star at the centre.
    All nodes at the same network distance from the root node, are placed at the same distance from the root node in this visualisation,
    as indicated by the green dashed circles. The red triangle, node $v$, is the root of a \tsedef{branch} $\Tcal(v,r)$, a smaller tree containing all the nodes $u$ which lie on a path from the root through $v$ (which is also a shortest path from $r$ to $u$ in the full graph $\Gcal$). These nodes $u$ in the branch are therefore further from the root than $v$, $d_{ur} \geq d_{vr}$. The branch $\Tcal(v,r)$ illustrated here includes $v$ (red triangle), the nodes indicated with purple squares and the edges between these nodes. The degree of a node is the number of neighbours so here node $v$ has degree $4$.}
	\label{f:rootedtree}
    \end{minipage}
\end{figure}

We will work with simple graphs with one component, so every vertex is connected by a path to every other vertex. We will analyse this in terms of \tsedef{trees} (for example see section 4.2.10 of Wasserman \& Faust \cite{WF94}, section 6.7 of Newman \cite{N10} or section 1.4 of Latora et al.\ \cite{LNR17}), subgraphs with no closed loops and one component so the number of edges in a tree is one less than the number of vertices in the tree.
Our trees $\Tcal(r)$ are \tsedef{rooted trees} in that they are defined by starting at a special node, the root node $r$ for that tree. An example is shown in \figref{f:rootedtree}.

\begin{figure}[htb]
  \begin{minipage}[c]{0.5\textwidth}
		\includegraphics[width=\textwidth]{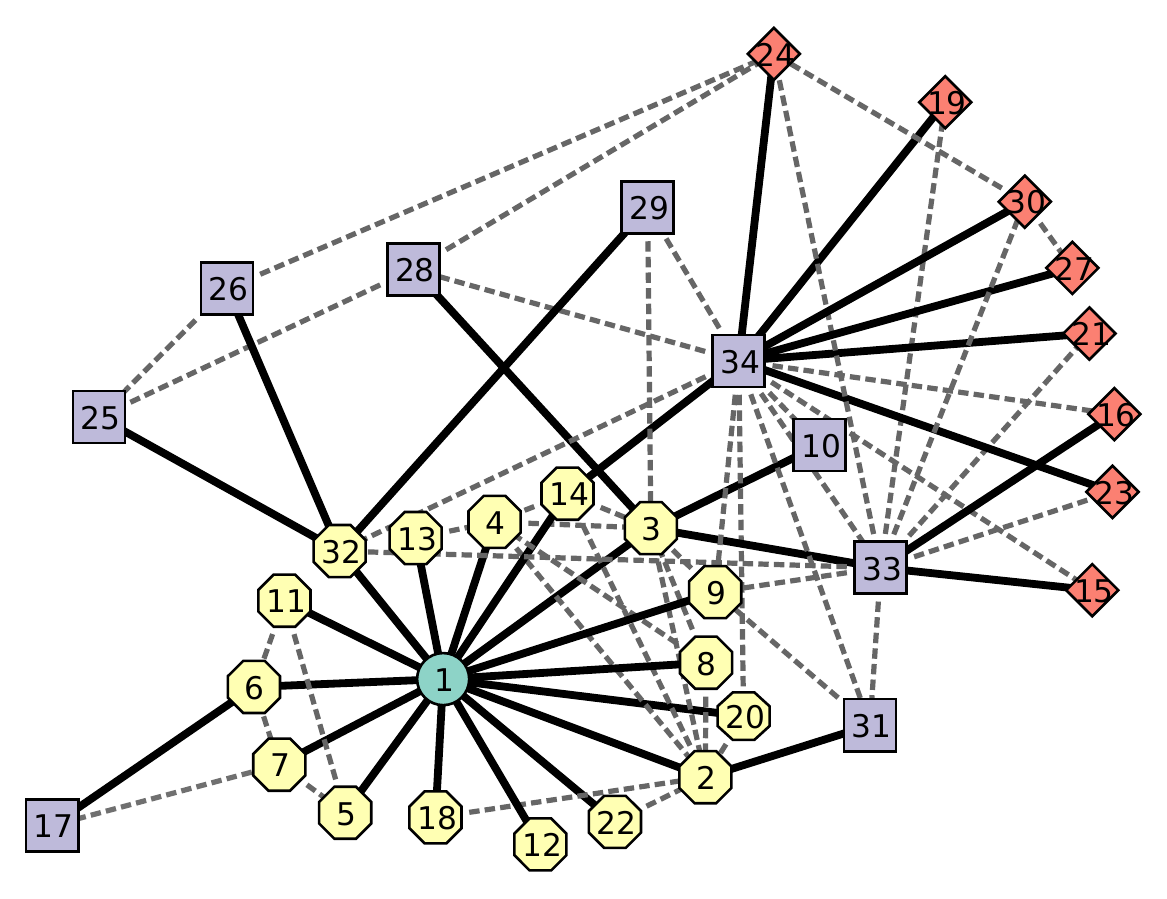}
  \end{minipage}
  \hfill
  \begin{minipage}[c]{0.5\textwidth}
	\caption{The Zachary Karate club network \cite{Z77}.  The nodes of the same colour and shape have shortest paths to node 1 of the same length.
    The thick black lines are edges which are part of one possible shortest-path tree $\Tcal(1)$ with node index $1$ as the root node of the tree.
    The dashed grey lines indicate edges which are not in the tree.
    These shortest-path trees are not unique as can be seen here since we can include edge $(7,17)$ in $\Tcal(1)$ instead of the edge $(6,17)$ used here.
		Node labels correspond to those used in \cite{Z77}.}
	\label{f:Zachary}
  \end{minipage}
\end{figure}

We are interested in the closeness $c_r$ of a vertex $r$ which is defined to be\footnote{Care is needed as at least one high impact paper \cite{BE06} defines closeness to be the sum of distances to all other nodes, $\sum_v d_{vr}=(N-1)/c_r$, something known elsewhere as ``status'' \cite{H59,HH95,WS03} and which when normalised is proportional to what we call the ``farness'' $f_r=1/c_r$ of vertex $r$. The definition of closeness in \cite{BE06} is the inverse of the expression used here \eqref{a:cdef} and in all other papers we have found. The paper \cite{BE06} derives it's definition from two other high impact papers: \cite{S66} (eq.2, pp.583) and \cite{F78} (pp.225). These last two papers also use what we call farness but those authors call their measure ``point centrality'' so avoiding confusion.}
\bea
    \frac{1}{c_r}
    &=&
    \frac{1}{(N-1)} \sum_{v \in \Vcal \setminus r} d_{vr}
    \label{a:cdef}
\eea
where $d_{uv}$ is the length of the shortest path between any pair of vertices $u$ and $v$ and the sum is over all vertices except $r$.
To study the closeness $c_r$ of a vertex $r$ we defined the \href{https://en.wikipedia.org/wiki/Shortest-path_tree}{\tsedef{shortest-path trees}} which are rooted on node $r$. These trees contain one shortest path from a root node $r$ to every other node in the full network $\Gcal$. At least one shortest-path tree exists for every root node $r$ and these trees can be defined using a \href{https://en.wikipedia.org/wiki/Breadth-first_search}{breadth-first search} algorithm. The shortest-path trees are not unique, one root node can have several different shortest-path trees, but any one of these trees will be sufficient for our purposes. The key property of these trees is that the distance from every node in the tree to the root node is the same in both the shortest-path tree $\Tcal(r)$ and in the full graph $\Gcal$. We give more details on how such trees may be found and a proof of the properties we have stated in \appref{a:sptree}. We show an example of a shortest-path tree in \figref{f:Zachary}.

The picture we have is as follows.
Suppose we start at node $r$ of degree  $k_r$ and we look a few steps always from the root at all the nodes $\{v_\ell\}$ which are distance $\ell$ from the root node.
Each of these nodes $v_\ell$ will be the root of a branch\footnote{This branch $\Tcal(v_\ell,r)$  containing all the nodes $u$ which are further from the root $v_\ell$ (i.e.\ $d_{ur} \geq \ell$) and these nodes $u$ lie on a path from the root node $r$ passing through $v_\ell$ (so the branch $\Tcal(v_\ell,r)$ includes $v_\ell$ itself) which is also a shortest path from $r$ to $u$ in the both the full graph $g$ and in the tree $\Tcal(r)$.  The edges in the branch $\Tcal(v_\ell,r)$ are all the edges from the full tree $\Tcal(r)$ which run between nodes in the branch. So all nodes in the branch $\Tcal(v_\ell,r)$ will be at least distance $\ell$ from the root node and the branch itself is also a tree, a subgraph of the original tree $\Tcal(v_\ell,r) \subset \Tcal(r) \subset \Gcal$.}
$\Tcal(v_\ell,r)$ of our tree $\Tcal(r)$.

One key assumption we make is that \emph{all} these branches have similar statistical properties because the vast majority of such branches are in the `bulk' of the network. To start with, we will assume that in terms of our measurements $\Tcal(v_\ell,r) \equiv \Tcal(v_\ell^\prime,r)$ for any two nodes $v_\ell$ and $v_\ell^\prime$ distance $\ell$ from our root.  For the same reason, in most graphs we might expect these subgraphs to be similar whatever the root node was\footnote{To be more precise, if our measurement $M$ is a map from a branch to some real numbers, then  we require that $M(\Tcal(v_\ell,r)) \approx  M(\Tcal(v_\ell^\prime,r') ) \approx M_\ell$ so that the average value of these measurements for any given $\ell$, say $M_\ell$, is a fair representation of the individual measurements.} so $\Tcal(v_\ell,r) \equiv \Tcal_\ell$.

In particular, we can look at the number of nodes at distance $\ell$ from the root  $r$ which we denote as $n(\ell;r)$ ask how this grows with distance $\ell$. If the  average number of child nodes (neighbours which are one step further out) is $\zbar(v_\ell,r)$, then our assumption that branches are similar statistically means we are assuming that this $\zbar$ only depends on the distance from the root node so $\zbar(v_\ell,r) = \zbar_\ell$.
Hence, we estimate that the number of nodes $\ell$ steps away from our root vertex $r$ as
\bea
	n_\ell (r)  \approx  \zbar_{\ell} \, n_{\ell-1}(r)
	\, , &&
	n_\ell (r)  \approx \prod_{\ell'=1}^\ell \zbar_{\ell'}
	\, , \quad
	\text{for} \quad \ell \geq 1 \, , \quad n_0(r)=1 \, .
\eea

It is immediately possible to improve on this approximation. For a start, we usually know, or can easily find, the degree $k_v$ of each node $v$ so we will assume that we know the values $n_1 (r)=k_r$ since neighbours are the nodes at distance one from the root.
So at this stage we have a model with $n_0(r)=1$, the $N$ local values $n_1 (r)=k_r$, and then  global (independent of root vertex $r$ chosen) parameters $\zbar_\ell$ for $\zbar\geq 2$.
However, our goal is to find a simple statistical relationship between closeness and degree, so we will make a further approximation. If the graph looks the statistically similar once we are looking at nodes in the bulk, then assuming that $\zbar_\ell \approx \zbar$ independent of $\ell$ for $\ell\geq 2$ is consistent with our picture. This then leaves us with
\bea
	n_\ell (r)  \approx  \zbar^{\ell-1} \, k_r
	\, , \quad
	\text{for} \quad \ell \geq 1 \, .
	\label{a:nellexp}
\eea
The exponential growth in the number of nodes distance $\ell$ from \emph{any} root node, as encoded in \eqref{a:nellexp} is our second key assumption. This will not be true for networks embedded on a plane or other Euclidean spaces as there we expect $n(\ell)$ to measure the surface area of a shape of radius $\ell$ which would follow a power-law $n_\ell \sim \ell^{D-1}$ for a $D$-dimensional Euclidean space.


Clearly, to get a network of significant size we need $\zbar_\ell \geq 1$. However, the total number of nodes $N$ is given by
\beq
    N = \sum_{\ell=0}^\infty n_\ell (r)
    \label{a:Nfromn}
\eeq
so a model with constant $\zbar_\ell = \zbar \geq 1$ gives an infinite graph which is of little use for the finite graphs found in real data sets.
So we know that in practice $\zbar_\ell<1$ for larger $\ell$ in a graph of a finite number of nodes $N$.
This contradicts our need for $\zbar_{\ell}>1$ for small $\ell$ in order to get a network of any size.

The crudest solution, and the one we will follow here, is to assume that $\zbar_{\ell}=\zbar$ for $2 \leq \ell \leq \ellmax_r$ where $\ellmax_r$ is some long distance cutoff which may depend on the root vertex $r$ considered with $\zbar_{\ell}=0$ for larger $\ell$. So we work with the following model
\beq
    n_\ell(r)
    =
    \begin{cases}
    	1                    & \text{if} \quad \ell=0 \, , \\
    	\zbar^{\ell-1} k_r   & \text{if} \quad 1 \leq \ell \leq \ellmax_r \, , \\
    	0                    & \text{if} \quad \ellmax_r < \ell  \, .
    \end{cases}
    \label{a:nell}
\eeq
To improve clarity of the expressions, we will now drop the explicit dependence on the root vertex $r$ chosen
and write $c \equiv c_r$, $k \equiv k_r$, and $\ellmax \equiv \ellmax_r$.

We will determine the distance cutoff $\ellmax$ by imposing \eqref{a:Nfromn} which given our model for $\zbar_{\ell}$ now becomes
\beq
    N =  1 + \sum_{\ell=1}^{\ellmax} \zbar^{\ell-1} k  = 1 + k \frac{(\zbar^{\ellmax}-1)}{(\zbar-1)}\, .
    \label{a:Nfromn2}
\eeq
Inverting this we see that the distance cutoff $\ellmax$ we need is given by, for large $N$,
\beq
    {\ellmax}(N,k)  \approx \frac{\ln\left(N(\zbar-1)/k \right) }{\ln(\zbar)} \, .
    \label{a:lmaxexp}
\eeq
Even in this simplest approximation, it is clear that the distance cutoff $\ellmax$ depends on our choice of root vertex through the degree of the root node.

In principle $\ellmax$ in \eqref{a:nell} and \eqref{a:Nfromn2} is an integer but it is clear from the form in \eqref{a:lmaxexp} that we need $\ellmax$ to be a real number, in some sense an average over the actual distances from the root to the leaves (nodes with degree one) of the tree. So the real number valued $\ellmax$ given by \eqref{a:lmaxexp} sets the scale of the distance beyond which the terms in these sums become negligible. We also note as an aside that $\ellmax$ depends on the size of the network through a $\ln(N)$ factor, not as a power $N^{1/D}$, and this is the classic ``small-world'' effect seen in many network length scales such as diameter and average distance, see \secref{a:asp} for further discussion.

We will assume our network is connected so a shortest path exists between all node pairs in the network.
We can now rewrite the closeness $c \equiv c_r$ \eqref{a:cdef} of a root vertex $r$ using \eqref{a:nell} and \eqref{a:lmaxexp} to gives us
\bea
 \frac{1}{c}
 &=&
 \frac{1}{\Omega}\sum_{\ell=1}^{\ellmax} \ell n_\ell
 \, , \quad
 \Omega = \sum_{\ell=1}^{\ellmax} n_\ell \, .
\eea
Here for simplicity we have dropped the explicit dependence on the root vertex $r$ so $c \equiv c_r$, $\ellmax \equiv \ellmax_r$, and $\Omega \equiv \Omega_r$.

For the normalisation $\Omega$ we have that $\Omega=(N-1)$ and using \eqref{a:Nfromn2} we can express this in terms of our other variables
\bea
\Omega
&=&
\sum_{\ell=1}^{\ellmax} k \zbar^{\ell-1}
=
k \frac{(\zbar^{\ellmax} -1)}{(\zbar-1)}
=
N-1 \, .
\label{a:Omegaexp}
\eea
Note this gives us a link between $\ellmax$, $N$ and $\zbar$. We will eventually use \eqref{a:Omegaexp} to eliminate $\ellmax$ as we assume $N$ is known. However the expressions are simpler in terms of $\ellmax$ so here we will use \eqref{a:Omegaexp} to eliminate $N$ and $\Omega$.

Using \eqref{a:nell} and \eqref{a:lmaxexp} gives us that 
\bea
  \frac{1}{c}
  &=&
  \frac{1}{\Omega} \sum_{\ell=1}^{\ellmax} k \ell \zbar^{\ell-1}
  \label{a:fcalc1}
  =
  \frac{k}{\Omega}  \frac{d}{d\zbar} \sum_{\ell=0}^{\ellmax} \zbar^{\ell}
  =
  \frac{k}{\Omega}  \frac{d}{d\zbar} \left( \frac{\zbar^{\ellmax + 1}-1}{\zbar-1} \right)
  \\   &=&
  \frac{k}{\Omega}
  \left(
  \frac{(\ellmax +1)\zbar^{\ellmax}}{\zbar-1}
  -
  \frac{(\zbar^{\ellmax + 1}-1)}{(\zbar-1)^2}
  \right)
\eea

Using \eqref{a:Omegaexp} to eliminate $\Omega$ (i.e.\ $N$) in terms of $\ellmax$ and $\zbar$, we have that
\bea
  \frac{1}{c}
  &=&
  \frac{(\zbar-1)}{(\zbar^{\ellmax} -1)}   \left[
  \frac{(\ellmax +1)\zbar^{\ellmax}}{\zbar-1}
  -
  \frac{(\zbar^{\ellmax + 1}-1)}{(\zbar-1)^2}
  \right]
  \\
  &=&
  \frac{\ellmax \zbar^{\ellmax}}{(\zbar^{\ellmax} -1)}
  +
  \frac{\zbar^{\ellmax}}{(\zbar^{\ellmax} -1)}
  -
  \frac{(\zbar^{\ellmax + 1}-1)}{(\zbar^{\ellmax} -1)}
  \frac{1}{(\zbar-1)}
  \\
  &=&
  \ellmax
  \left( 1 -
  \frac{1}{(\zbar^{\ellmax }-1)}
  \right)
  +
  \frac{1}{(\zbar-1)}
  \frac{1}{(\zbar^{\ellmax} -1)}
  \left(
  \zbar^{\ellmax}(\zbar-1)
  -
  (\zbar^{\ellmax + 1}-1)
  \right)
  \\
  &=&
  \ellmax
  \left( 1 -
  \frac{1}{(\zbar^{\ellmax }-1)}
  \right)
  +
  \frac{1}{(\zbar-1)}
  \label{a:farness3}
\eea

Now we can use \tref{a:lmaxexp} in \eqref{a:farness3} to produce a prediction of the relationship between the closeness of a node and its degree, also showing how closeness should vary with the size of the network and we find that
\bea
\frac{1}{c}
&\approx&
\left( \frac{1}{(\zbar-1)}
+ \frac{\ln(\zbar-1)}{\ln(\zbar)} \right)
+ \frac{1}{\ln(\zbar)} \ln(N )
- \frac{1}{\ln(\zbar)} \ln(k)
+ O\left(\frac{\ln(N)}{N} \right)
\, .
\label{a:farness2}
\eea

We now restore the dependence on the root vertex in our notation to emphasises which quantities depend on this choice, and which are fixed network values. The prediction is that the inverse of closeness $c_v$ of any node $v$ should show a linear dependence on the logarithm of the degree $k_v$ of that node with a slope that is the inverse of the log of the branching ratio parameter, that is
\beq
\frac{1}{c_r}  = -\frac{1}{\ln(\zbar)}\ln(k_r) + \beta \, .
\label{a:farnesskform}
\eeq
Our calculation suggests that the parameter $\beta$ is a function of other known parameters but that it is also independent of the vertex $v$ chosen, so that
\beq
  \beta
  =
  \beta(\zbar,N)
  =
  \left( \frac{1}{(\zbar-1)}
  + \frac{\ln(\zbar-1)}{\ln(\zbar)} \right)
  + \frac{1}{\ln(\zbar)} \ln(N ) \, .
  \label{a:betadef}
\eeq
In our analysis we will assume that the number of nodes $N$ and degrees of the nodes $k_v$ are known as such information is often available. Then in principle we have one unknown global parameter, $\zbar$, which are fixed whatever vertex $v$ we consider.

However, our calculation is fairly crude. The key assumptions are the statistical similarity of the branches and the exponential growth in the number of nodes at distance $\ell$ from any root. These are the idea which lead to the form \eqref{a:farnesskform}. The details of the implementation, such as the precise form for the cutoff\footnote{A final note is that there are some formal issues here.  The calculation went through a parameter $\ellmax$ which was initially an integer yet later it became a real valued parameter.  We are of course making a particular analytic continuation of the results of sums of integers such as \eqref{a:cdef} and \eqref{a:Omegaexp}. Technically these analytic continuations are not even unique without an additional criterion but the `natural' forms  given here define the continuation chosen.}
of our sums, here \eqref{a:nell} summarised by our single parameter $\ellmax$, will alter the detail form of $\beta$ but not the broad dependence of closeness $c_r$ on degree $k_r$ and the number of nodes $N$. For that reason, we can regard $\beta$ along with $\zbar$ as two global parameters (i.e.\ the same for all root nodes) to be determined.

\subsection{Determining $\zbar$ and $\beta$}\label{a:detzbarbeta}

There are a number of ways of looking this relation between closeness and degree \eqref{a:farnesskform} when determining $\zbar$ and $\beta$.

First we could set these parameters based on \eqref{a:farness2}. That means we would find shortest-path trees for all vertices $v$, find their average degree $\zbarnum$, and use this to set the value of $\zbar$. Logically we would then choose  $\betanum$ using $N$ and $\zbar=\zbarnum$ in \eqref{a:betadef}.

Another approach would be to use the result for the average degree of a neighbour in a random graph \cite{MR95} (also see section 13.3 \cite{N10}, section 5.4 \cite{LNR17} and the discussion in \secref{a:asp}) which leads to the suggestion that
\beq
 \zbarrnd = \frac{\texpect{k^2}}{\texpect{k}} -1\,.
 \label{a:randomz}
\eeq
Again we can then use \eqref{a:betadef} with $N$ and $\zbar=\zbarrnd$ to suggest a value for $\betarnd$.
Here the expectation values are averages over the degree distribution in the full graph $\Gcal$.
Random graphs do become very similar to trees close to their percolation transition and this approach for $\zbarnum$ and $\betanum$ ought to work well in that region.
However, a typical shortest-path tree has many fewer edges than the full graph and many of those edges are involved in short loops so in practice it is not clear that these averages on the original graph averages are going to be of much relevance to our shortest-paths.   Nevertheless, \eqref{a:randomz} provides us with a useful reference point.


Given the very simple minded approximations, neither of the previous approaches is likely to be very effective for most cases.
The driving force behind the form \eqref{a:farnesskform} is the idea that the number of nodes at distance $\ell$ from any one chosen node rises exponentially, something found in most networks. The precise link between the parameters $\zbar$ and $\beta$ and properties of the network is not going to be as simple or universal as the the simple derivation given here. So the most effective approach may be to treat $\zbar$ and $\beta$ as two \emph{independent} parameters. That is we ignore \eqref{a:betadef} and just do a linear fit of inverse closeness values $1/c_v$ to the logarithm of degree $k_v$ using data from as many vertices $v$ as we can to give $\zbarfit$ and $\betafit$ \eqref{a:farnesskform}. We lose little predictive power in using the data to fix these two model parameters rather than one.  We can then turn this around.  The fitted values $\zbarfit$ and $\betafit$ give us two new global network measurements.  Looking at differences between the fitted values and the alternative values suggested above can give us insights into the complexity of our network.

Indeed we can take this a step further and define new network vertex measures $\zbar_v$ and $\beta_v$ for each vertex $v$ by taking the closeness and degree values for that vertex and inverting \eqref{a:farnesskform} and \eqref{a:betadef}. The  $\beta_v$ parameter is hard to interpret but the $\zbar_v$ tells us what sort of shortest-path tree that vertex sees, independent of its degree. Our assumptions state that such a value $\zbar_v$ will be roughly constant but individual variations could give insights into the network structure.

There is another way we can look at the effective branching ratio parameter $\zbar$ and that is to actually measure it in actual shortest-path trees.
Calculating the average degree of the nodes in a finite tree does not tell us much as this is close to one by definition. What we really want is to look away from the outer edges of the tree, away from the degree one leaf nodes, to look at the degree of nodes in the central part of the tree as it grows in size moving away from the root node. One way might be to look at the modified average degree where we average over all nodes that have degree larger than one (so excluding all leaf nodes) and we also exclude the root node. However, for our artificial networks in our examples, where the average degree was $10.0$, we find values which are much lower (typically between 2.2--2.3 for the Erd\H{o}s-R\'{e}yni networks, 4.5--6.6 for the Barab\'{a}si-Albert networks and 4.0--4.1  for the Configuration Barab\'{a}si-Albert networks) than the degree of the original network and our $\zbarfit$ values. What is happening is that even without out the leaf nodes, close to the edge the branches are often made up of low degree node and these pull the average measured down.  The growth of the tree as we move away from the root is driven by the presence of large degree nodes closer to the root node in the tree. By definition there are many fewer nodes close to the root node so these large degree nodes do not have much effect on the average degree measured in non-leaf nodes of the tree.

\subsection{Average Shortest Path, Closeness and Random Graphs}\label{a:asp}


Finding a typical length scale of a network was one of the earliest challenges in Network Science. The length $d_{uv}$ of the shortest path between two nodes $u$ and $v$ provides the natural measure of distance, as it satisfies both the mathematical criteria for a distance function and our own intuition about the importance of short paths to minimise the costs and losses in communication in real networks.

Characteristic length scales are important in any system and for networks the average distance between all node pairs $\ellav$ is such a length scale where
\beq
 \ellav
 =
 \frac{1}{N(N-1)} \sum_{u \in \Vcal} \sum_{v \in \Vcal \setminus u} d_{uv}
 \, .
 \label{a:ellavdef}
\eeq
This is the mathematical quantity that represents Milgram's ``six-degrees of separation'' \cite{M67,TM69} and investigated on large scales in modern data sets, for instance \cite{LH08,BBRUV12,BV12}. It has been the focus of great interest in some of the earliest theoretical papers such as \cite{WS98,CL02,DMS03a}.

The average distance of a network $\ellav$ has a simple relationship to the closeness considered in this paper as average distance $\ellav$ is simply the half the average over all nodes of the inverse closeness
\beq
\ellav
 =
 \frac{1}{N} \sum_{r \in \Vcal} (c_r )^{-1} \, .
 \label{a:ellav}
\eeq
If we insert our expression \eqref{a:farnesskform} we have that
\bea
 \ellav
 &=&
 - \frac{1}{N} \frac{1}{\ln(\zbar)} \sum_{r\in\Vcal}  \ln(k_r) + \beta
 \label{a:lavb}
\eea
giving us
\bea
 \ellav
 &=&
 +  \frac{1}{\ln(\zbar)} \ln(N )
 -  \frac{1}{\ln(\zbar)} \texpect{\ln(k)}
 +  \frac{1}{(\zbar-1)}
 +  \frac{\ln(\zbar-1)}{\ln(\zbar)}
 \label{a:lav}
\eea
where we have used $\beta$ from \eqref{a:betadef}. This gives for large $N$ and fixed degree distribution that
\bea
 \lim_{N \to \infty}
 \ellav
 &=&
 \frac{1}{\ln(\zbar)} \ln(N ) \, .
 \label{a:lavlim}
\eea

This is the same dependence of average path length $\ellav$  on $N$ found for large random graphs \cite{CL02,DMS03a}. That is for an ensemble of networks defined by configuration model, so no node-node correlations, where the degree distribution has a finite second moment in the infinite graph limit, the leading term in $N$ is found to be \cite{CL02,DMS03a}
\bea
 \lim_{N \to \infty}
 \ellav
 & = &
    \frac{\ln ( N) }{\ln(\zbarrnd)}
    \label{a:lavrnd}
\eea
where $\zbarrnd$ is defined in \eqref{a:randomz}.

We can, however, repeat our calculation for closeness with rather more certainty for random graphs, that is those defined by the configuration model. The necessary results are all contained in equation (13.73) of \cite{N10} or equation (5.41) of \cite{LNR17} (and in several other places) which gives that for random graphs we have
\bea
 \frac{n(\ell)}{n(\ell-1)} &=& \zbarrnd
 \, .
\eea
It is interesting to note that this result is used to look at average distance, e.g.\ equation (13.73) of \cite{N10} and (5.43) of \cite{LNR17}, and so gives an \emph{average} closeness value of \eqref{a:ellav}. This contrasts with our focus on the average result for an individual node of a given degree. Indeed, as far as we can tell, the literature in this area has always focussed on average closeness, on the average path length\footnote{This is also a normalised version of the \tsedef{Wiener index} $W$ of a graph which is the sum of the distances between all node pairs, so $W= [ N(N-1)/2 ] \ellav$.}.

In terms of our calculation for closeness, we can see that  our form for $n(\ell)$ in \eqref{a:nell}, an assumption in most cases, is also found in the results for \emph{ensemble averages} given by the random graph model for large graphs provided $\zbar=\zbarrnd$.
From here we can follow the rest of our calculation with its additional assumptions, as given in \secref{a:closenessest}, to see that our formula for closeness of a random graph will have the same form but now there is a very specific prediction for $\zbar$, namely that for a random graph  $\zbar=\zbarrnd=(\texpect{k^2}/\kav)-1$ of \eqref{a:randomz}.

While the exact result \eqref{a:randomz} for $\zbar$ in a large random graph is of interest, we do not expect the precise value to be informative for real networks. As Newman says about random graph models and simple network models in general (p.448 \cite{N10})
\begin{quote}
they can give you a feel for the types of effect one might expect to see, or the general directions of changes in quantities. But they don't usually give quantitative predictions for the behaviour of real networks.
\end{quote}
So the value of random graph models here is that they give an example where the assumed form for $n(\ell)$ of \eqref{a:nell} is to be expected (on average) giving some further justification for its use in general. In fact our work on the Erd\H{o}s-R\'{e}yni model and the randomised Barab\'{a}si-Albert model, both examples of random graph models, show there are still clear differences between the value of $\zbar$ found from fits even in these finite size cases and the value $\zbarrnd$ suggested by the random graph work for infinite graphs.  However the overall form, the prediction of dependence of closeness on the logarithm of degree $\ln(k)$, does work well in these models. The lack of a perfect match for random graph models is presumably because other approximations we have made come into play for finite graphs suggesting improvements could be made in future work on closeness in random graph models.

The random graph example also highlights another issue, that of networks with
power-law degree distributions with exponents between two and three, $p(k) \sim k^{-\gamma}$ for $2<\gamma<3$. In network models with this distribution, the average degree $\kav$ is well defined for infinite graphs but the second moment $\texpect{k^2}$ diverges as we take $N \to \infty$.
The result on p15879 of \cite{CL02} for such random graphs is that the average path length $\ellav$ is simply $\lim_{N \to \infty} {\ellav}_\mathrm{CL} = O(\ln(N)/\ln(\ln(N)) )$ for some constant $A$. This is because random graphs with such power-law distributions are arranged like ``an `octopus' with a dense subgraph having small diameter as the core'' (from p.15881 of Chung \& Lu \cite{CL02} but also see discussion in Bollob\'{a}s \cite{B03c}). This result for the average path length points to the need for a different theoretical approach to closeness in these cases.  So the failure of our closeness-degree relationship in a few networks constructed from real-world data maybe due to the nature of their large scale structure which will require a different theoretical approach.

\begin{figure}[htb]
        \centering
		\includegraphics[width=1\textwidth]{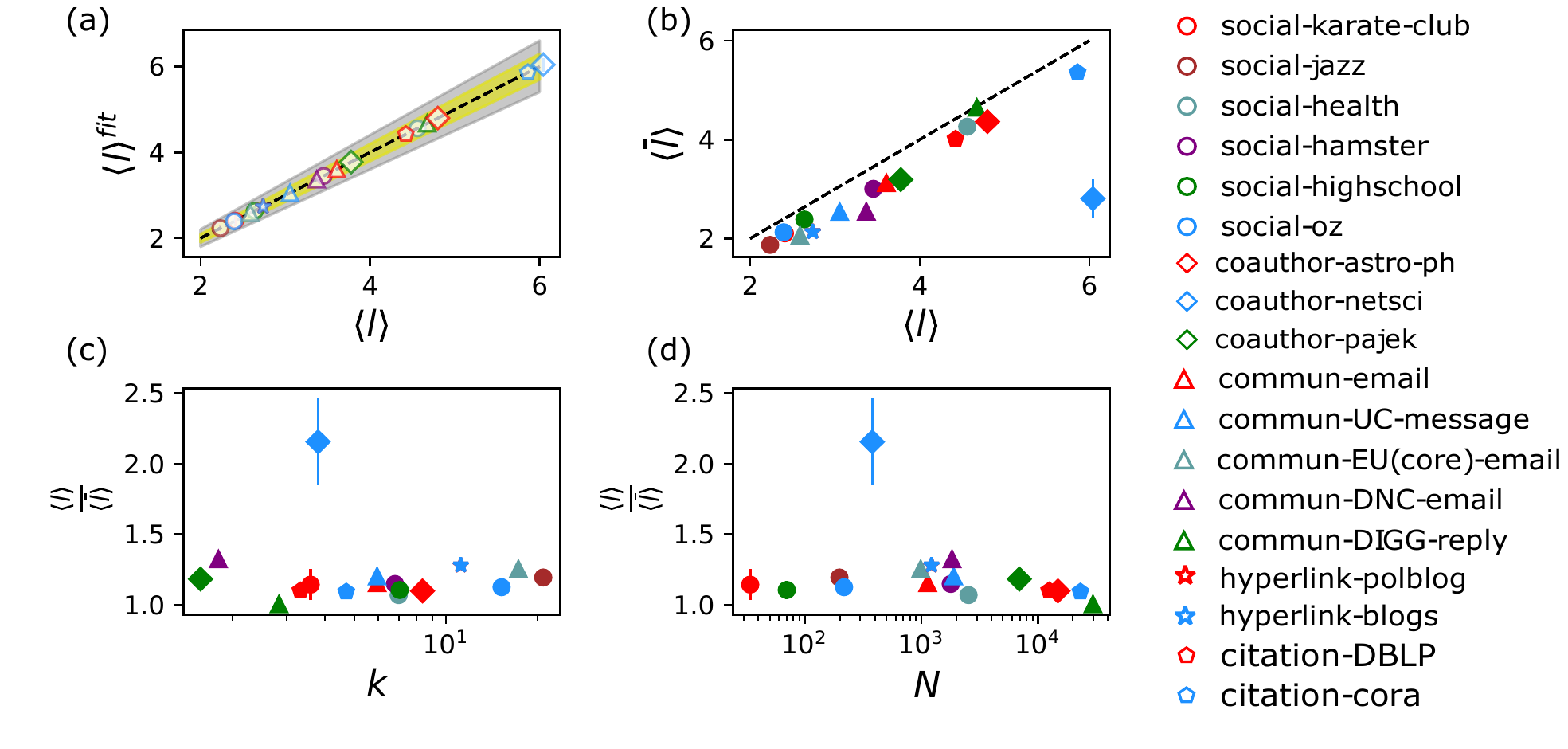}
		\caption{A plot comparing the predicted average path $\left<l\right>$ from fitted value $\betafit$ and theoretical value $\beta$ for the eighteen Konect-SNAP real-world datasets. The empty fill markers correspond to predictions based on the fitted parameter $\betafit$ using \eqref{a:lavb} and solid fill markers correspond to predictions based on \eqref{a:lav}. In (a), we used fitted value $\betafit$ to compute average path length
    	using \eqref{a:lavb}. Our model can predict average shortest path in all 19 datasets within 5\% deviation (shade area correspond to deviation between theoretical value and real value: 5\% for yellow region and 10\% for grey). From (b), we notice the result from the theoretical calculation  using \eqref{a:lav} are consistently lower than measured value from the network. This may suggest some other constant appears, however, the average shortest path in the large $N$ limit will converge to  $\ln N/\ln \bar{z} $. To reveal the large limit behaviours predicted by \eqref{a:lavlim}, we investigate the relation between average shortest path $\left<l\right>$ in (c) and $k$, and  $\left<l\right>$ and $\ln N$ in (d).
    }
		\label{f:ellav2}
\end{figure}

A plot comparing the predicted average path $\left<l\right>$ from fitted value $\betafit$ and theoretical value $\beta$ for the eighteen Konect-SNAP real-world datasets is shown in \figref{f:ellav2}.



\clearpage
\subsection{Eccentricity and $L(N,k)$}\label{a:ecc}


So far we have focussed our attention on closeness.  The inverse of closeness, farness $f_r = 1/c_r$, is the average distance from a given node $r$ to all other nodes. So farness is a node-dependent length scale which we can contrast against the average path length which sets a network length scale. However, we have another node-dependent length scale present in our calculation, $L(N,k)$ of \eqref{a:lmaxexp}. This $L(N,k)$ is the cutoff we use in our sum so in some sense it represents the \emph{typical} largest distance from any node $r$ of degree $k$ to any other node. One can think of $L(N,k)$ as being some type of \emph{average} of the distance to end every branch in the shortest-path tree.

The actual largest distance from a node $r$ to any other node is known as the \tsedef{eccentricity} $e_r$ of node $r$ \cite{WF94,WS03}
\beq
 e_r = \max\{ d_{rv} | v \in \Vcal\}
 \, .
 \label{a:eccentricity}
\eeq
Clearly as both $L(N,k)$ and $e_r$ are dealing with largest distances there might be some merit in comparing the two length scales and thinking they could scale in similar ways or at least be strongly correlated. Certainly we would expect $L(N,k) \leq e_r$. However the cutoff parameter $L(N,k)$ is an average over a set of maximum values while the eccentricity $e_r$ is simply a maximum value of a set so we expect far more fluctuations in eccentricity values even if there was a strong correlation with $L(N,k)$.

We have done a quick comparison for the theoretical models we have used with results given in \tabref{t:eccresults}. What we see is that there is a good linear fit of eccentricity to $\ln(k)$, $e_r = a + b \ln(k_r)$, so this aspect of the prediction is good.  Note that a good linear fit of eccentricity to the logarithm of degree was shown in Figure~1 of \cite{WS03} for one example of each of the  Erd\H{o}s-R\'{e}yni and Barab\'{a}si-Albert models. So our results confirm this observation.
\begin{table}[htb!]
\begin{center}
	\begin{tabular}{cc|c|ccc}
        \hline
        Network type               &  N   & $1/\ln(\zbarfit)$   &  $\chisqred$  & gradient $b$ & intercept $a$ \\ \hline
        \multirow{3}{*}{ER}        & 1000 & 0.46 & 1.12 & -0.64 & 6.17  \\
                                   & 2000 & 0.42 & 1.09 & -0.19 & 5.48  \\
                                   & 4000 & 0.43 & 1.07 & -0.70 & 6.90  \\ \hline
        \multirow{3}{*}{BA}        & 1000 & 0.30 & 1.03 & -0.20 & 4.50  \\
                                   & 2000 & 0.32 & 1.05 & -0.42 & 5.28  \\
                                   & 4000 & 0.31 & 1.04 & -0.43 & 5.67  \\
        \multirow{3}{*}{Config-BA} & 1000 & 0.35 & 1.04 & -0.22 & 4.55  \\ \hline
                                   & 2000 & 0.36 & 1.05 & -0.43 & 5.32  \\
                                   & 4000 & 0.35 & 1.05 & -0.43 & 5.68  \\ \hline
	\end{tabular}
\end{center}
	\caption{Table of results for showing how the linear fit of eccentricity to the logarithm of degree behaves for simple graphs with average degree $10.0$ produced using one of three artificial models with the same average degree $\texpect{k}=10.0$ but with a different number of nodes, $N$. Each `ER' network is a standard Erd\H{o}s-R\'{e}yni network, a `BA' network is produced using pure preferential attachment in the Barab\'{a}si-Albert model, and the `Config-BA' network is a configuration model version of a Barab\'{a}si-Albert model network. The results for $1/\ln(\zbarfit)$ (the uncertainty is $0.01$) come from linear fits of inverse closeness, taken from a table in the main text.  We have fitted $e_r = a + b \ln(k_r)$ to the same models. The values of the intercept $a$ and gradient $b$  given are the mean over fits to each of 100 realisations (standard deviation is roughly 0.01). The reduced chi-square $\chisqred$ is for the eccentricity linear fit.
	}
	\label{t:eccresults}
\end{table}

However, the results in \tabref{t:eccresults} shown no clear link between the gradient $b$ of this eccentricity fit to the $-1/\ln(\zbarfit)$ predicted if $e_r \propto L(N,k_r)$. As eccentricity is an extremal value unlike the average used to find closeness, it could be we need to look at much larger networks to see any link between eccentricity and our simple cutoff scale $L(N,k_r)$. As there is no clear signal here, we have not pursued this further.

\subsection{Ring Calculations of First-Passage Times}\label{a:ring}


Random walks play an important role in both physical processes and as a theoretical and computational tool to probe systems. The use of random walks on networks is widespread as can be seen in any textbook such as \cite{WF94,N10,LNR17,C21}. This section is prompted by the study of first-passage times on a network using a ``ring'' approximation as suggested in \cite{BL06a}. We aim to show how the ``ring'' approximation in \cite{BL06a} is closely related to the shortest-path tree approximation used here. These are types of mean-field approximations, see section 3.2.5 of \cite{MPL17} for an extensive review of these and alternative approaches as well as for known results.

The ring approximation represents a network as a sequence of rings, where the ring labelled $\ell$ contains all $n_r(\ell)$ nodes which lie at a distance $\ell$ from a given root node $r$. Then we represent the probability of finding a random walker at a node at distance $\ell$ from the root node $r$ at time $t$ as $w_\ell(t)$. The master equation is then
\beq
 w_i(t+1)
 =
 \begin{cases}
 p_{01} w_1(t) & \text{if} \quad i=0
 \\
 \sum_{j=1}^{e(r)} B_{i j}(r)   w_j(t)
 \end{cases}
 \label{a:fprweqn}
\eeq
where $e(r)$ is the eccentricity of node $r$, the largest distance from $r$ to any other vertex.
Here random walkers which arrive at the root node $r$ at time $t$ are removed at time $(t+1)$. So the mean first-passage time $\tau(r)$ for a node $r$ is
\beq
 \tau(r) = \sum_{t=1}^\infty t \, w_{0}(t)
\eeq
The parameter $p_{01}$ in \eqref{a:fprweqn} is the probability of moving a neighbour of the root node to the root node, from ring one to ring zero. From the master equation \eqref{a:fprweqn}, it is clear that to find the mean first-passage time $\tau(r)$ it is sufficient to know $w_1(t)$, the probability of finding a random walker at ring one at any time,
\beq
 \tau(r) = \sum_{t=0}^\infty t \, p_{01} \, w_1(t) \, .
\eeq
This means we need to analyse the matrix $B_{i j}(r)$ which is the probability of a walker moving from ring $j$ to ring $i$ when we are considering random walkers moving towards root node $r$. This matrix $\Bmat(r)$ is a constant tridiagonal stochastic matrix.
Assuming a strongly connected network, standard properties of such matrices tell us that the largest eigenvalue $\lambda_1$ is therefore constrained to be $1 \geq \lambda_1 \geq 1-B_{01}(r)$, and it is linked to the only right-eigenvector $\vvec^{(1)}$ whose entries can be chosen to be all positive.
We can the write the average time for a random walker to first reach the root node in terms of some initial distribution of random walkers $w_i(t=0)$ as
\beq
 \tau(r; \wvec(t=0))
 =
 \sum_{t=1}^\infty t \sum_{i=1}^{e(r)} p_{01} [(\Bmat(r))^{t-1}]_{1i} w_i(t=0)
\eeq
If the starting point of the random walk is chosen with equal probability, then in terms of the number of nodes in each ring we have that $w_i(t=0) =n_r(i)/(N-1)$. The mean first-passage time $\tau(r)$ for a root node $r$ is then
\bea
 \tau(r)
 &=&
 \sum_{t=1}^\infty t \sum_{\ell=1}^{e(r)} B_{01}(r) [(\Bmat(r))^{t-1}]_{1\ell} \frac{n_r(\ell)  }{ (N-1) }
 \label{a:mfptring}
\eea
This represents the most general form of the mean first-passage time within the mean-field ring approximation. It is hard to make progress as we still need to know $(3e(r)-2)$ entries for each of the $N$ $\Bmat(r)$ matrices. In most cases, results are obtained by working with a specific simple model, such as the Erd\H{o}s-R\'{e}yni model considered in \cite{BL06a}, so the matrices $\Bmat(r)$ are known expression in terms of a few parameters.

This is where we can make contact with the very simple approximations made when considering closeness using the shortest-path tree. While simpler than many approaches in the literature, our approximations work well for closeness in many real networks so it is interesting to see what they give here for first-passage times. One part of our approximation is simply that $n_r(\ell) = k_r \zbar^{\ell-1}$ and with that we find
\bea
 \tau(r)
 &=&
 \frac{k_r  p_{01} }{(N-1)}
 \sum_{t=1}^\infty t \sum_{\ell=1}^{L(N,k)} [\Bmat^{t-1}]_{1\ell} \zbar^{\ell-1} \, .
\eea
Note we have to use our cutoff $L(N,k)$ rather than the eccentricity $e(r)$. Also note that inherent in our  approach is that away from the root node, the network looks the same statistically.  So another important aspect of our approximation is that the matrix $\Bmat$ is independent of the root node, but this is something that will also occur in most models found in the literature.

The key point is that the main dependence on the root node is through the degree $k_r$ used to start the exponential growth in nodes encoded in our expression $n_r(\ell) = k_r \zbar^{\ell-1}$. There is some logarithmic dependence coming through the cutoff $L(N,k)$ but we assume this will be a smaller correction in any practical finite size network.  Thus one simple conclusion is that on average, the mean first-passage time $\tau(r)$ will be proportional to degree. This is something which was seen in \cite{BL06a} for a real network and in the Erd\H{o}s-R\'{e}yni model but not explained there.

Assuming the dominant contribution is from the largest eigenvalue $\lambda_1$, with normalised left- and right-eigenvectors $\uvec^{(1)}$ and $\vvec^{(1)}$ respectively, we have that
\bea
 \tau(r)
 &=&
 \sum_{t=1}^\infty t \sum_{\ell=1}^{L(N,k_r)} p_{01} [\Bmat^{t-1}]_{1\ell} \frac{k_r \zbar^{\ell-1}  }{ (N-1) }
 \\
 &=&
 \frac{k_r p_{01}  }{(N-1)}
 c_1 \sum_{t=1}^\infty t \sum_{\ell=1}^{L(N,k_r)} (\lambda_1)^{t-1} v_\ell^{(1)}
\eea
where $c_1$ gives the overlap of the initial condition with the left-eigenvector $\uvec^{(1)}$, namely
\beq
 c_1 = \sum_{\ell=1}^{L(N,k_r)} u_\ell^{(1)}  \zbar^{\ell-1}
\, .
\eeq
This gives us
\bea
 \tau(r)
 &=&
 k_r \frac{c_1 p_{01}  }{(N-1)}
  \frac{1}{(1-\lambda_1)^2} \sum_{\ell=1}^{L(N,k_r)}  v_\ell^{(1)} \, .
\eea
If we have a simple model for the entries in the matrix $\Bmat$, it is relatively easy to find the leading eigenvalue and its eigenvectors of this tridiagonal matrix so a more sophisticated result is possible.  However
the main point for this paper is that the dominant dependence of the mean first-passage time $\tau(r)$ on the choice of root vertex $r$ is through the factor of the degree $k_r$ and this factor appeared simply through our assumption that the number of nodes grew exponentially as $n_r(\ell) = k_r \zbar^{\ell-1}$.

\clearpage
\section{Shortest-Path Tree Algorithm}\label{a:sptree}

Our algorithm to find shortest-path trees is a slight variation on well known depth-first search algorithms used to find shortest paths between nodes. We have included it because our focus is rather different from those standard algorithms and because we will use our algorithm to deduce a key property of shortest-path trees. That is the property that every node in a connected network is the root node of at least one shortest-path tree. We can always find a shortest path from each node to any other node (in the same connected component). However, if we take a set of shortest paths, one from a root node to each other node,  the union of such a set of paths does not in general give a tree. The success of the algorithm below shows that shortest-path trees can always be constructed.

This shortest-path tree algorithm generates an example of a shortest-path tree starting from a given root node $v$. It is simply a \href{https://en.wikipedia.org/wiki/Breadth-first_search}{breadth-first search} algorithm \cite{S21b} (\href{https://www.baeldung.com/cs/minimum-spanning-vs-shortest-path-trees}{Dijstra's algorithm}  for unweighted networks) where we record which edge was used to reach each node for the first time in the breadth first search as these edges form a shortest-path tree. In practice, there are various ways to optimise this implementation, for example see section 10.3 \cite{N10} and \cite{S21b}, but this version serves as a simple example which we will then use to highlight some properties of shortest-path trees that we use in our work.

\begin{enumerate}
	\item Label all nodes $v$ with distance to the root node as $-1$, \texttt{distance[v]=-1}.
	\item Label all nodes $v$ with inner neighbour $-1$, \texttt{inner\_neighbour[v]=-1}.
	\item Start from root node $r$, set \texttt{current\_distance=0},  \texttt{distance[r]=current\_distance}.
	\item Create a set \texttt{next\_set} containing just node $r$.
	\item \label{en:nextshell} Increment current distance, \texttt{current\_distance+=1}.
	\item Copy the contents of \texttt{next\_set} into \texttt{current\_set}.
	\item Remove the contents of \texttt{next\_set} so it is now an empty set.
	\item \label{en:loop} Loop through all nodes $u$ in  \texttt{current\_set}. For each $u$ do the following.
	\begin{enumerate}
		\item \label{en:addnextlevel}  Add a neighbouring node $v$ to  \texttt{next\_set} if the distance from $v$ to the root has not been set, i.e.\ add if \texttt{distance[v]=-1} and set  \texttt{inner\_neighbour[v]=u}.
		\item On the other hand if \texttt{distance[v]=current\_distance} then you may choose to change to the new node $u$ using \texttt{inner\_neighbour[v]=u}. This node {v} has already been found and is in a shortest-path tree.
		\\
		Note that at this point we could change the tree defined by using this new neighbour, the current $u$, instead of the existing node $v$ already found.  This might be done with a random number, say 50\% of the time.
	\end{enumerate}
	\item Once the loop in \ref{en:loop} has finished, if \texttt{next\_set} is not empty, then loop back to \ref{en:nextshell}.
	\item The edges in the tree $(v,u)$ are given by \texttt{u=inner\_neighbour[v]} where $u$ is one step closer to the root than $v$. Note the one exception is the root node which has no inner neighbour and \texttt{inner\_neighbour[r]=-1}.
\end{enumerate}

This algorithm also gives us a proof that in a single component simple graph, there always exists at least one such shortest-path tree $\Tcal(r)$ for every node $r$. The proof can be expressed as follows where we set $N(u) \equiv \mathtt{inner\_neighbour[u]}$.
\begin{enumerate}
	\item Every node in the graph is visited by this algorithm as we are assuming a single component. So $N(u)$ is always defined for every node except for the root node.
	\item The edge set of the tree $\Tcal(r)$ is $\Ecal_r = \{ (u, N(u) ) \, | \, u \in \Vcal \setminus r \}$.
	\item This edge set $\Ecal_v$ contains all the vertices in $\Vcal$, the vertex set of the original graph, so $\Vcal$ is also the vertex set of the tree.
	\item Our tree is then the subgraph $\Tcal(v) = \{ \Vcal, \Ecal_v\}$ of the original graph.
	\item\label{en:inexists} For all non-root nodes $u$, all nodes in $N(u)$ are exactly one step closer to the root node than $u$.
	\item For any node $u$, the sequence $\{ u_i\}$, where $u_0=r$ is the root node, $u_i = N(u_{i-1})$ for $i=1,2,\ldots,\ell$, and $v_\ell=u$, always exists.
	\\
	Note that $u_i$ in the sequence is $i$ distance from the root node from \ref{en:inexists}.
	\\
	That this path exists then follows from \ref{en:inexists}, since $N(u)$ is always defined so we can always start from $v_\ell$ and iterate down the sequence. The iteration terminates at $u_0=r$, the root node, as the node $u_1$ will always be one step away from the root so we must have $N(u_1)=r$.
	\\
	It then follows that this defines a path between any given node $u$ and the root $r$.
	\item  This path $\{ u_i\}$ must be a shortest path because the edge from $u_{i-1}$ to $u_{i}$ would always be visited in the algorithm before any edge between $u_{i}$ and nodes further away from the root node than $u_{i-1}$ as this is what the breadth first search guarantees. Each edge is visited when we are studying the neighbours of a node in step \ref{en:loop} of the algorithm.
	\item The edge set $\Ecal_v$ therefore contains paths from every vertex to the root. Hence, all vertices in are connected in the $\Tcal(v)$ subgraph so this is a single component subgraph.
	\item The edge set $\Ecal_v$ has one less edge than the the total number of vertices in the graph which is a necessary and sufficient condition for a single component graph to be a tree.
	\item Thus this algorithm defines a spanning tree $\Tcal(v) = \{ \Vcal, \Ecal_v\}$ that contains a shortest path from every vertex to the root node. That is it is a shortest-path spanning tree.
\end{enumerate}

\newpage

\section{Data Sets}\label{a:datasets}

We used a variety of networks for which data is openly available. These were analysed in two groups.  The first group of eighteen networks we refer to as the ``Konect-SNAP networks'' were analysed in greater detail and we provide further information on each data set here.  The second set of one hundred and twelve networks we describe as our ``Netzschleuder'' networks. We provide have provided independent copies of the data used in \cite{EC22}.

\subsection{Description of Konect-SNAP networks}\label{a:ksdata}

For these networks, all but one can be found on \href{http://konect.cc/}{KONECT} \cite{K13c,KONECT} with the
\href{https://snap.stanford.edu/data/email-Eu-core.html}{\texttt{commun-EU(core)-email}} coming from the \href{https://snap.stanford.edu/data/}{Stanford Large Network Dataset Collection} \cite{SNAP}.
However, many of these networks can be found on other repositories of network data. Our aim was to find networks of different sizes representing contrasting types of interaction which we break down into five broad categories:
social networks (\texttt{social-\ldots}),
communication networks (\texttt{commun-\ldots}), citation networks (\texttt{citation-\ldots}), co-author networks (\texttt{coauth-\ldots}), and hyperlink networks (\texttt{hyperlink-\ldots}). These networks have been used in many contexts in other publications but we will only give a brief summary of each one.

In each case we created a simple graph, ignoring edge directions and weights, node types, time stamps, and any other such information. We took the largest connected component (LCC) of the graph and performed our analysis on this.  Some basic statistics on each graph is given in \tref{t:datastats} and then more detailed information on each data set follows.

\begin{table}[htb!]
	\begin{center}
\begin{tabular}{lrrc}
\hline
Network Name                   & Number of nodes           & Number of edges               & Mean distance \\ \hline
\texttt{social-karate-club}    & 34                        & 78                            & 2.44          \\
\texttt{social-jazz}           & 198                       & 2742                          & 2.21          \\
\texttt{social-hamster}        & 1858                      & 12534                         & 3.39          \\
\texttt{social-oz}             & 217                       & 2672                          & 2.33          \\
\texttt{social-highschool}     & 70                        & 366                           & 2.66          \\
\texttt{social-health}         & 2539                      & 12969                         & 4.52          \\ \hline
\texttt{commun-email}          & 1133                      & 5451                          & 3.65          \\
\texttt{commun-UC-message}     & 1899                      & 59835                         & 3.07          \\
\texttt{commun-EU(core)-email} & 1005                      & 25571                         & 2.59          \\
\texttt{commun-DNC-email}      & 2029                      & 39264                         & 3.37          \\
\texttt{commun-DIGG-reply}     & 30398                     & 87627                         & 4.68          \\ \hline
\texttt{citation-DBLP-cite}    & 12590                     & 49759                         & 4.37          \\
\texttt{citation-Cora}         & 23166                     & 91500                         & 5.74          \\ \hline
\texttt{coauthor-astro-ph}     & 16046                     & 121251                        & 5.10          \\
\texttt{coauthor-netscience}   & 1461                      & 2742                          & 6.28          \\
\texttt{coauthor-pajek}        & 6927                      & 11850                         & 3.79          \\ \hline
\texttt{hyperlink-polblog}     & 1224                      & 33430                         & 2.75          \\
\texttt{hyperlink-blogs}       & 1224                      & 19025                         & 2.72
\end{tabular}
\end{center}
\caption{Summary statistics for the original data sets used in this paper, for the whole graph not just the largest connected component. The mean distance is the average length of the shortest paths between all pairs of connected nodes.}
\label{t:datastats}
\end{table}

\subsubsection*{Social networks}

Social networks capture the social interactions between actors, such as friends, colleagues, clients and students. We used five data sets, the size of networks ranged from 34 to 2539 nodes. On average, we find the mean shortest distance are quite small compare other type of networks (apart from \texttt{social-health} dataset).

The \href{http://konect.cc/networks/ucidata-zachary/}{\texttt{social-karate-club}} is the well-known and much-used Zachary karate club dataset. The original data was collected from the members of a university karate club by Wayne Zachary in 1977 \cite{Z77} and each edge represents some type of social interaction between two members of the club.

The \href{http://konect.cc/networks/moreno_highschool/}{\texttt{social-highschool}} network represents friendships between boys in a small highschool in Illinois, USA. Each boy was asked once in the fall of 1957 and the spring of 1958. This dataset aggregates the results from both dates. A node represents a boy and an edge between two boys shows that at least one boy chose the other as a friend. The original network \cite{C64} is directed, weighted and allows multiple edges.

The \href{http://www.konect.cc/networks/petster-hamster-household/}{\texttt{social-hamster}} network comes from the Koblenz Network Collection (KONECT) \cite{KONECT} where it is described as the ``Hamsterster households network dataset''	but no further information is provided.

The \href{http://www.konect.cc/networks/arenas-jazz/}{\texttt{social-jazz}} network is the collaboration network between Jazz musicians. Each node is a Jazz musician and an edge denotes that two musicians have played together in a band \cite{GD03}.

The \href{http://konect.cc/networks/moreno_oz/}{\texttt{social-oz}} network is a network recording the friendships between 217 residents living at a residence hall located on the Australian National University campus \cite{FWK98}. A node represents a person and edge represent the friendship between them.

The \href{http://konect.cc/networks/moreno_health/}{\texttt{social-health}} network is a network created from a survey of students in 1994/1995 \cite{M01}. Each student was asked to list their five best female and five best male friends. A node represents a student and an edge between two students shows at least one chose the other as a friend.

\subsubsection*{Communication networks}

Communication networks describe the individual messages exchanged between people. Communication networks are often directed and typically contain multiple edges each with distinct time stamps so we are neglecting a lot of information when working with simple graphs representations.

The \href{http://konect.cc/networks/arenas-email/}{\texttt{commun-email}} network is the based on emails sent between members of the University Rovira i Virgili in Tarragona in the south of Catalonia in Spain \cite{GDDGA03}. Nodes are users and each edge represents that at least one email was sent between two users.


The \href{http://konect.cc/networks/dnc-temporalGraph/}{\texttt{commun-DNC-email}} network is built from the emails from the Democratic National Committee, the formal governing body for the United States Democratic Party. A dump of emails of the Democratic National Committee was leaked in 2016. Nodes in the network correspond to persons in the dataset. An edge in the dataset denotes that at least one email has been sent between the two linked nodes.

The \href{http://konect.cc/networks/opsahl-ucsocial/}{\texttt{commun-UC-message}} network represents messages sent between the users of an online community of students from the University of California, Irvine \cite{OP09}. An edge connects two users if they exchanged at least one message.


The \href{https://snap.stanford.edu/data/email-Eu-core.html}{\texttt{commun-EU(core)-email}} is a network representing email sent between members of a large European research institution \cite{LKF07}. An edge represents an email sent between members of the institution (nodes). This data was downloaded from sourced from the \href{https://snap.stanford.edu/data/}{Stanford Large Network Dataset Collection} \cite{SNAP}.

The \href{http://konect.cc/networks/munmun_digg_reply/}{\texttt{commun-DIGG-reply}} data \cite{CSJS09} gives a network of users of the social news website Digg. Each node is a user of the site two users are connected by an edge is one of those users replied to another user at any point.

\subsubsection*{Citation networks}

Citation networks represent documents as nodes in the network, with two nodes linked if one document cites another. These are direct acyclic graphs in principle but here we use a simple graph representation.

The \href{http://konect.cc/networks/dblp-cite/}{\texttt{citation-DBLP-cite}} is the citation network built from the DBLP database of computer science publications \cite{L02}.


The \href{http://konect.cc/networks/subelj_cora/}{\texttt{citation-Cora}} network uses another database of computer science papers, CORA \cite{MNRS00,SB13}. Our simple network is constructed as for the DBLP network.

\subsubsection*{Co-authorship network}

Co-authorship networks are networks connecting authors who have written articles together. Co-authorship networks are normally weighted but we ignore that here.

The \href{http://konect.cc/networks/dimacs10-astro-ph/}{\texttt{coauthor-astro-ph}} network is the co-authorship network from the astrophysics section (\texttt{astro-ph}) of arXiv preprint archive constructed in \cite{N01}. Nodes are authors and an edge denotes a collaboration on at least one paper.

The \href{http://konect.cc/networks/dimacs10-netscience/}{\texttt{coauthor-netscience}} network is a network of co-authors in the area of network science \cite{N06}. Nodes represent authors and edges denote collaborations.

The \href{http://konect.cc/networks/pajek-erdos/}{\texttt{coauthor-pajek}} is the co-authorship graph around Paul Erd\H{o}s \cite{pajek} which can be used to is used to define the ``Erd\H{o}s number''.

\subsubsection*{Hyperlink networks}

In hyperlink networks the nodes are pages or documents. These are linked by an edge if there is at least one hyperlink between these two documents in either direction as here we ignore the direction inherent to hyperlinks.

We use two examples from hyperlinks between blogs about politics during the U.S. Presidential Election of 2004 \cite{AG05}, \href{http://konect.cc/networks/moreno_blogs/}{\texttt{hyperlinks-blogs}} and  \href{http://konect.cc/networks/dimacs10-polblogs/}{\texttt{hyperlink-polblog}}.

\subsection{Description of Netzschleuder networks}\label{a:Netzdata}

We have used an additional 112 networks taken from the \href{https://networks.skewed.de/}{Netzschleuder} repository \cite{P20} though again, many of these are available elsewhere. We made use of simple interface to these network data sets provided by the \texttt{graph-tool} package \cite{P14b} and some examples of our code, along with copies of the datasets, doing this can be found in \cite{EC22}. Our sole selection criteria was that the network would download and run without additional work on our existing code. For this reason some excessively large data sets or those with unusual data structures were not included in our analysis. This means that some of the 112 \href{https://networks.skewed.de/}{Netzschleuder} networks that we analysed are not good candidates for our relationship, e.g.\ they have bipartite structure. We did not exclude such cases to ensure that we have as unbiased a sample of network data sets as possible.
For the 112 networks that we looked at, some basic properties along with additional results are given in \appref{a:netzres} below.
As there are too many datasets to discuss in detail here, please refer to the \href{https://networks.skewed.de/}{Netzschleuder} repository \cite{P20} for further information on individual examples.

\newpage
\section{Additional Results}

These results are not central to the work in the main paper but they are mentioned briefly in the Discussion section. In this section we show these results in more detail.

\subsection{Dependence of fit on $N$}\label{a:depfitN}

\begin{figure}[htb]
	\centering
	\includegraphics[width=1 \textwidth]{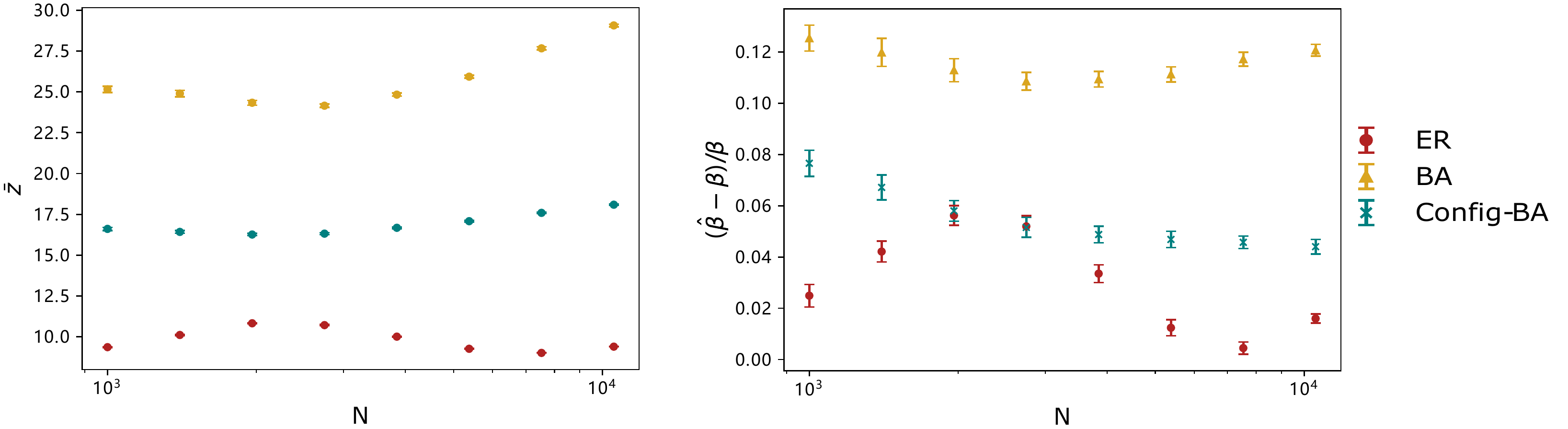}
	\caption{Plots showing the dependence of the best fit values $\zbarfit$ and $\betafit$ on the number of nodes $N$. These are shown for networks from three artificial models: the Erd\H{o}s-R\'{e}yni model (ER, red data), the Barab\'{a}si-Albert model (BA, yellow data), and the randomised Barab\'{a}si-Albert model (Config-BA, blue data), all for average degree $10.0$.
	Data points are the mean values with error bars showing the standard error of the mean estimated from $100$ realisations. 	
	On the left we can see the best fit value of the growth factor $\zbarfit$ has some non-linear dependence on system size $N$. On the right we compare the fitted value $\betafit=\hat{\beta}$ to the value $\beta=\beta(\zbarfit,N)$ predicted from \eqref{a:betadef} using the fitted value $\zbarfit$.
	It is clear that the predicted value $\beta=\beta(\zbarfit,N)$ from \eqref{a:betadef} is 5\% to 15\% below the best fit value though no strong trends are visible on this small range of $N$ values.
}
	\label{f:zbarbetaN}
\end{figure}

For the artificial models we have looked at how the parameters $\zbarfit$ and $\betafit$, found by fitting to \eqref{a:farnesskform},
depend on the number of nodes $N$. For $\betafit$ we compare the fitted value against the value $\beta=\beta(\zbarfit,N)$ predicted from \eqref{a:betadef} so are factoring out the expected $\ln(N)$ contribution to $\beta$. Results are shown in \figref{f:zbarbetaN}. In all cases there is no strong variation and while there are signs of some systematic variation, it is at a small scale and no clear pattern emerges.

\subsection{Higher-order polynomial fit}\label{a:higherpolyfit}

An interesting feature of our work is the failure of four of our eighteen real-world networks to give us a good fit: \texttt{social-jazz}, \texttt{commun-UC-message}, \texttt{coauthor-astro-ph} and \texttt{coauthor-pajek}. This is clear from their values of reduced chi-square which are all greater than ten. Several of the plots, particularly those showing the fractional error, also show issues with these data sets but also some clear trends in some other data sets even if this is within statistical fluctuation for each individual point. In the case of the \texttt{social-jazz} network we could dismiss this as this is such a small network, though we note that our relationship has worked well for several smaller networks. However the other three networks with high $\chisqred$ one to fifteen thousand nodes and these show our relationship \eqref{a:farnesskform} is not the last word. Even when the chi-square measure looks good, the plots of fractional error show a convincing trend that is not captured by our relationship.  Again we stress that these deviation are not that large, no worse than $5\%$ in most cases. Nevertheless this points to the need to go beyond our simple derivation.

One way to get get a better fit is to try to fit a higher order polynomial in $\ln(k)$ to the  inverse closeness values, that is
\begin{equation}
	1/c=\sum_{i=0}^{m}p_i (\ln k)^{i}
	\label{a:higherorder}
\end{equation}
where $m$ is the maximum number of the parameters in the polynomial fit, $p_i$ is the coefficient for $i$-th power of $\ln k$.
Working with $m>1$ is not motivated by any theoretical consideration but we work with it here because it is easy to implement. We have not found this to be particularly effective except in one case, \texttt{commun-DIGG-reply}, where we already had a good fit but could see clear trends in the fractional deviation plots. Results for some examples are shown below. The data on first order ($m=1$) corresponds to what was used in the main text.

\begin{figure}[htb!]
	\centering
	\includegraphics[width=0.49 \textwidth]{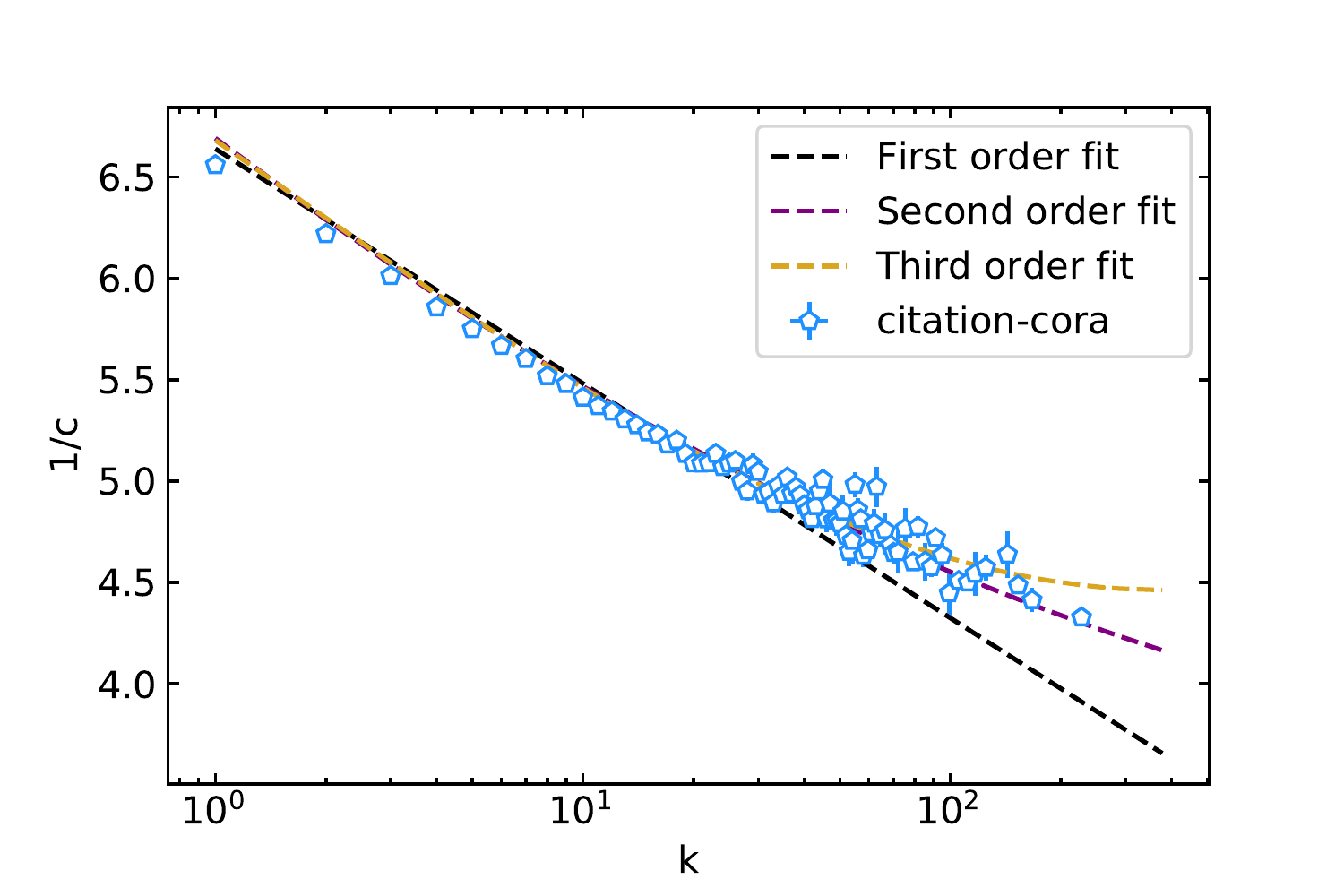}
	\hspace*{0.04\textwidth}
	\includegraphics[width=0.45 \textwidth]{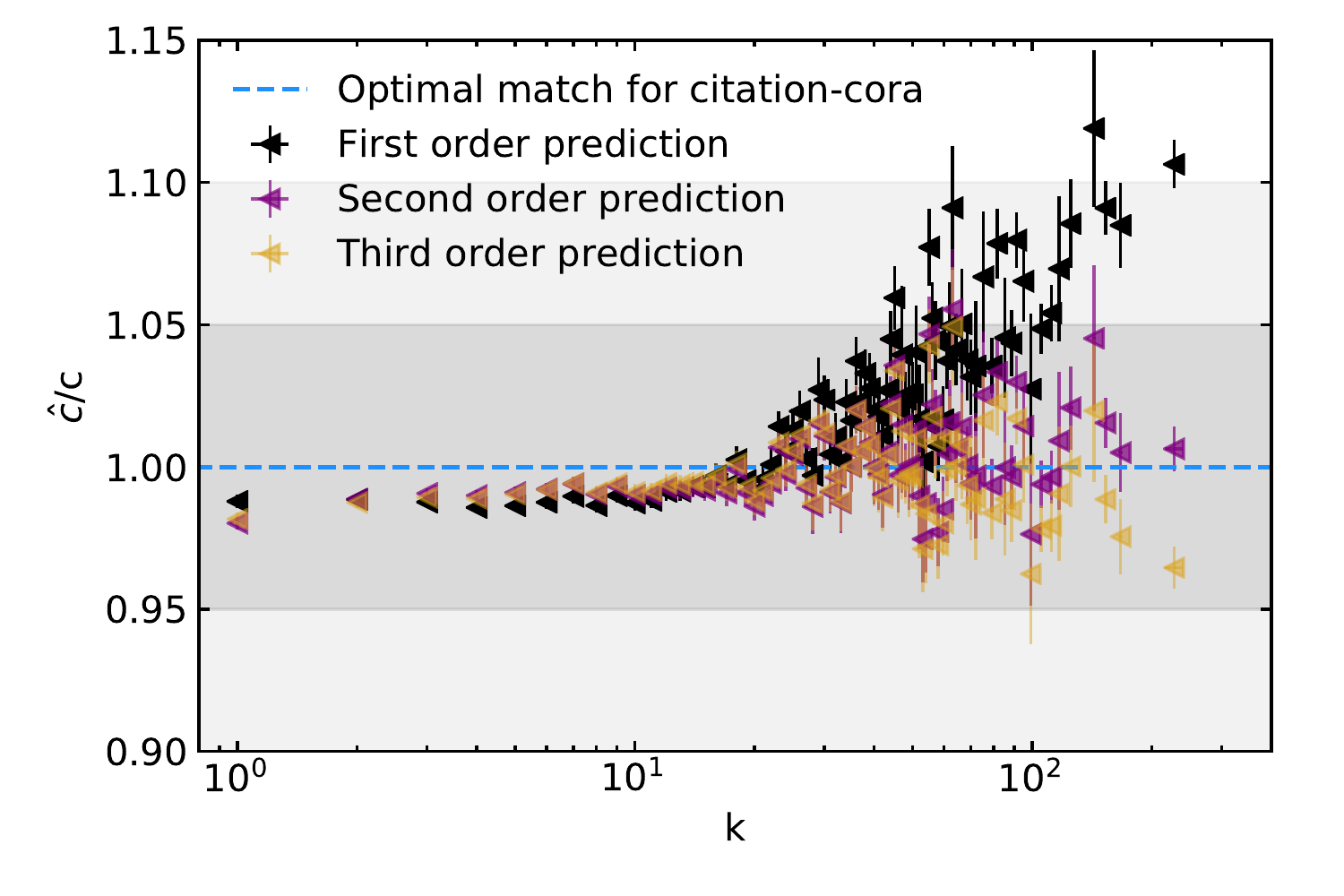}
	\caption{The effect of fitting inverse closeness to higher order polynomials in the logarithm of degree $\ln(k)$ \eqref{a:higherorder} for the CORA citation data \texttt{citation-Cora}.  On the left we show the data points (means with standard error of mean for error bars) against the dashed lines for different polynomial fits. On the right we show the fitted value $\hat{c}$ divided by the data $c$ against degree $k$. The shaded bands mark $5\%$ and $10\%$ deviations.}
	\label{f:corahigherorder}
\end{figure}

\begin{figure}[htb!]
	\centering
	\includegraphics[width=0.49 \textwidth]{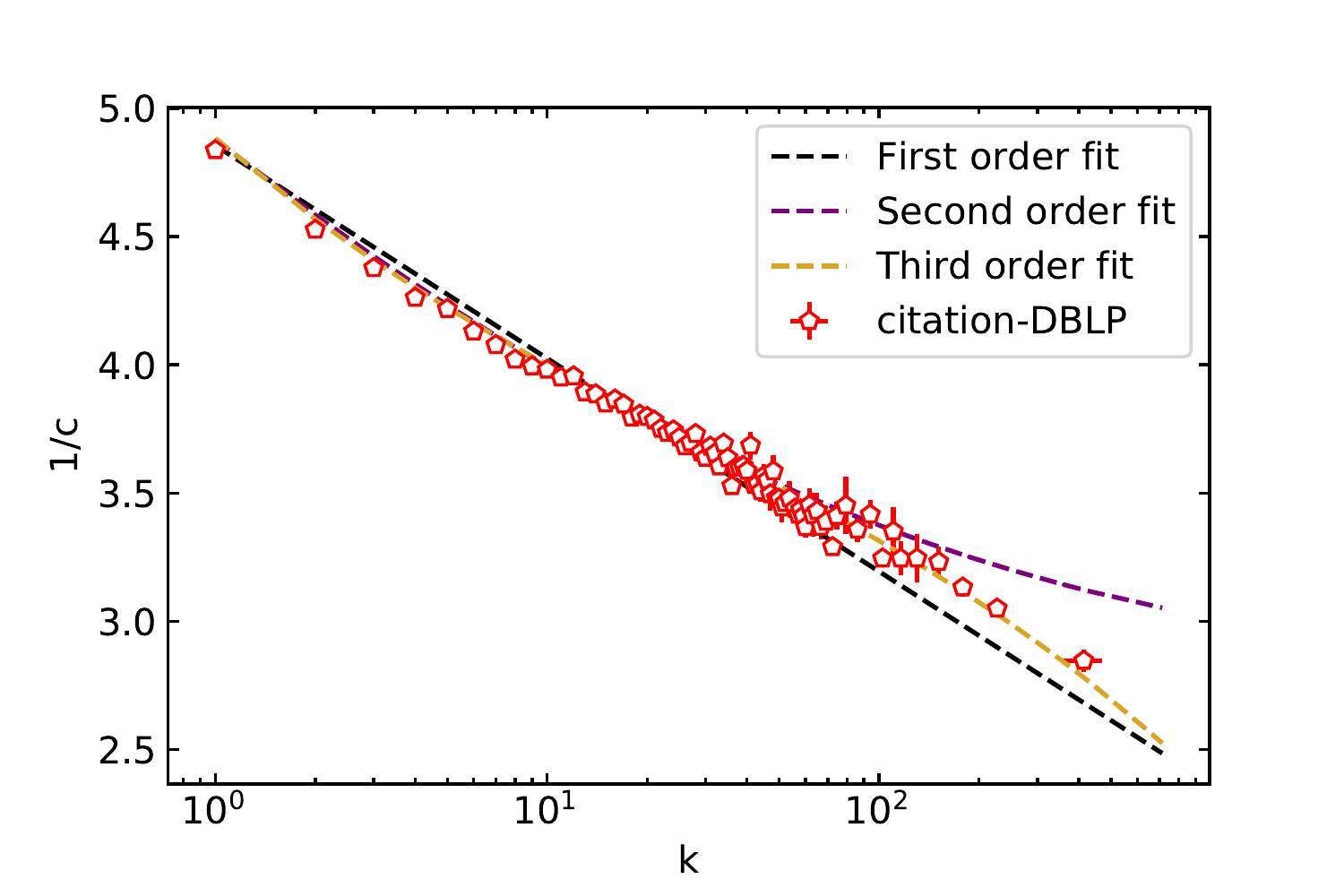}
	\hspace*{0.04\textwidth}
	\includegraphics[width=0.45 \textwidth]{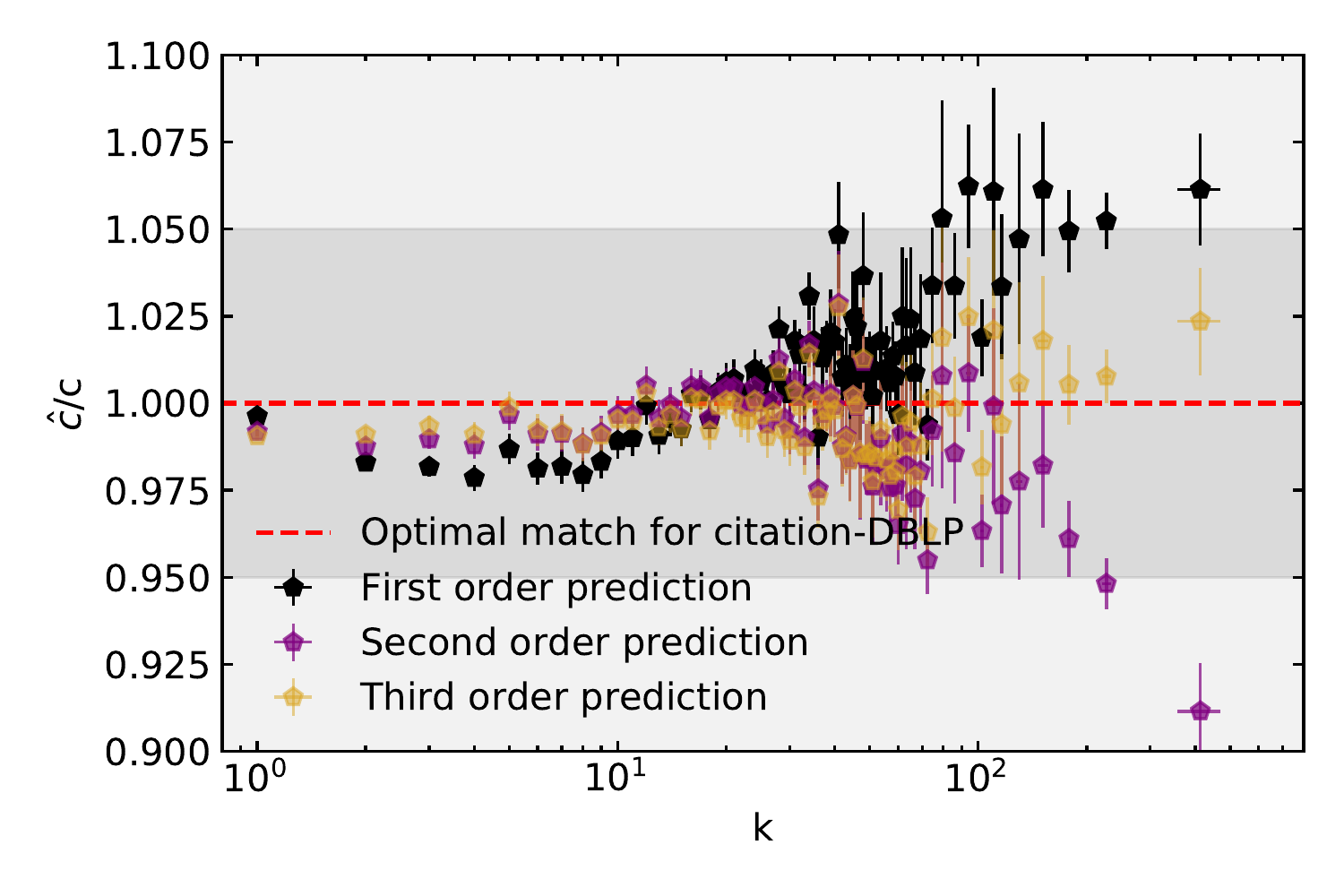}
	\caption{The effect of fitting inverse closeness to higher order polynomials in the logarithm of degree $\ln(k)$ \eqref{a:higherorder} for the DBLP citation data \texttt{citation-DBLP-cite}.  On the left we show the data points (means with standard error of mean for error bars) against the dashed lines for different polynomial fits. On the right we show the fitted value $\hat{c}$ divided by the data $c$ against degree $k$.The shaded bands mark $5\%$ and $10\%$ deviations.}
	\label{f:dblphigherorder}
\end{figure}

\begin{figure}[htb!]
	\centering
	\includegraphics[width=0.49 \textwidth]{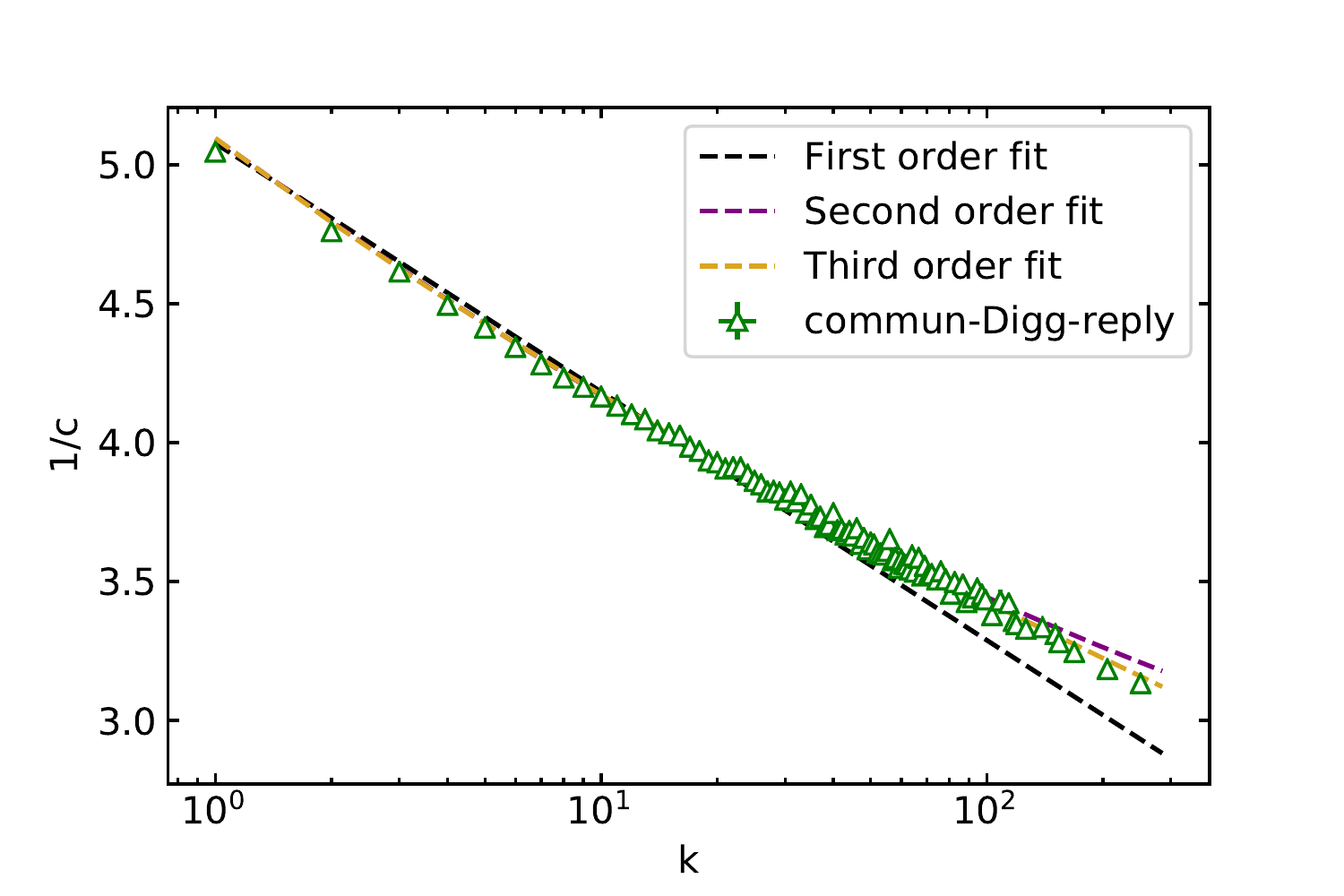}
	\hspace*{0.04\textwidth}
	\includegraphics[width=0.45 \textwidth]{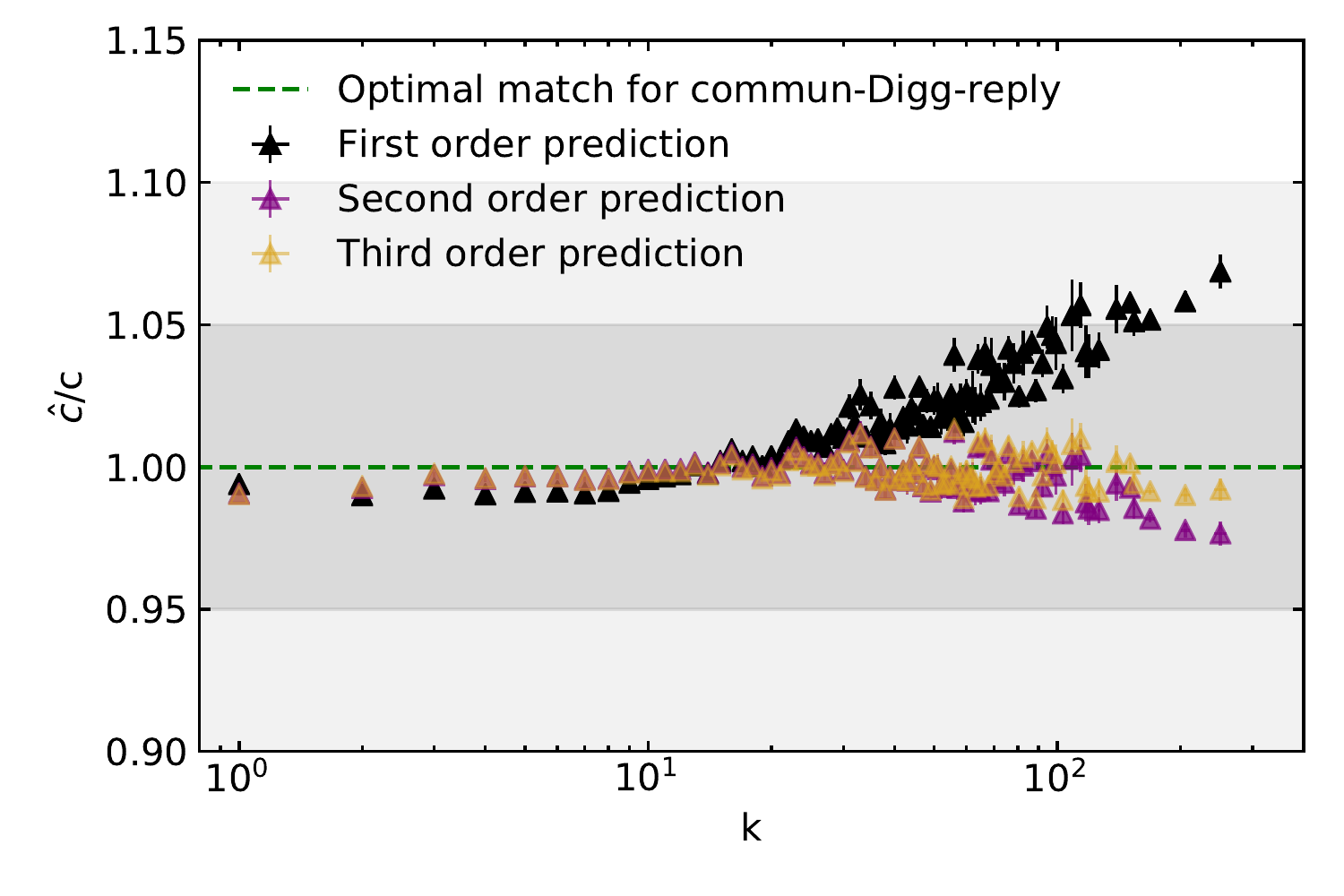}
	\caption{The effect of fitting inverse closeness to higher order polynomials in the logarithm of degree $\ln(k)$ \eqref{a:higherorder} for the DIGG communication network \texttt{commun-DIGG-reply} .  On the left we show the data points (means with standard error of mean for error bars) against the dashed lines for different polynomial fits. On the right we show the fitted value $\hat{c}$ divided by the data $c$ against degree $k$.The shaded bands mark $5\%$ and $10\%$ deviations.}
	\label{f:digghigherorder}
\end{figure}

\begin{figure}[htb!]
	\centering
	\includegraphics[width=1 \textwidth]{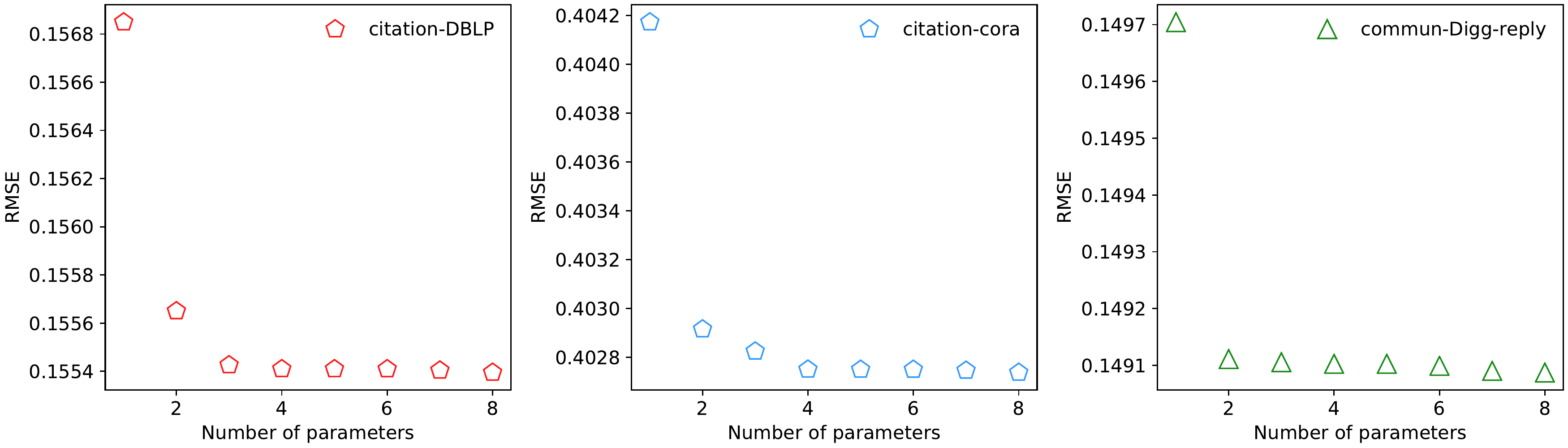}
	\caption{Higher order correction revealed by comparing root mean squared error (RMSE). For those datasets suggesting a dependency of higher order relation between $\ln(k)$ and $1/c$, we use the higher order polynomial fit of \eqref{a:higherorder} but the improvement is not significant.}
	\label{f:higherorder}
\end{figure}


\clearpage

\subsection{Fluctuations in Closeness}\label{a:cfluct}

\begin{figure}[htb!]
	\begin{center}
		\includegraphics[width=0.8\linewidth]{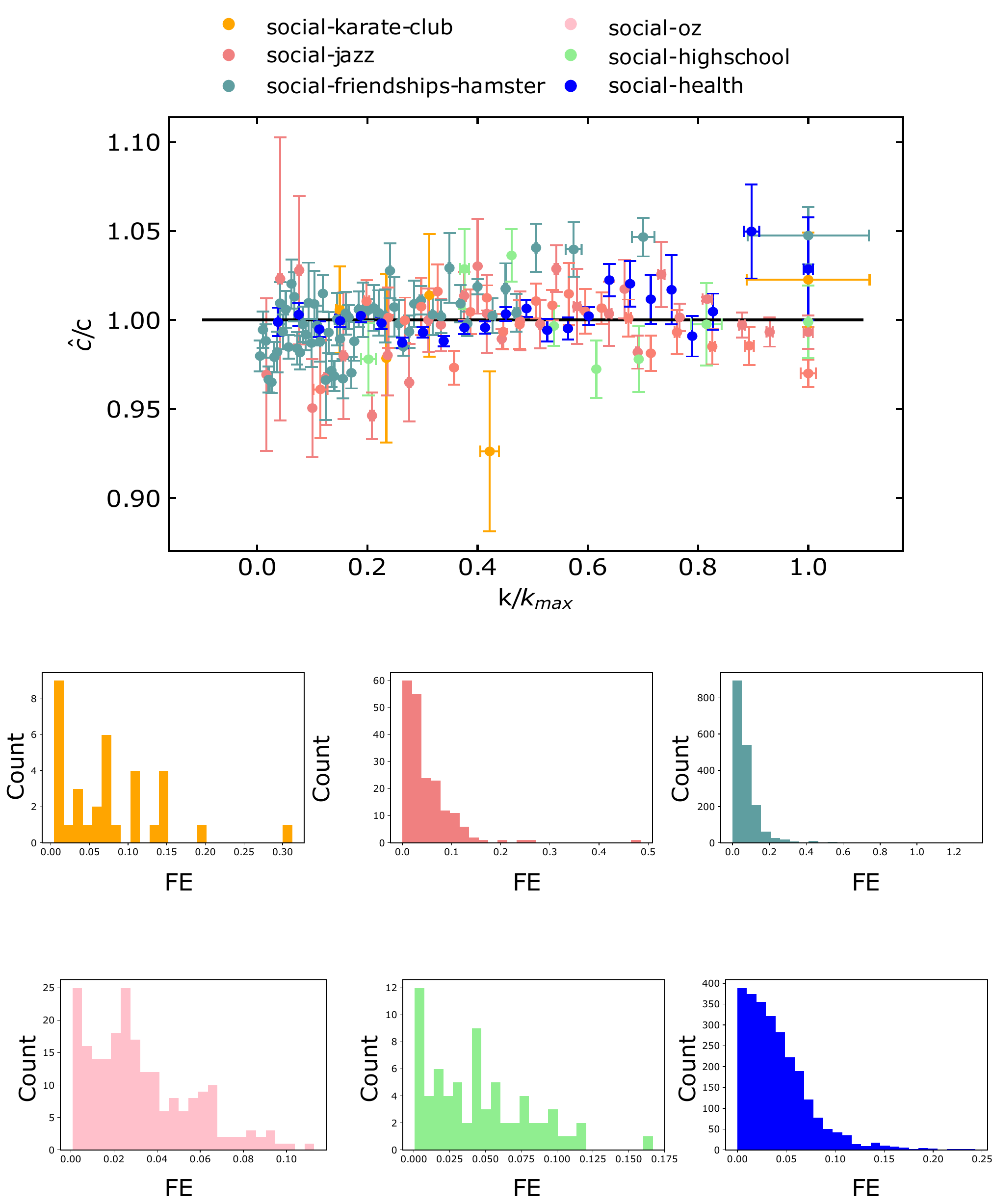}
	\end{center}
	\caption{Results for six friendship networks derived from real-world data,
            see \tabref{t:datastats} for the statistics of each dataset.
            In the top plot, the horizontal axis is the degree divided by the largest degree in each data set. The vertical axis is the predicted value $\hat{c}$ for degree $k$ divided by the equivalent measured value $c=\texpect{c}_k$ averaged over nodes with the same degree $k$. The predicted value comes from \eqref{a:farnesskform} using values $\zbarfit$ and $\betafit$ obtained by fitting the data to \eqref{a:farnesskform}.   The error bars show the standard error of the mean. The histograms show the number of data points in each data set with a specified  absolute value of the fractional error where $\mathrm{FE}= (|c^{-1}-\hat{c}^{-1}) / \hat{c}^{-1}$. We can see that even for a small network, such as the the Karate club data set, our conjecture \eqref{a:farnesskform} is successful.}
	\label{f:friendnetwork}
\end{figure}

\begin{figure}[htb!]
    \centering
    \includegraphics[width=0.7 \textwidth]{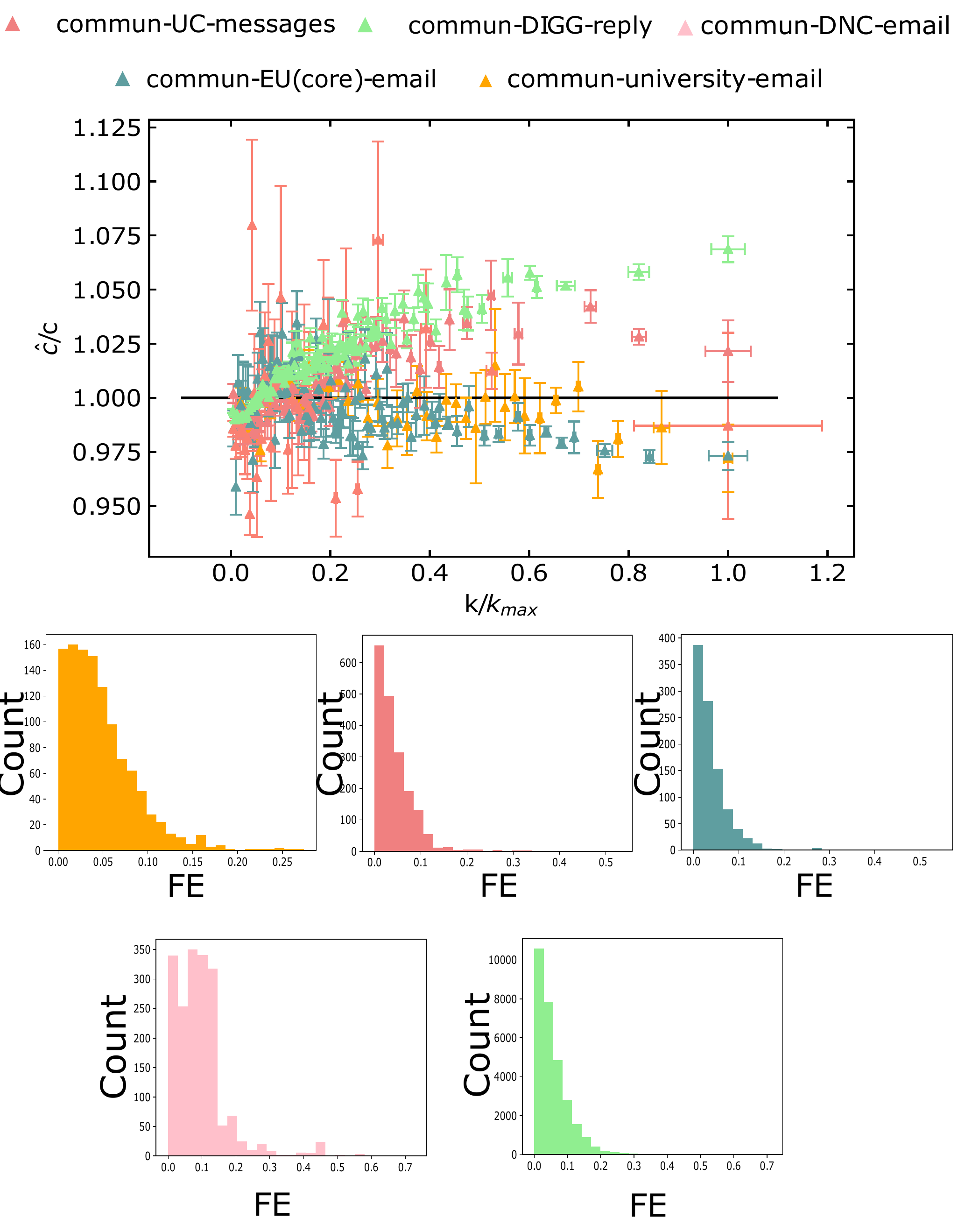}
    \caption{Results for five communication networks derived from real-world data,
    see \tabref{t:datastats} for the statistics of each dataset.
    In the top plot, the horizontal axis is the degree divided by the largest degree in each data set. The vertical axis is the predicted value $\hat{c}$ for degree $k$ divided by the equivalent measured value $c=\texpect{c}_k$ averaged over nodes with the same degree $k$. The predicted value comes from \eqref{a:farnesskform} using values $\zbarfit$ and $\betafit$ obtained by fitting the data to \eqref{a:farnesskform}.   The error bars show the standard error of the mean. The histograms show the number of data points in each data set with a specified  absolute value of the fractional error where $\mathrm{FE}= (|c^{-1}-\hat{c}^{-1}) / \hat{c}^{-1}$. The poor fit of \texttt{commun-DIGG-reply} reflected in the bad $\chisqred$ value is clearly visible here.}
    \label{f:commnet}
\end{figure}

We can take a closer look at some of the fluctuations in closeness around the predicted value for the six social networks in \figref{f:friendnetwork} and the five communication networks \figref{f:commnet}. The distribution of the number of nodes for a given range of fractional error in their closeness value, the closeness measured compared to the predicted value from the fit, shows variations between data sets but generally confirms that most individual nodes have a closeness that is reasonably similar to the prediction.

\clearpage

\subsection{Additional Results for the Netzschleuder Networks}\label{a:netzres}

Further information on 112 networks taken from the \href{https://networks.skewed.de/}{Netzschleuder} repository \cite{P20}. These were the networks which would download and run automatically using our code without any additional work. The information is provided here is available as a spreadsheet file provided \texttt{extra\_result220208tidy.xlsx} in \cite{EC22}. We have split the results into three tables: in \tabref{t:extraresult1} are the 49  networks with $\chi^2_r$ of $2.0$ or more, in \tabref{t:extraresult2} are the 49  networks with an excellent fit to our degree-closeness relationship \eqref{a:farnesskform}, and finally in \tabref{t:extraresult3} are the 12  networks where the topology in the simple graph representation used in our analysis means their results are of little relevance here.

The columns of tables \ref{t:extraresult1}, \ref{t:extraresult2} and \ref{t:extraresult3} are labelled as follows:
\begin{center}
\begin{tabular}{r l}
ID                                & internal identification number. \\
Name                              & name of the Network \href{https://networks.skewed.de/}{Netzschleuder} repository \cite{P20}.\\
$N$                               & Number of nodes \\
$\texpect{k}$                     & Average degree \\
$\rho$                            & Density of network, $\rho = \texpect{k}/(N-1)$ \\
Rk                                & Rank of network by  $\chi^2_r$ value \\
$\chi^2_r$                        & Reduced chi-square value for fit of closeness to \eqref{a:farnesskform} \\
$\zbarfit$                        & Growth parameter value found by a fit to \eqref{a:farnesskform} \\
$\zbarfit_\mathrm{err}$           & Error in $\bar{z}$ \\
$\betafit$                        & Value of $\beta$ parameter found by a fit to \eqref{a:farnesskform} \\
$\betafit_\mathrm{err}$           & Error in $\hat{\beta}$  \\
$\beta$                           & Value of beta obtained using $\bar{z}$  value in formula \eqref{a:betadef} \\
$\beta_\mathrm{err}$              & Error in ${\beta}$  \\
$\rho^\mathrm{P}$                 & Pearson correlation measure \\
$\rho^\mathrm{IP}$                & Improved Pearson correlation measure \\
$\ellav$                          & Average path length in network\\
$\ellav^\mathrm{FP}$              & avgPathLengthFitPredicted \\
$\ellav^\mathrm{FP}_\mathrm{err}$ & Error in $\ellav^\mathrm{FP}$  \\
$\ellav^\mathrm{zP}$              & avgPathLengthzbarPredicted \\
$\ellav^\mathrm{zP}_\mathrm{err}$ & Error in $\ellav^\mathrm{zP}$ \\
\end{tabular}
\end{center}

\begin{sidewaystable}[htpb]
\tiny
    \centering
    \begin{tabular}{|l|l|l|l|l|l|l|l|l|l|l|l|l|l|l|l|l|l|l|l|}
    \hline
        ID & Name & $N$ & $\texpect{k}$ & $\rho$ density & Rk & $\chi^2_r$ & $\zbarfit$ & $\zbarfit_\mathrm{err}$ & $\betafit$ & $\betafit_\mathrm{err}$ & $\beta$ & $\beta_\mathrm{err}$ & $\rho^\mathrm{P}$ &  $\rho^\mathrm{IP}$ & $\ellav$ & $\ellav^\mathrm{FP}$ & $\ellav^\mathrm{FP}_\mathrm{err}$ & $\ellav^\mathrm{zP}$ & $\ellav^\mathrm{zP}_\mathrm{err}$ \\ \hline \hline
        47 & un\_migrations & 231 & 72.7 & 0.32 & 1 & 3187.1 & 37 & 4 & 2.81 & 0.03 & 2.53 & 0.05 & 1.0 & -0.9 & 1.7 & 1.68 & 0.05 & 1.412 & 0.014 \\ \hline
        77 & fediverse & 816 & 280.9 & 0.34 & 2 & 2941.2 & 1170 & 150 & 2.370 & 0.014 & 1.95 & 0.02 & 1.0 & -0.9 & 1.7 & 1.671 & 0.019 & 1.251 & 0.005 \\ \hline
        67 & collins\_yeast & 1004 & 16.6 & 0.0166 & 3 & 162.2 & 8 & 0.9 & 6.60 & 0.06 & 4.39 & 0.18 & 0.7 & -0.5 & 5.5 & 5.55 & 0.09 & 3.34 & 0.13 \\ \hline
        72 & dnc & 849 & 24.5 & 0.029 & 4 & 60.1 & 1200 & 500 & 3.01 & 0.02 & 1.95 & 0.05 & 0.7 & -0.5 & 2.8 & 2.76 & 0.03 & 1.69 & 0.04 \\ \hline
        29 & sp\_hospital & 75 & 30.4 & 0.41 & 5 & 37.3 & 15.9 & 1.3 & 2.78 & 0.04 & 2.61 & 0.05 & 1.0 & -1 & 1.6 & 1.60 & 0.05 & 1.423 & 0.015 \\ \hline
        58 & plant\_pol\_kato & 768 & 3.1 & 0.0040 & 6 & 34.1 & 29 & 5 & 4.040 & 0.015 & 2.99 & 0.11 & 0.5 & -0.6 & 3.9 & 3.903 & 0.017 & 2.85 & 0.10 \\ \hline
        50 & contact & 274 & 15.5 & 0.057 & 7 & 28.0 & 88 & 6 & 2.756 & 0.007 & 2.26 & 0.02 & 1.0 & -1 & 2.4 & 2.424 & 0.009 & 1.929 & 0.014 \\ \hline
        27 & law\_firm & 71 & 28.4 & 0.41 & 8 & 20.7 & 16.1 & 1.3 & 2.78 & 0.03 & 2.58 & 0.05 & 1.0 & -1 & 1.6 & 1.60 & 0.05 & 1.395 & 0.013 \\ \hline
        32 & polbooks & 105 & 8.4 & 0.081 & 9 & 19.5 & 12 & 4 & 3.86 & 0.11 & 2.9 & 0.3 & 0.6 & -0.6 & 3.1 & 3.08 & 0.15 & 2.14 & 0.16 \\ \hline
        14 & montreal & 29 & 5.2 & 0.186 & 10 & 14.1 & 8.3 & 0.9 & 2.77 & 0.04 & 2.67 & 0.09 & 0.9 & -1 & 2.2 & 2.15 & 0.05 & 2.06 & 0.06 \\ \hline
        42 & jazz\_collab & 198 & 27.7 & 0.141 & 11 & 13.1 & 15.3 & 1.7 & 3.35 & 0.05 & 2.98 & 0.08 & 0.9 & -0.9 & 2.2 & 2.24 & 0.07 & 1.87 & 0.04 \\ \hline
        24 & terrorists\_911 & 62 & 4.9 & 0.080 & 12 & 12.8 & 8.6 & 2 & 3.55 & 0.08 & 3.0 & 0.2 & 0.8 & -0.8 & 2.9 & 2.95 & 0.10 & 2.39 & 0.16 \\ \hline
        89 & anybeat & 8518 & 9.6 & 0.00113 & 13 & 9.8 & 6600 & 2000 & 2.976 & 0.006 & 2.03 & 0.04 & 0.2 & -0.3 & 2.9 & 2.856 & 0.007 & 1.91 & 0.03 \\ \hline
        80 & jung & 77 & 7.6 & 0.100 & 14 & 9.6 & 8.2 & 1.6 & 3.28 & 0.08 & 3.1 & 0.2 & 0.8 & -0.8 & 2.5 & 2.46 & 0.11 & 2.32 & 0.13 \\ \hline
        82 & jdk & 77 & 7.6 & 0.100 & 15 & 9.6 & 8.2 & 1.6 & 3.28 & 0.08 & 3.1 & 0.2 & 0.8 & -0.8 & 2.5 & 2.46 & 0.11 & 2.32 & 0.13 \\ \hline
        49 & sp\_primary\_school & 242 & 68.7 & 0.29 & 16 & 8.2 & 24.3 & 0.7 & 3.032 & 0.013 & 2.75 & 0.02 & 1.0 & -1 & 1.7 & 1.732 & 0.018 & 1.450 & 0.005 \\ \hline
        109 & epinions\_trust & 32223 & 21.2 & 0.00066 & 17 & 7.7 & 31.5 & 0.4 & 4.198 & 0.003 & 4.03 & 0.01 & 0.5 & -0.8 & 3.7 & 3.676 & 0.004 & 3.510 & 0.011 \\ \hline
        37 & foodweb\_baywet & 103 & 30.1 & 0.30 & 18 & 7.6 & 22 & 1.6 & 2.80 & 0.03 & 2.53 & 0.04 & 1.0 & -1 & 1.7 & 1.72 & 0.04 & 1.459 & 0.013 \\ \hline
        56 & celegans\_metabolic & 453 & 8.9 & 0.0197 & 19 & 7.0 & 20 & 4 & 3.25 & 0.04 & 3.06 & 0.14 & 0.6 & -0.6 & 2.7 & 2.66 & 0.06 & 2.48 & 0.10 \\ \hline
        16 & windsurfers & 43 & 15.6 & 0.37 & 20 & 6.0 & 10.5 & 1.2 & 2.80 & 0.06 & 2.67 & 0.09 & 1.0 & -1 & 1.7 & 1.67 & 0.08 & 1.53 & 0.03 \\ \hline
        46 & residence\_hall & 214 & 17.1 & 0.080 & 21 & 5.1 & 12.2 & 1.1 & 3.47 & 0.04 & 3.20 & 0.08 & 0.9 & -0.9 & 2.4 & 2.38 & 0.06 & 2.11 & 0.04 \\ \hline
        71 & plant\_pol\_robertson & 1882 & 16.2 & 0.0086 & 22 & 5.0 & 68 & 5 & 3.718 & 0.009 & 2.80 & 0.03 & 0.7 & -0.8 & 3.3 & 3.306 & 0.011 & 2.39 & 0.02 \\ \hline
        43 & cintestinalis & 201 & 25.3 & 0.127 & 23 & 4.7 & 15.1 & 1.1 & 3.27 & 0.03 & 3.00 & 0.05 & 0.9 & -0.9 & 2.2 & 2.15 & 0.04 & 1.89 & 0.03 \\ \hline
        18 & macaque\_neural & 47 & 13.3 & 0.29 & 24 & 4.2 & 9.7 & 1.2 & 2.93 & 0.06 & 2.76 & 0.10 & 1.0 & -0.9 & 1.9 & 1.85 & 0.08 & 1.68 & 0.04 \\ \hline
        92 & google & 12354 & 23 & 0.00186 & 25 & 4.2 & 1E+20 & 7E+20 & 2.173 & 0.006 & 1.20 & 0.02 & 0.1 & -0.1 & 2.1 & 2.118 & 0.008 & 1.148 & 0.015 \\ \hline
        3 & kangaroo & 17 & 10.7 & 0.67 & 26 & 4.1 & 12.3 & 2 & 2.24 & 0.06 & 2.18 & 0.08 & 1.0 & -1 & 1.4 & 1.36 & 0.08 & 1.30 & 0.02 \\ \hline
        63 & crime & 1263 & 2.2 & 0.00174 & 27 & 3.8 & 3.1 & 0.4 & 13.41 & 0.08 & 7.4 & 0.7 & 0.3 & -0.2 & 13 & 12.96 & 0.10 & 7.0 & 0.7 \\ \hline
        81 & reactome & 5973 & 48.8 & 0.0082 & 28 & 3.8 & 14.6 & 0.7 & 5.27 & 0.02 & 4.29 & 0.06 & 0.6 & -0.6 & 4.2 & 4.21 & 0.03 & 3.23 & 0.04 \\ \hline
        13 & dutch\_criticism & 35 & 4.6 & 0.135 & 29 & 3.7 & 3.9 & 0.5 & 3.66 & 0.10 & 3.8 & 0.3 & 0.9 & -0.9 & 2.7 & 2.71 & 0.13 & 2.81 & 0.16 \\ \hline
        28 & blumenau\_drug & 75 & 4.8 & 0.065 & 30 & 3.6 & 10.6 & 1.7 & 3.32 & 0.04 & 2.89 & 0.13 & 0.8 & -0.9 & 2.8 & 2.81 & 0.06 & 2.38 & 0.10 \\ \hline
        55 & eu\_airlines & 417 & 14.2 & 0.034 & 31 & 3.5 & 39 & 6 & 3.25 & 0.02 & 2.66 & 0.07 & 0.8 & -0.8 & 2.8 & 2.76 & 0.03 & 2.17 & 0.05 \\ \hline
        0 & sa\_companies & 11 & 2.4 & 0.24 & 32 & 3.4 & 2.27 & 0.23 & 3.56 & 0.14 & 4.0 & 0.3 & 0.9 & -0.9 & 2.6 & 2.62 & 0.18 & 3.1 & 0.2 \\ \hline
        25 & train\_terrorists & 64 & 7.6 & 0.121 & 33 & 3.4 & 10.5 & 2.6 & 3.40 & 0.08 & 2.8 & 0.2 & 0.9 & -0.8 & 2.7 & 2.69 & 0.11 & 2.12 & 0.12 \\ \hline
        17 & elite & 44 & 4.5 & 0.105 & 34 & 3.1 & 5 & 0.8 & 3.59 & 0.09 & 3.5 & 0.2 & 0.9 & -0.8 & 2.8 & 2.76 & 0.12 & 2.62 & 0.16 \\ \hline
        84 & elec & 1300 & 56.2 & 0.043 & 35 & 3.1 & 76 & 4 & 3.087 & 0.011 & 2.67 & 0.02 & 0.8 & -0.9 & 2.2 & 2.243 & 0.015 & 1.823 & 0.011 \\ \hline
        6 & november17 & 22 & 6 & 0.29 & 36 & 3.0 & 9.7 & 2.1 & 2.66 & 0.07 & 2.43 & 0.14 & 0.9 & -0.9 & 2 & 1.98 & 0.10 & 1.75 & 0.08 \\ \hline
        107 & facebook\_wall & 30793 & 9.2 & 0.00030 & 37 & 2.9 & 6.52 & 0.05 & 6.187 & 0.005 & 6.61 & 0.03 & 0.7 & -0.8 & 5.3 & 5.288 & 0.006 & 5.71 & 0.02 \\ \hline
        110 & slashdot\_zoo & 26997 & 21.3 & 0.00079 & 38 & 2.5 & 38.1 & 0.5 & 4.019 & 0.003 & 3.82 & 0.01 & 0.6 & -0.9 & 3.4 & 3.446 & 0.003 & 3.249 & 0.009 \\ \hline
        104 & inploid & 4542 & 13.4 & 0.0030 & 39 & 2.5 & 400 & 70 & 2.843 & 0.010 & 2.41 & 0.04 & 0.2 & -0.5 & 2.5 & 2.544 & 0.013 & 2.11 & 0.03 \\ \hline
        64 & polblogs & 793 & 34 & 0.043 & 40 & 2.4 & 123 & 12 & 3.036 & 0.013 & 2.39 & 0.03 & 0.8 & -0.9 & 2.4 & 2.424 & 0.018 & 1.781 & 0.016 \\ \hline
        21 & cs\_department & 61 & 11.6 & 0.193 & 41 & 2.4 & 9.2 & 1.1 & 3.10 & 0.06 & 2.92 & 0.10 & 0.9 & -0.9 & 2.1 & 2.06 & 0.08 & 1.88 & 0.05 \\ \hline
        59 & unicodelang & 858 & 2.9 & 0.0034 & 42 & 2.4 & 13.6 & 2.2 & 4.41 & 0.02 & 3.64 & 0.17 & 0.4 & -0.5 & 4.2 & 4.21 & 0.03 & 3.44 & 0.16 \\ \hline
        75 & openflights & 3147 & 11.9 & 0.0038 & 43 & 2.3 & 20 & 1.3 & 4.411 & 0.014 & 3.72 & 0.06 & 0.6 & -0.6 & 3.9 & 3.924 & 0.018 & 3.24 & 0.05 \\ \hline
        87 & wiki\_rfa & 2449 & 78.6 & 0.032 & 44 & 2.2 & 102 & 5 & 3.151 & 0.009 & 2.70 & 0.02 & 0.7 & -0.9 & 2.3 & 2.292 & 0.013 & 1.837 & 0.009 \\ \hline
        51 & celegansneural & 239 & 14.4 & 0.061 & 45 & 2.2 & 12 & 1.2 & 3.48 & 0.04 & 3.26 & 0.09 & 0.8 & -0.9 & 2.5 & 2.49 & 0.06 & 2.27 & 0.05 \\ \hline
        73 & interactome\_figeys & 8 & 3 & 0.43 & 46 & 2.0 & 5.2 & 2 & 2.30 & 0.16 & 2.4 & 0.3 & 0.9 & -0.9 & 1.7 & 1.7 & 0.2 & 1.75 & 0.18 \\ \hline
        95 & internet\_as & 22963 & 4.2 & 0.00018 & 47 & 2.0 & 24.3 & 1.1 & 4.061 & 0.004 & 4.18 & 0.04 & 0.2 & -0.4 & 3.8 & 3.842 & 0.005 & 3.96 & 0.04 \\ \hline
        35 & football & 115 & 10.7 & 0.094 & 48 & 2.0 & 26 & 28 & 3.2 & 0.2 & 2.5 & 0.5 & 0.3 & -0.3 & 2.5 & 2.5 & 0.3 & 1.8 & 0.3 \\ \hline
        36 & football\_tsevans & 115 & 10.7 & 0.094 & 49 & 2.0 & 26 & 28 & 3.2 & 0.2 & 2.5 & 0.5 & 0.3 & -0.3 & 2.5 & 2.5 & 0.3 & 1.8 & 0.3 \\ \hline
    \end{tabular}
    \caption{Further information on 49 of the 112 networks taken from the \href{https://networks.skewed.de/}{Netzschleuder} repository \cite{P20}. These have $\chi^2_r \geq 2.0$ when fitting the closeness-degree data to the relationship \eqref{a:farnesskform}. Taken from the file \texttt{extra\_result220208tidy.xlsx} provided in \cite{EC22}.}
    \label{t:extraresult1}
\end{sidewaystable}

\begin{sidewaystable}[htpb]
\tiny
    \centering
    \begin{tabular}{|l|l|l|l|l|l|l|l|l|l|l|l|l|l|l|l|l|l|l|l|}
    \hline
        ID & Name & $N$ & $\texpect{k}$ & $\rho$ density & Rk & $\chi^2_r$ & $\zbarfit$ & $\zbarfit_\mathrm{err}$ & $\betafit$ & $\betafit_\mathrm{err}$ & $\beta$ & $\beta_\mathrm{err}$ & $\rho^\mathrm{P}$ &  $\rho^\mathrm{IP}$ & $\ellav$ & $\ellav^\mathrm{FP}$ & $\ellav^\mathrm{FP}_\mathrm{err}$ & $\ellav^\mathrm{zP}$ & $\ellav^\mathrm{zP}_\mathrm{err}$ \\ \hline         \hline
        103 & email\_enron & 33696 & 10.7 & 0.00032 & 50 & 1.8 & 132 & 7 & 4.297 & 0.004 & 3.14 & 0.02 & 0.4 & -0.4 & 4 & 4.025 & 0.005 & 2.87 & 0.02 \\ \hline
        65 & new\_zealand\_collab & 1463 & 5.8 & 0.0040 & 51 & 1.8 & 169 & 26 & 2.920 & 0.008 & 2.43 & 0.04 & 0.5 & -0.7 & 2.7 & 2.749 & 0.009 & 2.25 & 0.04 \\ \hline
        102 & topology & 34761 & 6.2 & 0.00018 & 52 & 1.7 & 39 & 1.3 & 3.989 & 0.003 & 3.87 & 0.03 & 0.2 & -0.5 & 3.8 & 3.768 & 0.004 & 3.65 & 0.03 \\ \hline
        48 & physician\_trust & 95 & 8 & 0.085 & 53 & 1.7 & 10.9 & 1.4 & 3.30 & 0.05 & 2.96 & 0.11 & 0.9 & -0.9 & 2.5 & 2.46 & 0.06 & 2.13 & 0.06 \\ \hline
        30 & lesmis & 77 & 6.6 & 0.087 & 54 & 1.6 & 23 & 8 & 3.10 & 0.06 & 2.41 & 0.16 & 0.8 & -0.7 & 2.6 & 2.64 & 0.08 & 1.95 & 0.11 \\ \hline
        31 & sp\_office & 92 & 16.4 & 0.180 & 55 & 1.6 & 17.5 & 2 & 2.90 & 0.04 & 2.62 & 0.07 & 0.9 & -0.9 & 2 & 1.96 & 0.05 & 1.68 & 0.03 \\ \hline
        53 & facebook\_friends & 329 & 11.9 & 0.036 & 56 & 1.6 & 17 & 4 & 4.29 & 0.07 & 3.07 & 0.18 & 0.6 & -0.5 & 3.6 & 3.58 & 0.09 & 2.36 & 0.12 \\ \hline
        88 & dblp\_cite & 240 & 7.2 & 0.030 & 57 & 1.6 & 11.5 & 2.4 & 3.78 & 0.06 & 3.3 & 0.2 & 0.6 & -0.6 & 3.1 & 3.08 & 0.09 & 2.61 & 0.14 \\ \hline
        86 & sp\_infectious & 410 & 13.5 & 0.033 & 58 & 1.6 & 7.8 & 0.9 & 4.78 & 0.06 & 4.00 & 0.16 & 0.7 & -0.7 & 3.6 & 3.63 & 0.09 & 2.86 & 0.10 \\ \hline
        34 & adjnoun & 112 & 7.6 & 0.068 & 59 & 1.5 & 11.7 & 1.5 & 3.22 & 0.04 & 2.97 & 0.10 & 0.8 & -0.9 & 2.5 & 2.54 & 0.05 & 2.28 & 0.07 \\ \hline
        1 & new\_guinea\_tribes & 16 & 7.3 & 0.49 & 60 & 1.5 & 6.6 & 0.6 & 2.57 & 0.05 & 2.56 & 0.08 & 1.0 & -1 & 1.5 & 1.54 & 0.07 & 1.53 & 0.03 \\ \hline
        57 & wiki\_science & 677 & 19.3 & 0.029 & 61 & 1.5 & 9.9 & 0.5 & 4.50 & 0.03 & 3.91 & 0.07 & 0.8 & -0.9 & 3.4 & 3.43 & 0.04 & 2.84 & 0.04 \\ \hline
        15 & ceo\_club & 40 & 4.8 & 0.123 & 62 & 1.4 & 14 & 8 & 2.99 & 0.13 & 2.5 & 0.4 & 0.7 & -0.6 & 2.4 & 2.45 & 0.18 & 1.9 & 0.2 \\ \hline
        44 & interactome\_pdz & 161 & 2.6 & 0.0163 & 63 & 1.3 & 5.1 & 1.1 & 5.70 & 0.08 & 4.2 & 0.4 & 0.5 & -0.5 & 5.3 & 5.33 & 0.10 & 3.9 & 0.4 \\ \hline
        85 & chess & 5643 & 18.9 & 0.0033 & 64 & 1.3 & 12.67 & 0.27 & 4.541 & 0.009 & 4.46 & 0.03 & 0.8 & -0.8 & 3.6 & 3.603 & 0.012 & 3.52 & 0.02 \\ \hline
        62 & faa\_routes & 792 & 4.3 & 0.0054 & 65 & 1.3 & 4.13 & 0.29 & 6.07 & 0.05 & 5.8 & 0.2 & 0.6 & -0.6 & 5.2 & 5.21 & 0.06 & 5.0 & 0.2 \\ \hline
        111 & prosper & 3513 & 21.2 & 0.0060 & 66 & 1.3 & 60.7 & 2.7 & 3.510 & 0.007 & 3.00 & 0.02 & 0.7 & -0.8 & 2.9 & 2.858 & 0.010 & 2.348 & 0.015 \\ \hline
        79 & bitcoin\_trust & 4709 & 8.2 & 0.00174 & 67 & 1.3 & 35.2 & 1.9 & 3.846 & 0.007 & 3.40 & 0.04 & 0.5 & -0.7 & 3.5 & 3.512 & 0.009 & 3.06 & 0.03 \\ \hline
        66 & netscience & 379 & 4.8 & 0.0127 & 68 & 1.3 & 14 & 8 & 6.55 & 0.12 & 3.3 & 0.5 & 0.3 & -0.2 & 6 & 6.04 & 0.16 & 2.8 & 0.4 \\ \hline
        26 & highschool & 67 & 8 & 0.121 & 69 & 1.3 & 5 & 0.6 & 3.83 & 0.09 & 3.7 & 0.2 & 0.8 & -0.9 & 2.6 & 2.60 & 0.13 & 2.48 & 0.11 \\ \hline
        91 & foldoc & 13274 & 13.7 & 0.00103 & 70 & 1.3 & 8.12 & 0.16 & 5.044 & 0.011 & 5.61 & 0.04 & 0.5 & -0.7 & 3.9 & 3.872 & 0.015 & 4.44 & 0.03 \\ \hline
        68 & interactome\_stelzl & 1493 & 4 & 0.0027 & 71 & 1.3 & 6.2 & 0.4 & 5.23 & 0.02 & 5.11 & 0.14 & 0.5 & -0.6 & 4.8 & 4.78 & 0.03 & 4.66 & 0.12 \\ \hline
        20 & contiguous\_usa & 49 & 4.4 & 0.092 & 72 & 1.2 & 1.85 & 0.15 & 6.4 & 0.3 & 7.2 & 0.7 & 0.7 & -0.7 & 4.2 & 4.2 & 0.4 & 5.0 & 0.4 \\ \hline
        8 & zebras & 23 & 9.1 & 0.41 & 73 & 1.2 & 6.4 & 0.9 & 2.95 & 0.09 & 2.78 & 0.14 & 0.9 & -0.9 & 1.9 & 1.86 & 0.12 & 1.70 & 0.06 \\ \hline
        74 & interactome\_vidal & 2783 & 4.3 & 0.00155 & 74 & 1.2 & 7.9 & 0.4 & 5.299 & 0.016 & 4.92 & 0.10 & 0.5 & -0.6 & 4.8 & 4.84 & 0.02 & 4.46 & 0.09 \\ \hline
        33 & game\_thrones & 107 & 6.6 & 0.062 & 75 & 1.2 & 8.9 & 1.3 & 3.58 & 0.05 & 3.21 & 0.15 & 0.8 & -0.8 & 2.9 & 2.90 & 0.07 & 2.54 & 0.11 \\ \hline
        96 & word\_assoc & 7751 & 56.8 & 0.0073 & 76 & 1.2 & 111.5 & 3.4 & 3.361 & 0.005 & 2.91 & 0.01 & 0.8 & -0.9 & 2.5 & 2.550 & 0.007 & 2.096 & 0.007 \\ \hline
        105 & pgp\_strong & 39796 & 9.9 & 0.00025 & 77 & 1.2 & 12.45 & 0.32 & 6.086 & 0.008 & 5.25 & 0.04 & 0.4 & -0.4 & 5.5 & 5.498 & 0.010 & 4.67 & 0.04 \\ \hline
        22 & dolphins & 62 & 5.1 & 0.084 & 78 & 1.2 & 4.9 & 0.9 & 4.24 & 0.12 & 3.7 & 0.3 & 0.7 & -0.7 & 3.4 & 3.36 & 0.15 & 2.8 & 0.2 \\ \hline
        94 & marvel\_universe & 19182 & 10 & 0.00052 & 79 & 1.2 & 22.2 & 0.7 & 4.957 & 0.006 & 4.21 & 0.03 & 0.3 & -0.6 & 4.5 & 4.450 & 0.008 & 3.71 & 0.03 \\ \hline
        108 & slashdot\_threads & 16377 & 7.9 & 0.00048 & 80 & 1.2 & 24.7 & 0.5 & 4.388 & 0.003 & 4.05 & 0.02 & 0.6 & -0.8 & 4 & 3.971 & 0.004 & 3.637 & 0.017 \\ \hline
        76 & bitcoin\_alpha & 3235 & 8.2 & 0.0025 & 81 & 1.1 & 27.4 & 1.6 & 3.878 & 0.009 & 3.47 & 0.04 & 0.6 & -0.7 & 3.5 & 3.514 & 0.011 & 3.10 & 0.04 \\ \hline
        101 & linux & 913 & 8.8 & 0.0096 & 82 & 1.1 & 30 & 4 & 3.41 & 0.02 & 3.02 & 0.08 & 0.6 & -0.7 & 2.9 & 2.93 & 0.03 & 2.55 & 0.06 \\ \hline
        5 & moreno\_taro & 22 & 3.5 & 0.167 & 83 & 1.1 & 5.2 & 2.6 & 3.2 & 0.2 & 3.0 & 0.6 & 0.6 & -0.6 & 2.5 & 2.5 & 0.3 & 2.2 & 0.4 \\ \hline
        83 & advogato & 3140 & 20.5 & 0.0065 & 84 & 1.1 & 42.7 & 1.8 & 3.564 & 0.008 & 3.16 & 0.03 & 0.6 & -0.8 & 2.9 & 2.941 & 0.011 & 2.540 & 0.018 \\ \hline
        98 & movielens\_100k & 23761 & 6 & 0.00025 & 85 & 1.1 & 17.24 & 0.31 & 5.095 & 0.003 & 4.58 & 0.02 & 0.5 & -0.7 & 4.8 & 4.778 & 0.004 & 4.26 & 0.02 \\ \hline
        69 & bible\_nouns & 1707 & 10.6 & 0.0062 & 86 & 1.1 & 13.2 & 0.9 & 4.09 & 0.02 & 3.93 & 0.08 & 0.6 & -0.7 & 3.4 & 3.38 & 0.03 & 3.22 & 0.06 \\ \hline
        38 & revolution & 141 & 2.3 & 0.0164 & 87 & 1.1 & 33 & 16 & 3.30 & 0.03 & 2.4 & 0.2 & 0.5 & -0.5 & 3.2 & 3.23 & 0.03 & 2.4 & 0.2 \\ \hline
        100 & nematode\_mammal & 26197 & 4.5 & 0.00017 & 88 & 1.1 & 7.8 & 0.22 & 7.047 & 0.009 & 6.03 & 0.07 & 0.4 & -0.4 & 6.7 & 6.662 & 0.010 & 5.65 & 0.07 \\ \hline
        54 & london\_transport & 369 & 2.3 & 0.0063 & 89 & 1.1 & 1.251 & 0.027 & 17.2 & 0.4 & 24.2 & 1.9 & 0.5 & -0.5 & 13.7 & 13.7 & 0.5 & 20.7 & 1.6 \\ \hline
        99 & digg\_reply & 6746 & 9.8 & 0.00145 & 90 & 1.1 & 17.3 & 0.4 & 4.367 & 0.005 & 4.13 & 0.02 & 0.8 & -0.9 & 3.7 & 3.714 & 0.007 & 3.480 & 0.019 \\ \hline
        52 & marvel\_partnerships & 181 & 2.5 & 0.0139 & 91 & 1.1 & 2.6 & 0.4 & 8.65 & 0.18 & 6.6 & 0.9 & 0.4 & -0.4 & 7.9 & 7.9 & 0.2 & 5.9 & 0.8 \\ \hline
        41 & student\_cooperation & 141 & 3.6 & 0.026 & 92 & 1.1 & 1.57 & 0.11 & 9.2 & 0.4 & 11.4 & 1.4 & 0.6 & -0.5 & 6.5 & 6.5 & 0.6 & 8.7 & 1.0 \\ \hline
        70 & interactome\_yeast & 1458 & 2.7 & 0.00185 & 93 & 1.1 & 4.6 & 0.4 & 7.22 & 0.04 & 5.9 & 0.3 & 0.4 & -0.4 & 6.8 & 6.81 & 0.05 & 5.5 & 0.3 \\ \hline
        60 & uni\_email & 1133 & 9.6 & 0.0085 & 94 & 1.1 & 12.6 & 0.5 & 4.309 & 0.014 & 3.83 & 0.05 & 0.8 & -0.9 & 3.6 & 3.606 & 0.018 & 3.12 & 0.04 \\ \hline
        93 & escorts & 15810 & 4.9 & 0.00031 & 95 & 1.0 & 9.05 & 0.25 & 6.190 & 0.008 & 5.46 & 0.06 & 0.4 & -0.5 & 5.8 & 5.785 & 0.009 & 5.05 & 0.05 \\ \hline
        90 & chicago\_road & 12978 & 3.2 & 0.00025 & 96 & 1.0 & 5.7 & 2.1 & 42.58 & 0.14 & 6.5 & 1.2 & 0.1 & 0 & 42 & 42.0 & 0.2 & 5.9 & 1.1 \\ \hline
        97 & cora & 3991 & 8.3 & 0.0021 & 97 & 1.0 & 5.65 & 0.21 & 6.71 & 0.03 & 5.89 & 0.11 & 0.5 & -0.6 & 5.7 & 5.65 & 0.03 & 4.84 & 0.08 \\ \hline
        78 & power & 4941 & 2.7 & 0.00055 & 98 & 1.0 & 2.69 & 0.17 & 19.80 & 0.06 & 9.7 & 0.5 & 0.2 & -0.2 & 19 & 18.99 & 0.08 & 8.9 & 0.5 \\ \hline
        61 & euroroad & 1039 & 2.5 & 0.0024 & 99 & 1.0 & 1.299 & 0.031 & 21.5 & 0.3 & 25.3 & 1.9 & 0.4 & -0.3 & 18.4 & 18.4 & 0.4 & 22.1 & 1.7 \\ \hline
        106 & paris\_transportation & 11 & 1.8 & 0.180 & 100 & 0.0 & 12.915 & 6E-15 & 1.9 & 0 & 1.99 & 0 & 1.0 & -1 & 1.8 & 1.80 & 0 & 1.91 & 0 \\ \hline
    \end{tabular}
    \caption{Further information on 51 of the 112 networks taken from the \href{https://networks.skewed.de/}{Netzschleuder} repository \cite{P20}. These networks fit the closeness-degree fitting to the relationship \eqref{a:farnesskform} extremely well with a reduced $\chi^2$ between 1.0 and 2.0. Taken from the file \texttt{extra\_result220208tidy.xlsx} provided in \cite{EC22}.}
    \label{t:extraresult2}
\end{sidewaystable}

\begin{sidewaystable}[htpb]
\tiny
    \centering
    \begin{tabular}{|l|l|l|l|l|l|l|l|l|l|l|l|l|l|l|l|l|l|l|l|}
    \hline
        ID & Name & $N$ & $\texpect{k}$ & $\rho$ density & Rk & $\chi^2_r$ & $\zbarfit$ & $\zbarfit_\mathrm{err}$ & $\betafit$ & $\betafit_\mathrm{err}$ & $\beta$ & $\beta_\mathrm{err}$ & $\rho^\mathrm{P}$ &  $\rho^\mathrm{IP}$ & $\ellav$ & $\ellav^\mathrm{FP}$ & $\ellav^\mathrm{FP}_\mathrm{err}$ & $\ellav^\mathrm{zP}$ & $\ellav^\mathrm{zP}_\mathrm{err}$ \\ \hline         \hline
        2 & rhesus\_monkey & 16 & 8.6 & 0.57 & 101 & 0.0 & 8.7 & 0.9 & 2.39 & 0.05 & 2.35 & 0.07 & 1.0 & -1 & 1.4 & 1.43 & 0.06 & 1.39 & 0.02 \\ \hline
        9 & cattle & 20 & 12.1 & 0.64 & 102 & 0.0 & 5.44 & 0.31 & 2.82 & 0.05 & 2.87 & 0.06 & 1.0 & -1 & 1.4 & 1.36 & 0.07 & 1.419 & 0.015 \\ \hline
        4 & high\_tech\_company & 21 & 15.1 & 0.76 & 103 & 0.0 & 4.24 & 0.16 & 3.11 & 0.05 & 3.23 & 0.06 & 1.0 & -1 & 1.2 & 1.24 & 0.07 & 1.362 & 0.009 \\ \hline
        10 & moreno\_sheep & 22 & 13.2 & 0.63 & 104 & 0.0 & 11.4 & 1.5 & 2.41 & 0.06 & 2.33 & 0.08 & 1.0 & -1 & 1.4 & 1.37 & 0.08 & 1.297 & 0.019 \\ \hline
        40 & foodweb\_little\_rock & 22 & 13.2 & 0.63 & 105 & 0.0 & 7.4 & 0.6 & 2.64 & 0.05 & 2.63 & 0.07 & 1.0 & -1 & 1.4 & 1.37 & 0.08 & 1.359 & 0.017 \\ \hline
        7 & bison & 26 & 17.1 & 0.68 & 106 & 0.0 & 6.5 & 0.5 & 2.81 & 0.06 & 2.84 & 0.07 & 1.0 & -1 & 1.3 & 1.32 & 0.08 & 1.347 & 0.014 \\ \hline
        11 & 7th\_graders & 29 & 17.2 & 0.61 & 107 & 0.0 & 7.2 & 0.6 & 2.80 & 0.06 & 2.80 & 0.07 & 1.0 & -1 & 1.4 & 1.38 & 0.08 & 1.381 & 0.017 \\ \hline
        12 & college\_freshmen & 31 & 26.8 & 0.89 & 108 & 0.0 & 3.6 & 0.08 & 3.67 & 0.05 & 3.81 & 0.05 & 1.0 & -1 & 1.1 & 1.11 & 0.06 & 1.253 & 0.003 \\ \hline
        23 & macaques & 38 & 26.3 & 0.71 & 109 & 0.0 & 4.67 & 0.17 & 3.40 & 0.05 & 3.48 & 0.06 & 1.0 & -1 & 1.3 & 1.29 & 0.07 & 1.368 & 0.008 \\ \hline
        19 & sp\_kenyan\_households & 47 & 21.4 & 0.47 & 110 & 0.0 & 10.5 & 0.9 & 2.81 & 0.05 & 2.70 & 0.06 & 1.0 & -1 & 1.5 & 1.53 & 0.07 & 1.426 & 0.018 \\ \hline
        39 & email\_company & 126 & 49.3 & 0.39 & 111 & 0.0 & 24.6 & 3.2 & 2.77 & 0.05 & 2.54 & 0.06 & 1.0 & -0.9 & 1.6 & 1.61 & 0.07 & 1.371 & 0.017 \\ \hline
        45 & fao\_trade & 133 & 97.8 & 0.74 & 112 & 0.0 & 5.45 & 0.18 & 3.93 & 0.05 & 3.99 & 0.06 & 1.0 & -1 & 1.3 & 1.26 & 0.08 & 1.315 & 0.007 \\ \hline
    \end{tabular}
    \caption{Further information on 12 of the 112 networks taken from the \href{https://networks.skewed.de/}{Netzschleuder} repository \cite{P20}. Our automated code was able to download and run on these networks but in the form used here these give no insight into the closeness-degree relationship \eqref{a:farnesskform}. This can be seen by the $0.0$ for the reduced $\chi^2_r$ and the average path length is less than 2.0. Typically the information in the topology of the simple graph representation of the data used in our code is trivial and all the interest in these examples lies in weights on the edges. This table is provided only for completeness. Taken from the file \texttt{extra\_result220208tidy.xlsx} provided in \cite{EC22}.}
    \label{t:extraresult3}
\end{sidewaystable}



%

%
%


\clearpage

\begin{thebibliography}{10}
\expandafter\ifx\csname url\endcsname\relax
  \def\url#1{\texttt{#1}}\fi
\expandafter\ifx\csname urlprefix\endcsname\relax\def\urlprefix{URL }\fi
\providecommand{\bibinfo}[2]{#2}
\providecommand{\eprint}[1]{\href{http://arXiv.org/abs/#1}{\texttt{arXiv:#1}}}
\providecommand{\DOI}[1]{DOI: \url{#1}}

\bibitem{BRMW13}
\bibinfo{author}{Brandes, U.}, \bibinfo{author}{Robins, G.},
  \bibinfo{author}{McCranie, A.} \& \bibinfo{author}{Wasserman, S.}
\newblock \bibinfo{title}{What is network science?}
\newblock \emph{\bibinfo{journal}{Network Science}}
  \textbf{\bibinfo{volume}{1}}, \bibinfo{pages}{1--15} (\bibinfo{year}{2013}).
\newblock \DOI{http://doi.org/10.1017/nws.2013.2}.

\bibitem{B50}
\bibinfo{author}{Bavelas, A.}
\newblock \bibinfo{title}{Communication patterns in task-oriented groups}.
\newblock \emph{\bibinfo{journal}{The Journal of the Acoustical Society of
  America}} \textbf{\bibinfo{volume}{22}}, \bibinfo{pages}{725--730}
  (\bibinfo{year}{1950}).

\bibitem{S66}
\bibinfo{author}{Sabidussi, G.}
\newblock \bibinfo{title}{The centrality index of a graph}.
\newblock \emph{\bibinfo{journal}{Psychometrika}}
  \textbf{\bibinfo{volume}{31}}, \bibinfo{pages}{581--603}
  (\bibinfo{year}{1966}).

\bibitem{F78}
\bibinfo{author}{Freeman, L.~C.}
\newblock \bibinfo{title}{Centrality in social networks conceptual
  clarification}.
\newblock \emph{\bibinfo{journal}{Social networks}}
  \textbf{\bibinfo{volume}{1}}, \bibinfo{pages}{215--239}
  (\bibinfo{year}{1978}).

\bibitem{HH95}
\bibinfo{author}{Hage, P.} \& \bibinfo{author}{Harary, F.}
\newblock \bibinfo{title}{Eccentricity and centrality in networks}.
\newblock \emph{\bibinfo{journal}{Social Networks}}
  \textbf{\bibinfo{volume}{17}}, \bibinfo{pages}{57--63}
  (\bibinfo{year}{1995}).

\bibitem{WS03}
\bibinfo{author}{Wuchty, S.} \& \bibinfo{author}{Stadler, P.~F.}
\newblock \bibinfo{title}{Centers of complex networks}.
\newblock \emph{\bibinfo{journal}{Journal of Theoretical Biology}}
  \textbf{\bibinfo{volume}{223}}, \bibinfo{pages}{45--53}
  (\bibinfo{year}{2003}).

\bibitem{HK04}
\bibinfo{author}{Hahn, M.~W.} \& \bibinfo{author}{Kern, A.~D.}
\newblock \bibinfo{title}{Comparative genomics of centrality and essentiality
  in three eukaryotic protein-interaction networks}.
\newblock \emph{\bibinfo{journal}{Molecular Biology and Evolution}}
  \textbf{\bibinfo{volume}{22}}, \bibinfo{pages}{803--806}
  (\bibinfo{year}{2004}).

\bibitem{KS08}
\bibinfo{author}{Kosch\"{u}tzki, D.} \& \bibinfo{author}{Schreiber, F.}
\newblock \bibinfo{title}{Centrality analysis methods for biological networks
  and their application to gene regulatory networks}.
\newblock \emph{\bibinfo{journal}{Gene Regulation and Systems Biology}}
  \textbf{\bibinfo{volume}{2}}, \bibinfo{pages}{GRSB.S702}
  (\bibinfo{year}{2008}).

\bibitem{KB08}
\bibinfo{author}{Kiss, C.} \& \bibinfo{author}{Bichler, M.}
\newblock \bibinfo{title}{Identification of influencers {\textemdash} measuring
  influence in customer networks}.
\newblock \emph{\bibinfo{journal}{Decision Support Systems}}
  \textbf{\bibinfo{volume}{46}}, \bibinfo{pages}{233--253}
  (\bibinfo{year}{2008}).

\bibitem{YD09}
\bibinfo{author}{Yan, E.} \& \bibinfo{author}{Ding, Y.}
\newblock \bibinfo{title}{Applying centrality measures to impact analysis: A
  coauthorship network analysis}.
\newblock \emph{\bibinfo{journal}{Journal of the American Society for
  Information Science and Technology}} \textbf{\bibinfo{volume}{60}},
  \bibinfo{pages}{2107--2118} (\bibinfo{year}{2009}).

\bibitem{LFH10}
\bibinfo{author}{Landherr, A.}, \bibinfo{author}{Friedl, B.} \&
  \bibinfo{author}{Heidemann, J.}
\newblock \bibinfo{title}{A critical review of centrality measures in social
  networks}.
\newblock \emph{\bibinfo{journal}{Business \& Information Systems Engineering}}
  \textbf{\bibinfo{volume}{2}}, \bibinfo{pages}{371--385}
  (\bibinfo{year}{2010}).

\bibitem{NSJ11}
\bibinfo{author}{Ni, C.}, \bibinfo{author}{Sugimoto, C.} \&
  \bibinfo{author}{Jiang, J.}
\newblock \bibinfo{title}{Degree, closeness, and betweenness: Application of
  group centrality measurements to explore macro-disciplinary evolution
  diachronically}.
\newblock In Noyons, E., Ngulube, P., \& Leta, J. (eds.) \emph{\bibinfo{booktitle}{Proceedings of ISSI 2013}},
  \bibinfo{pages}{605} (\bibinfo{year}{2011}).

\bibitem{WMWJ11}
\bibinfo{author}{Wang, J.}, \bibinfo{author}{Mo, H.}, \bibinfo{author}{Wang,
  F.} \& \bibinfo{author}{Jin, F.}
\newblock \bibinfo{title}{Exploring the network structure and nodal centrality
  of {China's} air transport network: A complex network approach}.
\newblock \emph{\bibinfo{journal}{Journal of Transport Geography}}
  \textbf{\bibinfo{volume}{19}}, \bibinfo{pages}{712--721}
  (\bibinfo{year}{2011}).

\bibitem{BH14}
\bibinfo{author}{Brandes, U.} \& \bibinfo{author}{Hildenbrand, J.}
\newblock \bibinfo{title}{Smallest graphs with distinct singleton centers}.
\newblock \emph{\bibinfo{journal}{Network Science}}
  \textbf{\bibinfo{volume}{2}}, \bibinfo{pages}{416--418}
  (\bibinfo{year}{2014}).
\newblock \DOI{http://doi.org/10.1017/nws.2014.25}.

\bibitem{DSP18}
\bibinfo{author}{Das, K.}, \bibinfo{author}{Samanta, S.} \&
  \bibinfo{author}{Pal, M.}
\newblock \bibinfo{title}{Study on centrality measures in social networks: a
  survey}.
\newblock \emph{\bibinfo{journal}{Social Network Analysis and Mining}}
  \textbf{\bibinfo{volume}{8}} (\bibinfo{year}{2018}).

\bibitem{WF94}
\bibinfo{author}{Wasserman, S.} \& \bibinfo{author}{Faust, K.}
\newblock \emph{\bibinfo{title}{Social Network Analysis: Methods and
  Applications (Structural Analysis in the Social Sciences)}}
  (\bibinfo{publisher}{Cambridge University Press}, \bibinfo{year}{1994}).

\bibitem{N10}
\bibinfo{author}{Newman, M.}
\newblock \emph{\bibinfo{title}{Networks: an introduction}}
  (\bibinfo{publisher}{Oxford University Press}, \bibinfo{year}{2010}).

\bibitem{LNR17}
\bibinfo{author}{Latora, V.}, \bibinfo{author}{Nicosia, V.} \&
  \bibinfo{author}{Russo, G.}
\newblock \emph{\bibinfo{title}{Complex Networks: Principles, Methods and
  Applications}} (\bibinfo{publisher}{Cambridge University Press},
  \bibinfo{year}{2017}).

\bibitem{C21}
\bibinfo{author}{Coscia, M.}
\newblock \emph{\bibinfo{title}{The Atlas for the Aspiring Network Scientist}}
  (\bibinfo{publisher}{Michele Coscia}, \bibinfo{year}{2021}).
\newblock \eprint{2101.00863}.

\bibitem{MZ03a}
\bibinfo{author}{Ma, H.-W.} \& \bibinfo{author}{Zeng, A.-P.}
\newblock \bibinfo{title}{The connectivity structure, giant strong component
  and centrality of metabolic networks}.
\newblock \emph{\bibinfo{journal}{Bioinformatics}}
  \textbf{\bibinfo{volume}{19}}, \bibinfo{pages}{1423--1430}
  (\bibinfo{year}{2003}).

\bibitem{S15a}
\bibinfo{author}{Schoch, D.}
\newblock \emph{\bibinfo{title}{A Positional Approach for Network Centrality}}.
\newblock Ph.D. thesis, \bibinfo{school}{Universit\"{a}t Konstanz}
  (\bibinfo{year}{2015}).

\bibitem{S16}
\bibinfo{author}{Schoch, D.}
\newblock \bibinfo{title}{Periodic table of network centrality}
  (\bibinfo{year}{2016}).
\\ \newblock \urlprefix\url{http://schochastics.net/sna/periodic.html}.

\bibitem{B88}
\bibinfo{author}{Bolland, J.~M.}
\newblock \bibinfo{title}{Sorting out centrality: An analysis of the
  performance of four centrality models in real and simulated networks}.
\newblock \emph{\bibinfo{journal}{Social networks}}
  \textbf{\bibinfo{volume}{10}}, \bibinfo{pages}{233--253}
  (\bibinfo{year}{1988}).

\bibitem{RPWDMK95}
\bibinfo{author}{Rothenberg, R.~B.} \emph{et~al.}
\newblock \bibinfo{title}{Choosing a centrality measure: epidemiologic
  correlates in the colorado springs study of social networks}.
\newblock \emph{\bibinfo{journal}{Social Networks}}
  \textbf{\bibinfo{volume}{17}}, \bibinfo{pages}{273--297}
  (\bibinfo{year}{1995}).

\bibitem{F97}
\bibinfo{author}{Faust, K.}
\newblock \bibinfo{title}{Centrality in affiliation networks}.
\newblock \emph{\bibinfo{journal}{Social networks}}
  \textbf{\bibinfo{volume}{19}}, \bibinfo{pages}{157--191}
  (\bibinfo{year}{1997}).

\bibitem{L06b}
\bibinfo{author}{Lee, C.-Y.}
\newblock \bibinfo{title}{Correlations among centrality measures in complex
  networks}.
\newblock \emph{\bibinfo{journal}{arXiv preprint physics/0605220}}
  (\bibinfo{year}{2006}).
\newblock \eprint{physics/0605220}.

\bibitem{VCLC08}
\bibinfo{author}{Valente, T.~W.}, \bibinfo{author}{Coronges, K.},
  \bibinfo{author}{Lakon, C.} \& \bibinfo{author}{Costenbader, E.}
\newblock \bibinfo{title}{How correlated are network centrality measures?}
\newblock \emph{\bibinfo{journal}{Connections (Toronto, Ont.)}}
  \textbf{\bibinfo{volume}{28}}, \bibinfo{pages}{16} (\bibinfo{year}{2008}).
\\ \newblock
  \urlprefix\url{https://www.ncbi.nlm.nih.gov/pmc/articles/PMC2875682/}.

\bibitem{BN14}
\bibinfo{author}{Batool, K.} \& \bibinfo{author}{Niazi, M.~A.}
\newblock \bibinfo{title}{Towards a methodology for validation of centrality
  measures in complex networks}.
\newblock \emph{\bibinfo{journal}{PloS one}} \textbf{\bibinfo{volume}{9}},
  \bibinfo{pages}{e90283} (\bibinfo{year}{2014}).

\bibitem{LLBM15}
\bibinfo{author}{Lozares, C.}, \bibinfo{author}{L{\'o}pez-Rold{\'a}n, P.},
  \bibinfo{author}{Bolibar, M.} \& \bibinfo{author}{Muntanyola, D.}
\newblock \bibinfo{title}{The structure of global centrality measures}.
\newblock \emph{\bibinfo{journal}{International Journal of Social Research
  Methodology}} \textbf{\bibinfo{volume}{18}}, \bibinfo{pages}{209--226}
  (\bibinfo{year}{2015}).

\bibitem{SVB17}
\bibinfo{author}{Schoch, D.}, \bibinfo{author}{Valente, T.~W.} \&
  \bibinfo{author}{Brandes, U.}
\newblock \bibinfo{title}{Correlations among centrality indices and a class of
  uniquely ranked graphs}.
\newblock \emph{\bibinfo{journal}{Social Networks}}
  \textbf{\bibinfo{volume}{50}}, \bibinfo{pages}{46--54}
  (\bibinfo{year}{2017}).

\bibitem{OFPASF19}
\bibinfo{author}{Oldham, S.} \emph{et~al.}
\newblock \bibinfo{title}{Consistency and differences between centrality
  measures across distinct classes of networks}.
\newblock \emph{\bibinfo{journal}{{PLOS} {ONE}}} \textbf{\bibinfo{volume}{14}},
  \bibinfo{pages}{e0220061} (\bibinfo{year}{2019}).

\bibitem{BEEKSWWS19}
\bibinfo{author}{Bringmann, L.~F.} \emph{et~al.}
\newblock \bibinfo{title}{What do centrality measures measure in psychological
  networks?}
\newblock \emph{\bibinfo{journal}{Journal of Abnormal Psychology}}
  \textbf{\bibinfo{volume}{128}}, \bibinfo{pages}{892--903}
  (\bibinfo{year}{2019}).

\bibitem{APB20}
\bibinfo{author}{Arnaudon, A.}, \bibinfo{author}{Peach, R.~L.} \&
  \bibinfo{author}{Barahona, M.}
\newblock \bibinfo{title}{Scale-dependent measure of network centrality from
  diffusion dynamics}.
\newblock \emph{\bibinfo{journal}{Physical Review Research}}
  \textbf{\bibinfo{volume}{2}}, \bibinfo{pages}{033104} (\bibinfo{year}{2020}).

\bibitem{OAS10}
\bibinfo{author}{Opsahl, T.}, \bibinfo{author}{Agneessens, F.} \&
  \bibinfo{author}{Skvoretz, J.}
\newblock \bibinfo{title}{Node centrality in weighted networks: Generalizing
  degree and shortest paths}.
\newblock \emph{\bibinfo{journal}{Social Networks}}
  \textbf{\bibinfo{volume}{32}}, \bibinfo{pages}{245--251}
  (\bibinfo{year}{2010}).

\bibitem{H59}
\bibinfo{author}{Harary, F.}
\newblock \bibinfo{title}{Status and contrastatus}.
\newblock \emph{\bibinfo{journal}{Sociometry}} \textbf{\bibinfo{volume}{22}},
  \bibinfo{pages}{23} (\bibinfo{year}{1959}).

\bibitem{S21b}
\bibinfo{author}{\u{S}ubelj, L.}
\newblock \bibinfo{title}{Algorithms for spanning trees of unweighted
  networks}.
\newblock \bibinfo{type}{Tech. Rep.}, \bibinfo{institution}{University of
  Ljubljana} (\bibinfo{year}{2021}).

\bibitem{ER59}
\bibinfo{author}{Erd\H{o}s, P.} \& \bibinfo{author}{R\'{e}yni, A.}
\newblock \bibinfo{title}{On random graphs. i}.
\newblock \emph{\bibinfo{journal}{Publicationes Mathematicae}}
  \textbf{\bibinfo{volume}{6}}, \bibinfo{pages}{290--297}
  (\bibinfo{year}{1959}).
\newblock \urlprefix\url{http://www.renyi.hu/~p_erdos/1959-11.pdf}.

\bibitem{BA99}
\bibinfo{author}{Barab\'{a}si, A.-L.} \& \bibinfo{author}{Albert, R.}
\newblock \bibinfo{title}{Emergence of scaling in random networks}.
\newblock \emph{\bibinfo{journal}{Science}} \textbf{\bibinfo{volume}{286}},
  \bibinfo{pages}{173} (\bibinfo{year}{1999}).

\bibitem{MR95}
\bibinfo{author}{Molloy, M.} \& \bibinfo{author}{Reed, B.}
\newblock \bibinfo{title}{A critical point for random graphs with a given
  degree sequence}.
\newblock \emph{\bibinfo{journal}{Random Structures and Algorithms}}
  \textbf{\bibinfo{volume}{6}}, \bibinfo{pages}{161--180}
  (\bibinfo{year}{1995}).
\newblock {citeseer.ist.psu.edu/molloy95critical.html}.

\bibitem{HSS08}
\bibinfo{author}{Hagberg, A.~A.}, \bibinfo{author}{Schult, D.~A.} \&
  \bibinfo{author}{Swart, P.~J.}
\newblock \bibinfo{title}{Exploring network structure, dynamics, and function
  using networkx}.
\newblock In \bibinfo{editor}{Varoquaux, G.}, \bibinfo{editor}{Vaught, T.} \&
  \bibinfo{editor}{Millman, J.} (eds.) \emph{\bibinfo{booktitle}{Proceedings of
  the 7th Python in Science Conference (SciPy2008)}}, \bibinfo{pages}{11--15}
  (\bibinfo{year}{2008}).

\bibitem{pajek}
\bibinfo{author}{Batagelj, V.}
\newblock \bibinfo{title}{Pajek datasets}.
\newblock
  \bibinfo{howpublished}{\url{http://vlado.fmf.uni-lj.si/pub/networks/data/}}
  (\bibinfo{year}{2017}).

\bibitem{KONECT}
\bibinfo{author}{Kunegis, J.}
\newblock \bibinfo{title}{The {KONECT} project}.
\newblock \bibinfo{howpublished}{\url{http://konect.cc/}}.

\bibitem{SNAP}
\bibinfo{author}{Leskovec, J.} \& \bibinfo{author}{Krevl, A.}
\newblock \bibinfo{title}{{SNAP Datasets}: {Stanford} large network dataset
  collection}.
\newblock \bibinfo{howpublished}{\url{http://snap.stanford.edu/data}}
  (\bibinfo{year}{2014}).

\bibitem{P20}
\bibinfo{author}{Peixoto, T.~P.}
\newblock \bibinfo{title}{The {Netzschleuder} network catalogue and
  repository}.
\\ \newblock \urlprefix\url{https://networks.skewed.de/}.

\bibitem{WS98}
\bibinfo{author}{Watts, D.~J.} \& \bibinfo{author}{Strogatz, S.~H.}
\newblock \bibinfo{title}{Collective dynamics of `small-world' networks.}
\newblock \emph{\bibinfo{journal}{Nature}} \textbf{\bibinfo{volume}{393}},
  \bibinfo{pages}{440--442} (\bibinfo{year}{1998}).
\newblock \DOI{http://doi.org/10.1038/30918}.

\bibitem{CL02}
\bibinfo{author}{Chung, F.} \& \bibinfo{author}{Lu, L.}
\newblock \bibinfo{title}{The average distances in random graphs with given
  expected degrees}.
\newblock \emph{\bibinfo{journal}{Proceedings of the National Academy of
  Sciences}} \textbf{\bibinfo{volume}{99}}, \bibinfo{pages}{15879--15882}
  (\bibinfo{year}{2002}).

\bibitem{DMS03a}
\bibinfo{author}{Dorogovtsev, S.~N.}, \bibinfo{author}{Mendes, J. F.~F.} \&
  \bibinfo{author}{Samukhin, A.~N.}
\newblock \bibinfo{title}{Metric structure of random networks}.
\newblock \emph{\bibinfo{journal}{Nuclear Physics B}}
  \textbf{\bibinfo{volume}{653}}, \bibinfo{pages}{307--338}
  (\bibinfo{year}{2003}).

\bibitem{BL06a}
\bibinfo{author}{Baronchelli, A.} \& \bibinfo{author}{Loreto, V.}
\newblock \bibinfo{title}{Ring structures and mean first passage time in
  networks}.
\newblock \emph{\bibinfo{journal}{Physical Review E}}
  \textbf{\bibinfo{volume}{73}}, \bibinfo{pages}{026103}
  (\bibinfo{year}{2006}).

\bibitem{BGHJ07}
\bibinfo{author}{Blondel, V.~D.}, \bibinfo{author}{Guillaume, J.-L.},
  \bibinfo{author}{Hendrickx, J.~M.} \& \bibinfo{author}{Jungers, R.~M.}
\newblock \bibinfo{title}{Distance distribution in random graphs and
  application to network exploration}.
\newblock \emph{\bibinfo{journal}{Physical Review E}}
  \textbf{\bibinfo{volume}{76}}, \bibinfo{pages}{066101}
  (\bibinfo{year}{2007}).

\bibitem{BR04}
\bibinfo{author}{Bollob\'{a}s, B.} \& \bibinfo{author}{Riordan, O.}
\newblock \bibinfo{title}{The diameter of a scale-free random graph}.
\newblock \emph{\bibinfo{journal}{Combinatorica}}
  \textbf{\bibinfo{volume}{24}}, \bibinfo{pages}{5--34} (\bibinfo{year}{2004}).

\bibitem{ECV20}
\bibinfo{author}{Evans, T.}, \bibinfo{author}{Calmon, L.} \&
  \bibinfo{author}{Vasiliauskaite, V.}
\newblock \bibinfo{title}{The longest path in the {Price} model}.
\newblock \emph{\bibinfo{journal}{Scientific Reports}}
  \textbf{\bibinfo{volume}{10}}, \bibinfo{pages}{10503} (\bibinfo{year}{2020}).
\newblock \eprint{1903.03667}.

\bibitem{CH03}
\bibinfo{author}{Cohen, R.} \& \bibinfo{author}{Havlin, S.}
\newblock \bibinfo{title}{Scale-free networks are ultrasmall}.
\newblock \emph{\bibinfo{journal}{Physical Review Letters}}
  \textbf{\bibinfo{volume}{90}}, \bibinfo{pages}{058701}
  (\bibinfo{year}{2003}).

\bibitem{WS03a}
\bibinfo{author}{White, S.} \& \bibinfo{author}{Smyth, P.}
\newblock \bibinfo{title}{Algorithms for estimating relative importance in
  networks}.
\newblock In \emph{\bibinfo{booktitle}{Proceedings of the ninth {ACM} {SIGKDD}
  international conference on Knowledge discovery and data mining - {KDD}
  {\textquotesingle}03}} (\bibinfo{publisher}{{ACM} Press}).

\bibitem{FLAOYMEC20}
\bibinfo{author}{Falkenberg, M.} \emph{et~al.}
\newblock \bibinfo{title}{Identifying time dependence in network growth}.
\newblock \emph{\bibinfo{journal}{{Physical Review Research}}}
  \textbf{\bibinfo{volume}{2}}, \bibinfo{pages}{023352} (\bibinfo{year}{2020}).
\newblock \eprint{2001.09118}.

\bibitem{ZMS20}
\bibinfo{author}{Zhou, B.}, \bibinfo{author}{Meng, X.} \&
  \bibinfo{author}{Stanley, H.~E.}
\newblock \bibinfo{title}{Power-law distribution of degree-degree distance: A
  better representation of the scale-free property of complex networks}.
\newblock \emph{\bibinfo{journal}{Proceedings of the National Academy of
  Sciences}} \textbf{\bibinfo{volume}{117}}, \bibinfo{pages}{14812--14818}
  (\bibinfo{year}{2020}).

\bibitem{BDL22}
\bibinfo{author}{Babul, S.}, \bibinfo{author}{Devriendt, K.} \&
  \bibinfo{author}{Lambiotte, R.}
\newblock \bibinfo{title}{Gromov centrality: A multi-scale measure of network
  centrality using triangle inequality excess}.
\newblock \bibinfo{type}{Tech. Rep.} (\bibinfo{year}{2022}).
\newblock \eprint{2205.04974}.

\bibitem{EC22}
\bibinfo{author}{Evans, T.~S.} \& \bibinfo{author}{Chen, B.}
\newblock \bibinfo{title}{Linking the network centrality measures closeness and
  degree: Additional data.}
\newblock \bibinfo{howpublished}{figshare}.
\newblock \DOI{https://doi.org/10.6084/m9.figshare.19216812}.

\bibitem{Z77}
\bibinfo{author}{Zachary, W.}
\newblock \bibinfo{title}{Information-flow model for conflict and fission in
  small-groups}.
\newblock \emph{\bibinfo{journal}{Journal Of Anthropological Research}}
  \textbf{\bibinfo{volume}{33}}, \bibinfo{pages}{452---473}
  (\bibinfo{year}{1977}).

\bibitem{BE06}
\bibinfo{author}{Borgatti, S.~P.} \& \bibinfo{author}{Everett, M.~G.}
\newblock \bibinfo{title}{A graph-theoretic perspective on centrality}.
\newblock \emph{\bibinfo{journal}{Social Networks}}
  \textbf{\bibinfo{volume}{28}}, \bibinfo{pages}{466--484}
  (\bibinfo{year}{2006}).

\bibitem{M67}
\bibinfo{author}{Milgram, S.}
\newblock \bibinfo{title}{The small world problem}.
\newblock \emph{\bibinfo{journal}{Psychology Today}}  (\bibinfo{year}{1967}).

\bibitem{TM69}
\bibinfo{author}{Travers, J.} \& \bibinfo{author}{Milgram, S.}
\newblock \bibinfo{title}{An experimental study of the small world problem}.
\newblock \emph{\bibinfo{journal}{Sociometry}} \textbf{\bibinfo{volume}{32}},
  \bibinfo{pages}{425} (\bibinfo{year}{1969}).

\bibitem{LH08}
\bibinfo{author}{Leskovec, J.} \& \bibinfo{author}{Horvitz, E.}
\newblock \bibinfo{title}{Planetary-scale views on an instant-messaging
  network}.
\newblock \bibinfo{type}{Tech. Rep.}, \bibinfo{institution}{Microsoft Research}
  (\bibinfo{year}{2007}).
\newblock \eprint{0803.0939v1}.

\bibitem{BBRUV12}
\bibinfo{author}{Backstrom, L.}, \bibinfo{author}{Boldi, P.},
  \bibinfo{author}{Rosa, M.}, \bibinfo{author}{Ugander, J.} \&
  \bibinfo{author}{Vigna, S.}
\newblock \bibinfo{title}{Four degrees of separation}.
\newblock In \emph{\bibinfo{booktitle}{Proceedings of the 3rd Annual {ACM} Web
  Science Conference on - {WebSci} {\textquotesingle}12}}
  (\bibinfo{publisher}{{ACM} Press}, \bibinfo{year}{2012}).

\bibitem{BV12}
\bibinfo{author}{Boldi, P.} \& \bibinfo{author}{Vigna, S.}
\newblock \bibinfo{title}{Four degrees of separation, really}.
\newblock In \emph{\bibinfo{booktitle}{2012 {IEEE}/{ACM} International
  Conference on Advances in Social Networks Analysis and Mining}}
  (\bibinfo{publisher}{{IEEE}}, \bibinfo{year}{2012}).

\bibitem{B03c}
\bibinfo{author}{Bollob\'{a}s, B.}
\newblock \bibinfo{title}{Mathematical results on scale-free random graphs}.
\newblock In \emph{\bibinfo{booktitle}{Handbook of Graphs and Networks}},
  \bibinfo{pages}{1--37} (\bibinfo{publisher}{Wiley}, \bibinfo{year}{2003}).

\bibitem{MPL17}
\bibinfo{author}{Masuda, N.}, \bibinfo{author}{Porter, M.~A.} \&
  \bibinfo{author}{Lambiotte, R.}
\newblock \bibinfo{title}{Random walks and diffusion on networks}.
\newblock \emph{\bibinfo{journal}{Physics Reports}}  (\bibinfo{year}{2017}).

\bibitem{K13c}
\bibinfo{author}{Kunegis, J.}
\newblock \bibinfo{title}{{KONECT} -- {The} {Koblenz} {Network} {Collection}}.
\newblock In \emph{\bibinfo{booktitle}{Proc. Int. Conf. on World Wide Web
  Companion}}, \bibinfo{pages}{1343--1350} (\bibinfo{year}{2013}).
\\ \newblock \DOI{https://doi.org/10.1145/2487788.2488173}.

\bibitem{C64}
\bibinfo{author}{Coleman, J.~S.}
\newblock \emph{\bibinfo{title}{Introduction to Mathematical Sociology}}
  (\bibinfo{publisher}{London Free Press Glencoe}, \bibinfo{year}{1964}).

\bibitem{GD03}
\bibinfo{author}{Gleiser, P.~M.} \& \bibinfo{author}{Danon, L.}
\newblock \bibinfo{title}{{Community} {Strucure} {in} {Jazz}}.
\newblock \emph{\bibinfo{journal}{Advances in Complex Systems}}
  \textbf{\bibinfo{volume}{06}}, \bibinfo{pages}{565--573}
  (\bibinfo{year}{2003}).
\newblock \urlprefix\url{https://doi.org/10.1142/S0219525903001067}.

\bibitem{FWK98}
\bibinfo{author}{Freeman, L.~C.}, \bibinfo{author}{Webster, C.~M.} \&
  \bibinfo{author}{Kirke, D.~M.}
\newblock \bibinfo{title}{Exploring social structure using dynamic
  three-dimensional color images}.
\newblock \emph{\bibinfo{journal}{Social Networks}}
  \textbf{\bibinfo{volume}{20}}, \bibinfo{pages}{109--118}
  (\bibinfo{year}{1998}).

\bibitem{M01}
\bibinfo{author}{Moody, J.}
\newblock \bibinfo{title}{Peer influence groups: Identifying dense clusters in
  large networks}.
\newblock \emph{\bibinfo{journal}{Soc. Netw.}} \textbf{\bibinfo{volume}{23}},
  \bibinfo{pages}{261--283} (\bibinfo{year}{2001}).

\bibitem{GDDGA03}
\bibinfo{author}{Guimer{\`{a}}, R.}, \bibinfo{author}{Danon, L.},
  \bibinfo{author}{D{\'{\i}}az-Guilera, A.}, \bibinfo{author}{Giralt, F.} \&
  \bibinfo{author}{Arenas, A.}
\newblock \bibinfo{title}{Self-similar community structure in a network of
  human interactions}.
\newblock \emph{\bibinfo{journal}{Physical Review E}}
  \textbf{\bibinfo{volume}{68}}, \bibinfo{pages}{065103}
  (\bibinfo{year}{2003}).

\bibitem{OP09}
\bibinfo{author}{Opsahl, T.} \& \bibinfo{author}{Panzarasa, P.}
\newblock \bibinfo{title}{Clustering in weighted networks}.
\newblock \emph{\bibinfo{journal}{Social Networks}}
  \textbf{\bibinfo{volume}{31}}, \bibinfo{pages}{155--163}
  (\bibinfo{year}{2009}).

\bibitem{LKF07}
\bibinfo{author}{Leskovec, J.}, \bibinfo{author}{Kleinberg, J.} \&
  \bibinfo{author}{Faloutsos, C.}
\newblock \bibinfo{title}{Graph evolution: Densification and shrinking
  diameters}.
\newblock \emph{\bibinfo{journal}{{ACM} Transactions on Knowledge Discovery
  from Data}} \textbf{\bibinfo{volume}{1}}, \bibinfo{pages}{2}
  (\bibinfo{year}{2007}).

\bibitem{CSJS09}
\bibinfo{author}{Choudhury, M.~D.}, \bibinfo{author}{Sundaram, H.},
  \bibinfo{author}{John, A.} \& \bibinfo{author}{Seligmann, D.~D.}
\newblock \bibinfo{title}{Social synchrony: Predicting mimicry of user actions
  in online social media}.
\newblock In \emph{\bibinfo{booktitle}{Proc. Int. Conf. on Comput. Science and
  Engineering}}, \bibinfo{pages}{151--158} (\bibinfo{year}{2009}).

\bibitem{L02}
\bibinfo{author}{Ley, M.}
\newblock \bibinfo{title}{The {DBLP} computer science bibliography: Evolution,
  research issues, perspectives}.
\newblock In \emph{\bibinfo{booktitle}{Proc. Int. Symposium on String Process.
  and Inf. Retr.}}, \bibinfo{pages}{1--10} (\bibinfo{year}{2002}).

\bibitem{MNRS00}
\bibinfo{author}{McCallum, A.~K.}, \bibinfo{author}{Nigam, K.},
  \bibinfo{author}{Rennie, J.} \& \bibinfo{author}{Seymore, K.}
\newblock \bibinfo{title}{Automating the construction of internet portals with
  machine learning}.
\newblock \emph{\bibinfo{journal}{Information Retrieval}}
  \textbf{\bibinfo{volume}{3}}, \bibinfo{pages}{127--163}
  (\bibinfo{year}{2000}).

\bibitem{SB13}
\bibinfo{author}{{\v S}ubelj, L.} \& \bibinfo{author}{Bajec, M.}
\newblock \bibinfo{title}{Model of complex networks based on citation
  dynamics}.
\newblock In \emph{\bibinfo{booktitle}{Proc. of the {WWW} Workshop on Large
  Scale Network Analysis}}, \bibinfo{pages}{527--530} (\bibinfo{year}{2013}).

\bibitem{N01}
\bibinfo{author}{Newman, M. E.~J.}
\newblock \bibinfo{title}{The structure of scientific collaboration networks}.
\newblock \emph{\bibinfo{journal}{Proc. Natl. Acad. Sci. U.S.A.}}
  \textbf{\bibinfo{volume}{98}}, \bibinfo{pages}{404--409}
  (\bibinfo{year}{2001}).

\bibitem{N06}
\bibinfo{author}{Newman, M. E.~J.}
\newblock \bibinfo{title}{Finding community structure in networks using the
  eigenvectors of matrices}.
\newblock \emph{\bibinfo{journal}{Phys. Rev. E}} \textbf{\bibinfo{volume}{74}}
  (\bibinfo{year}{2006}).

\bibitem{AG05}
\bibinfo{author}{Adamic, L.~A.} \& \bibinfo{author}{Glance, N.}
\newblock \bibinfo{title}{The political blogosphere and the 2004 us election:
  divided they blog}.
\newblock In \emph{\bibinfo{booktitle}{Proceedings of the 3rd international
  workshop on Link discovery}}, \bibinfo{pages}{36--43}
  (\bibinfo{organization}{ACM}, \bibinfo{year}{2005}).

\bibitem{P14b}
\bibinfo{author}{Peixoto, T.~P.}
\newblock \bibinfo{title}{The graph-tool python library}
  (\bibinfo{year}{2014}).
\\ \newblock \urlprefix\url{http://figshare.com/articles/graph_tool/1164194}.

\end{thebibliography}
\end{document}